\pdfoutput=1

\documentclass[11pt,twoside,a4paper,cmspaper,final,collab]{cms-tdr}
\def\svnVersion{433656P}\def\cmsCernNoTag{CERN-EP-2017-061}\def\cmsCernDate{\today}\def\cmsMessage{Published in the European Physical Journal C as \href{http://dx.doi.org/10.1140/epjc/s10052-017-5286-7}{\doi{10.1140/epjc/s10052-017-5286-7.}}}

\begin{document}\cmsNoteHeader{SMP-16-011}

\hyphenation{had-ron-i-za-tion}
\hyphenation{cal-or-i-me-ter}
\hyphenation{de-vices}
\RCS$Revision: 433475 $
\RCS$HeadURL: svn+ssh://svn.cern.ch/reps/tdr2/papers/SMP-16-011/trunk/SMP-16-011.tex $
\RCS$Id: SMP-16-011.tex 433475 2017-11-09 05:06:38Z krabbert $
\newlength\cmsFigWidth
\ifthenelse{\boolean{cms@external}}{\setlength\cmsFigWidth{0.48\textwidth}}{\setlength\cmsFigWidth{0.6\textwidth}}
\ifthenelse{\boolean{cms@external}}{\providecommand{\cmsLeft}{top}}{\providecommand{\cmsLeft}{left}}
\ifthenelse{\boolean{cms@external}}{\providecommand{\cmsRight}{bottom}}{\providecommand{\cmsRight}{right}}
\providecommand{\as}{\ensuremath{\alpS}\xspace}
\providecommand{\asmz}{\ensuremath{\alpS(M_\cPZ)}\xspace}
\providecommand{\asq}{\ensuremath{\alpS(Q)}\xspace}
\providecommand{\chisq}{\ensuremath{\chi^2}\xspace}
\providecommand{\chisqndof}{\ensuremath{\chi^2/n_\text{dof}}\xspace}
\providecommand{\chipsq}{\ensuremath{\chi^2_\text{p}}\xspace}
\providecommand{\chipsqndata}{\ensuremath{\chi^2_\text{p}/n_\text{data}}\xspace}
\providecommand{\ndata}{\ensuremath{n_{\mathrm{data}}}\xspace}
\providecommand{\ndof}{\ensuremath{n_{\mathrm{dof}}}\xspace}
\providecommand{\herafitter} {{\textsc{HERAFitter}}\xspace}
\providecommand{\xfitter}{{\textsc{xFitter}}\xspace}
\providecommand{\herwig}{{\textsc{herwig}}\xspace}
\providecommand{\herwigpp}{{\textsc{herwig++}}\xspace}
\providecommand{\herwigs}{{\textsc{herwig\,7}}\xspace}
\providecommand{\nlojetpp}{{\textsc{NLOJet++}}\xspace}
\providecommand{\fastnlo}{{\textsc{fastNLO}}\xspace}
\providecommand{\fastjet}{{\textsc{FastJet}}\xspace}
\providecommand{\GeVsq}{\ensuremath{\GeV^2}\xspace}
\providecommand{\GRV} {{\textsc{GRV}}\xspace}
\providecommand{\HOPPET}{{\textsc{HOPPET}}\xspace}
\providecommand{\LHAPDF}{{\textsc{LHAPDF}}\xspace}
\providecommand{\powhegbox}{{\textsc{powheg box}}\xspace}
\providecommand{\powheg}{{\textsc{powheg}}\xspace}
\providecommand{\pythia}{{\textsc{pythia}}\xspace}
\providecommand{\pythias}{{\textsc{pythia\,6}}\xspace}
\providecommand{\pythiae}{{\textsc{pythia\,8}}\xspace}
\providecommand{\powhegpluspythiae}{{\textsc{powheg}+\textsc{pythia\,8}}\xspace}
\providecommand{\QCDNUM}{{\textsc{QCDNUM}}\xspace}
\providecommand{\RooUnfold}{{\textsc{RooUnfold}}\xspace}
\providecommand{\mykt}{\kt}
\providecommand{\ptmax}{\ensuremath{p_{\mathrm{T,max}}}\xspace}
\providecommand{\avept}{\ensuremath{\langle p_\text{T1,2}\rangle}\xspace}
\providecommand{\etaabs}{\ensuremath{\abs{\eta}}\xspace}
\providecommand{\Ztwostar}{\ensuremath{\mathrm{Z2}^\star}\xspace}

\providecommand{\ptavg}{\ensuremath{p_{\mathrm{T,avg}}}\xspace}
\providecommand{\ptone}{\ensuremath{p_{\mathrm{T},1}}\xspace}
\providecommand{\pttwo}{\ensuremath{p_{\mathrm{T},2}}\xspace}
\providecommand{\ystar}{\ensuremath{y^{*}}\xspace}
\providecommand{\yboost}{\ensuremath{y_{\mathrm{b}}}\xspace}

\providecommand{\yabs}{\ensuremath{\abs{y}}\xspace}
\providecommand{\mun}{\ensuremath{\mu_0}\xspace}
\providecommand{\mur}{\ensuremath{\mu_\text{r}}\xspace}
\providecommand{\muf}{\ensuremath{\mu_\text{f}}\xspace}
\providecommand{\rbthm}{\rule[-2ex]{0ex}{5ex}}
\providecommand{\rbtrr}{\rule[-0.8ex]{0ex}{3.2ex}}

\providecommand{\epp}{\ensuremath{\text{e}^{+}\text{p}}\xspace}
\providecommand{\emp}{\ensuremath{\text{e}^{-}\text{p}}\xspace}

\newcommand{\x}{\ensuremath{\phantom{0}}}
\newcommand{\xx}{\ensuremath{\phantom{00}}}

\newcommand{\SPPS}{Sp$\bar{\text{p}}$S\xspace}
\newcommand{\hftwo}{\hspace*{\fill}}
\newcommand\todo[1]{\par\textcolor{red}{TODO: #1}}
\newcommand{\overbar}[1]{\mkern 1.5mu\overline{\mkern-1.5mu#1\mkern-1.5mu}\mkern 1.5mu}

\ifdefined\pdfsuppresswarningpagegroup
\pdfsuppresswarningpagegroup=1
\fi

\cmsNoteHeader{SMP-16-011}%
\title{Measurement of the triple-differential dijet cross section in
proton-proton collisions at $\sqrt{s}=8\TeV$ and constraints on
parton distribution functions}

\titlerunning{Triple-differential dijet cross section in pp collisions at 8 TeV and constraints on parton distributions functions}

\date{\today}

\abstract{A measurement is presented of the triple-differential dijet cross section at
a centre-of-mass energy of 8\TeV using
19.7\fbinv of data collected with the CMS detector in
proton-proton collisions at the LHC\@. The cross section is
measured as a function of the average transverse momentum, half the
rapidity separation, and the boost of the two leading jets in the event. The
cross section is corrected for detector effects and compared to
calculations in perturbative quantum chromodynamics at
next-to-leading order accuracy, complemented with electroweak and
nonperturbative corrections. New constraints on parton distribution
functions are obtained and the inferred value of the strong coupling constant
is $\alpha_S(M_\text{Z}) = 0.1199\,\pm{0.0015}\,(\mathrm{exp})\,
_{-0.0020}^{+0.0031}\,(\mathrm{theo})$, where $M_\text{Z}$ is the
mass of the Z boson.}

\hypersetup{%
pdfauthor={CMS Collaboration},%
pdftitle={Measurement of the triple-differential dijet cross section in
proton-proton collisions at sqrt(s) = 8 TeV and constraints on
parton distribution functions},%
pdfsubject={CMS},%
pdfkeywords={CMS, physics, QCD, PDF, jets, strong coupling constant, alpha-s}
}

\maketitle

\section{Introduction}
\label{sec:intro}

The pairwise production of hadronic jets is one of the fundamental
processes studied at hadron colliders. Dijet events with large transverse
momenta can be described by parton-parton scattering in the context of
quantum chromodynamics (QCD). Measurements of dijet cross sections can
be used to thoroughly test predictions of perturbative QCD (pQCD) at
high energies and to constrain parton distribution functions
(PDFs). Previous measurements of dijet cross sections in
proton-(anti)proton collisions have been performed as a function of
dijet mass at the \SPPS, ISR, and Tevatron
colliders~\cite{Banner:1982kt, Akesson:1985ms, Affolder:1999ua,
Abe:1993kb, Abe:1989gz, Abbott:1998wh}. At the CERN LHC, dijet
measurements as a function of dijet mass are reported in
Refs.~\cite{Aad:2010wv, Chatrchyan:2011qta, Aad:2011fc, Aad:2013tea,
Chatrchyan:2012bja}.  Also, dijet events have been
studied triple-differentially in transverse energy and
pseudorapidities $\eta_1$ and $\eta_2$ of the two leading
jets~\cite{Abe:1989te, Affolder:2000ew}.

\begin{figure}[h!tbp]
\centering
\includegraphics[width=\cmsFigWidth]{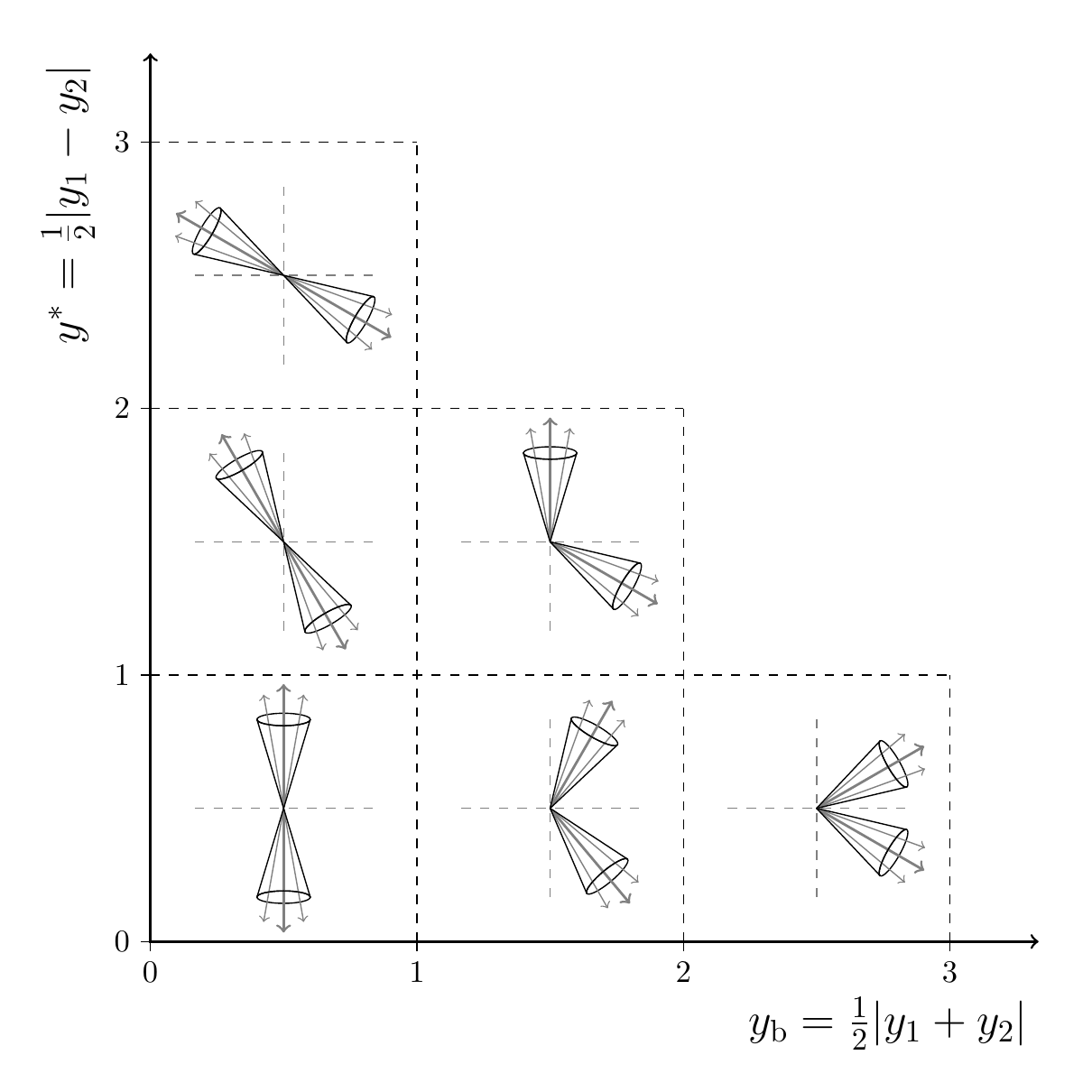}
\caption[Dijet event topologies]
{Illustration of the dijet event topologies in the \ystar and
\yboost kinematic plane. The dijet system can be classified as a
same-side or opposite-side jet event according to the boost \yboost
of the two leading jets, thereby providing insight into the parton
kinematics.}
\label{fig:ysyb_schema}
\end{figure}

In this paper, a measurement of the triple-differential dijet cross
section is presented as a function
of the average transverse momentum $\ptavg = (p_{\mathrm{T},1} +
p_{\mathrm{T},2}) / 2$ of the two leading jets, half of
their rapidity separation $\ystar = |y_1 - y_2| / 2$, and the
boost of the dijet system $\yboost = |y_1 + y_2| / 2$. The
dijet event topologies are illustrated in
Fig.~\ref{fig:ysyb_schema}.

The relation between the dijet rapidities and the parton momentum
fractions $x_{1,2}$ of the incoming protons at leading order (LO) is
given by $x_{1,2} = \frac{\pt}{\sqrt{s}} ( e^{\pm y_1} + e^{\pm
y_2})$, where $\pt = p_{\mathrm{T},1} = p_{\mathrm{T},2}$. For large
values of \yboost, the momentum fractions carried by the incoming
partons must correspond to one large and one small value, while for
small \yboost the momentum fractions must be approximately equal. In
addition, for high transverse momenta of the jets, $x$ values are
probed above~$0.1$, where the proton PDFs are less precisely known.

\begin{figure*}[h!tbp]
\centering
\includegraphics[width=0.4\textwidth]{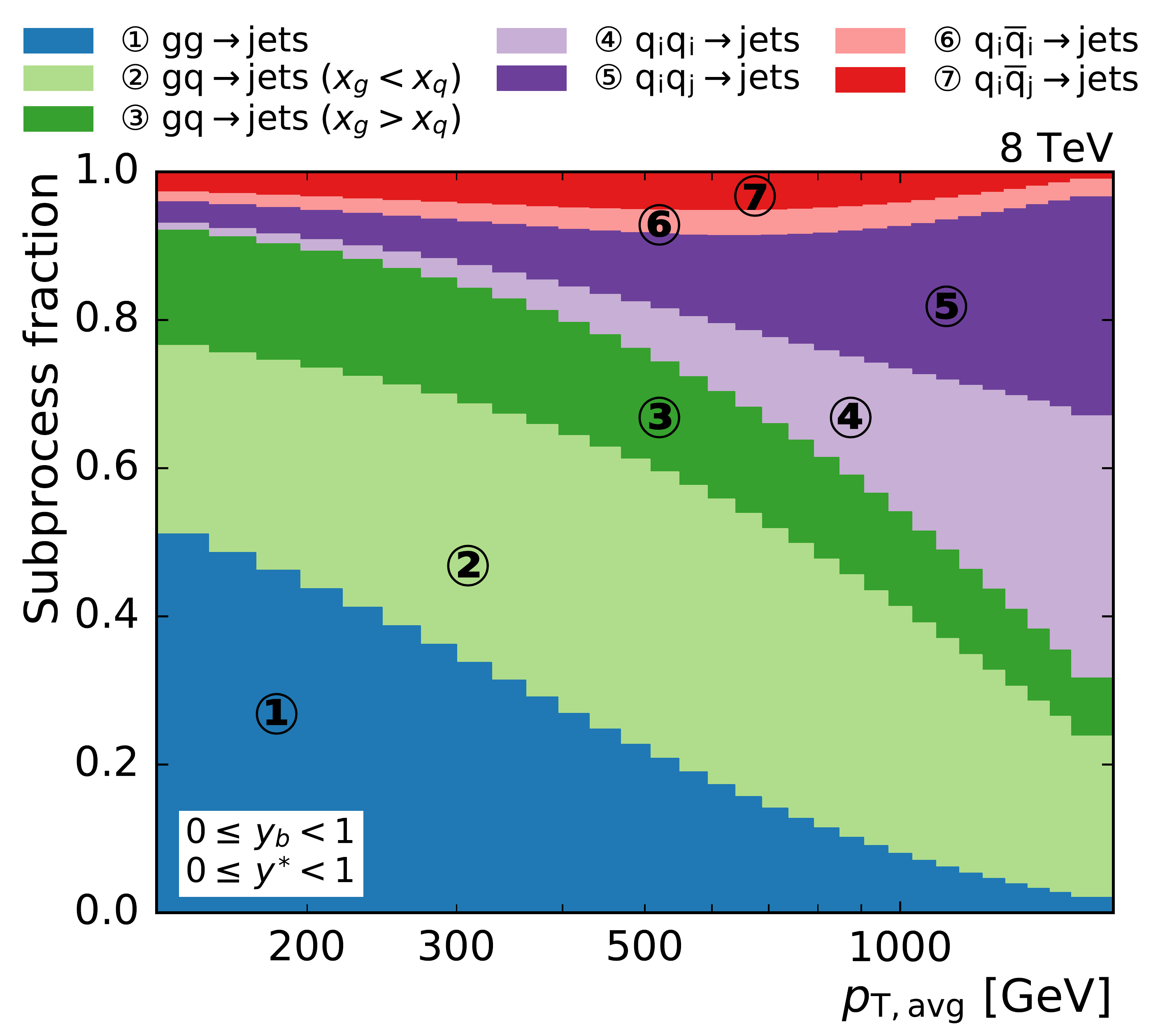}{ \hskip 0.8cm}
\includegraphics[width=0.4\textwidth]{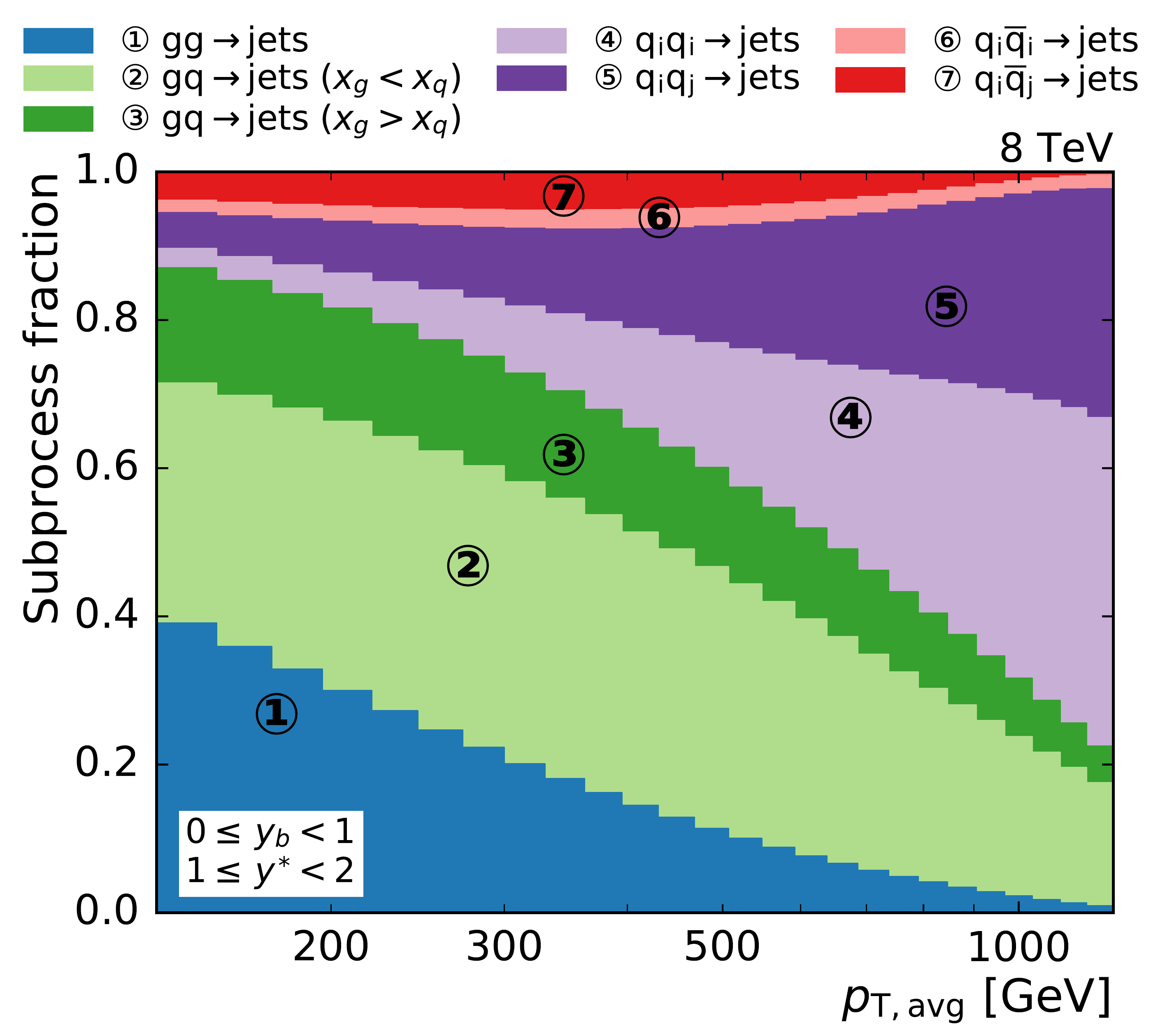}\\
\includegraphics[width=0.4\textwidth]{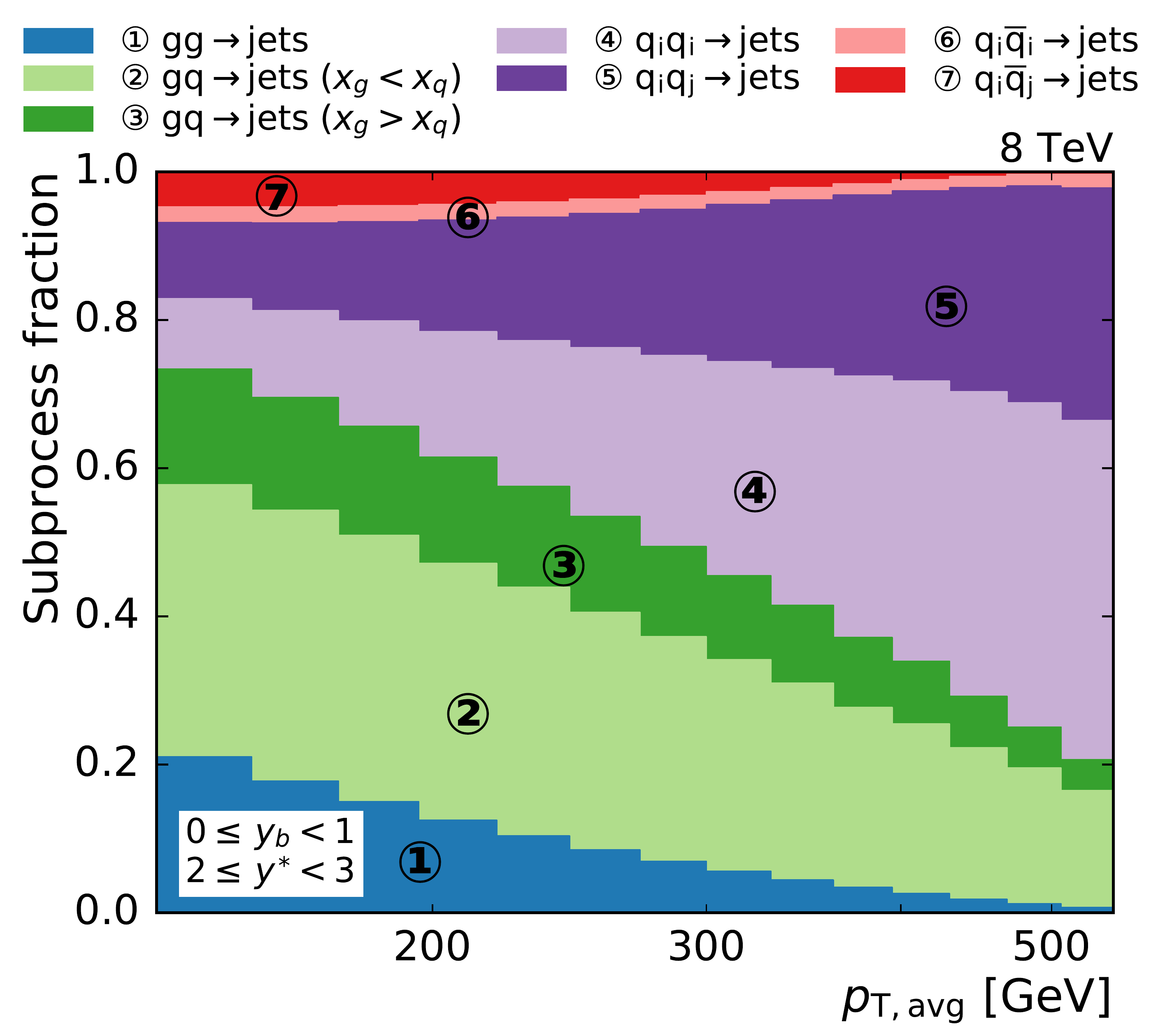}{ \hskip 0.8cm}
\includegraphics[width=0.4\textwidth]{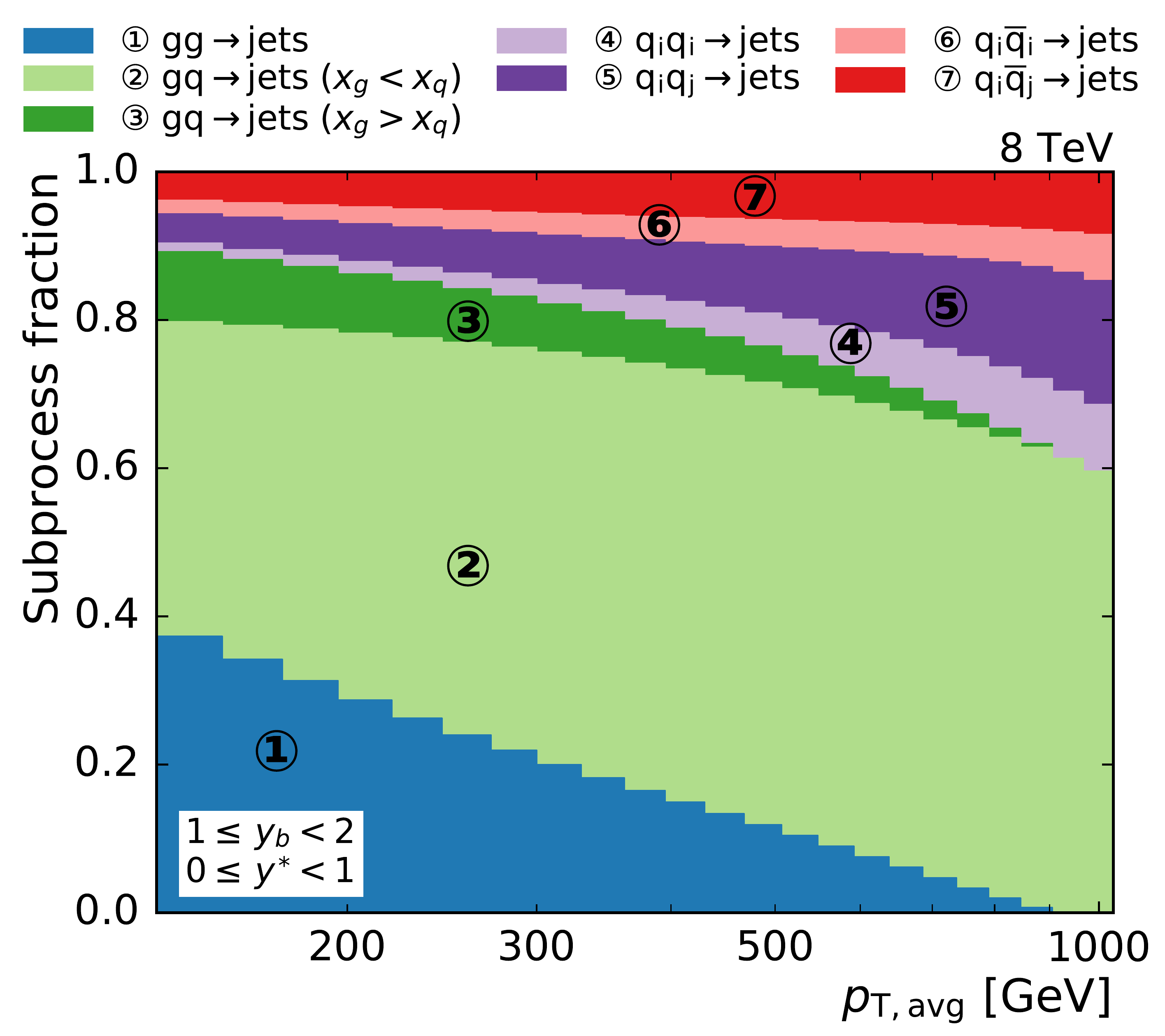}\\
\includegraphics[width=0.4\textwidth]{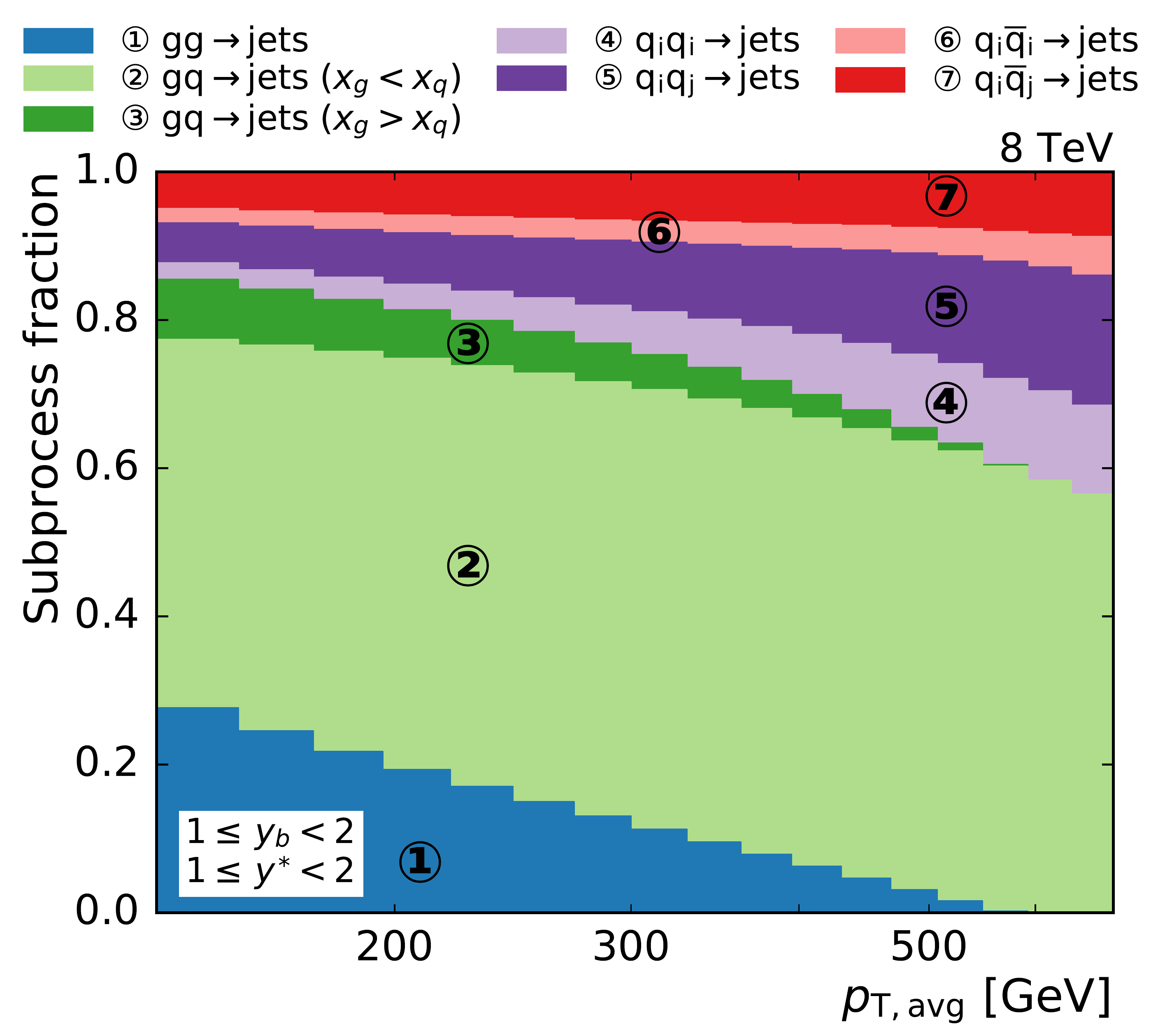}{ \hskip 0.8cm}
\includegraphics[width=0.4\textwidth]{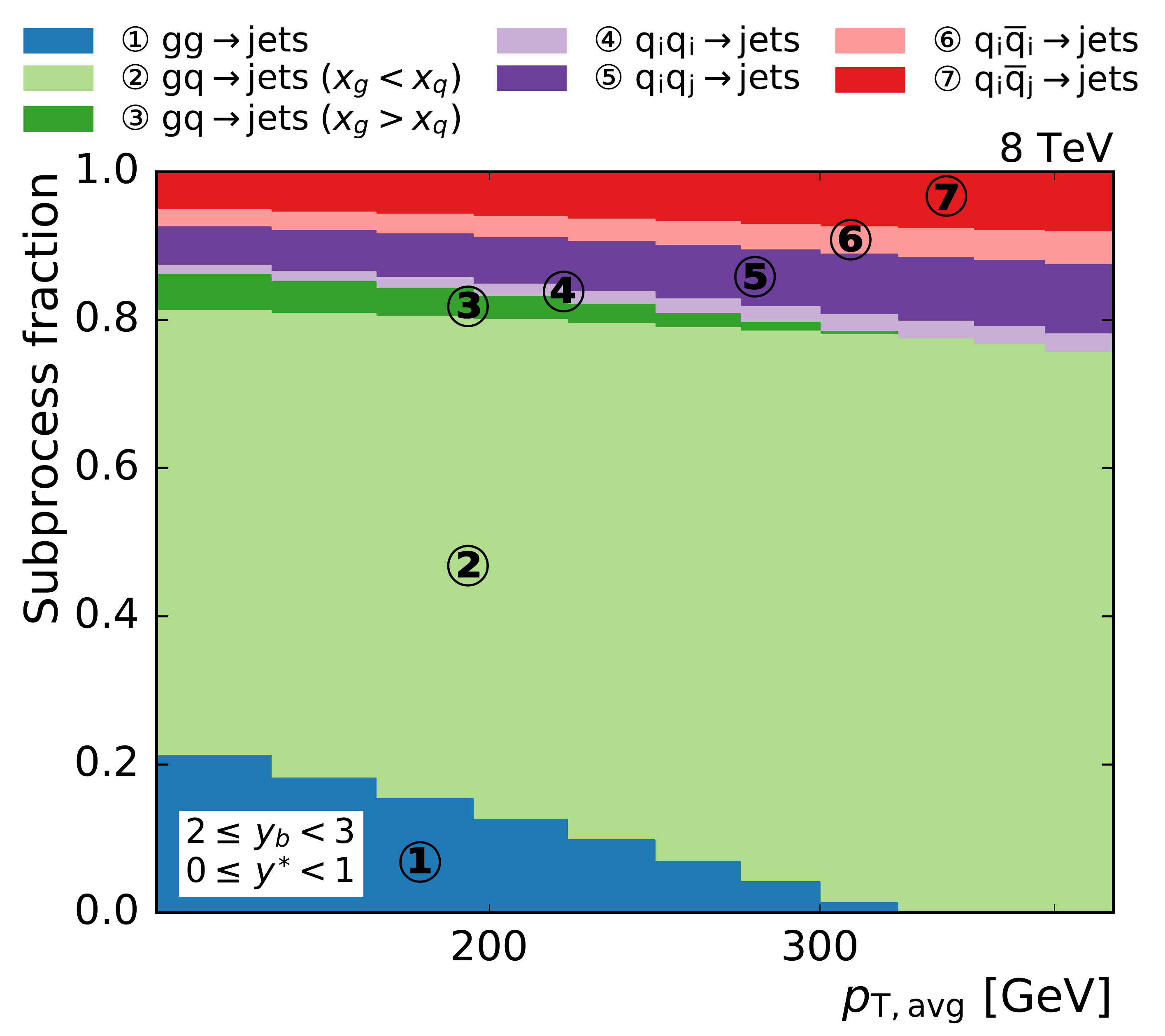}
\caption[Subprocess decomposition]{Relative contributions of all
subprocesses to the total cross section at NLO as a function of
\ptavg in the various \ystar and \yboost bins. The subprocess
contributions are grouped into seven categories according to the
type of the incoming partons. The calculations have been performed
with \nlojetpp. The notation implies the sum over
initial-state parton flavors as well as interchanged quarks and
antiquarks.}
\label{fig:subprocess_decomposition}
\end{figure*}

The decomposition of the dijet cross section into the contributing
partonic subprocesses is shown in Fig.~\ref{fig:subprocess_decomposition}
at next-to-leading order (NLO) accuracy, obtained using the
\nlojetpp program version~4.1.3~\cite{Nagy:2001fj,Nagy:2003tz}. At
small \yboost and large \ptavg a significant portion of the cross section
corresponds to quark-quark (and small amounts of anti\-quark-anti\-quark)
scattering with varying shares of equal- or unequal-type quarks.
In contrast, for large \yboost more than 80\% of the cross section
corresponds to partonic subprocesses with at least one gluon
participating in the interaction. As a consequence, new information about the PDFs
can be derived from the measurement of the triple-differential dijet
cross section.

The data were collected with the CMS detector at $\sqrt{s} = 8\TeV$
and correspond to an integrated luminosity of 19.7\fbinv.
The measured cross section is corrected
for detector effects and is compared to NLO calculations in pQCD,
complemented with electroweak (EW) and nonperturbative (NP)
corrections. Furthermore, constraints on the PDFs are studied and the
strong coupling constant \asmz is inferred.

\section{The CMS detector}
\label{sec:detector}

The central feature of the CMS apparatus is a superconducting solenoid
of 6\unit{m} internal diameter, providing a magnetic field of
3.8\unit{T}. Within the solenoid volume are a silicon pixel and strip
tracker, a lead tungstate crystal electromagnetic calorimeter (ECAL),
and a brass and scintillator hadron calorimeter (HCAL), each composed
of a barrel and two endcap sections. The silicon tracker measures
charged particles within the pseudorapidity range $\abs{\eta} <
2.5$. It consists of 1440 silicon pixel and 15\,148 silicon strip
detector modules. The ECAL consists of 75\,848 lead tungstate
crystals, which provide coverage in pseudorapidity $\abs{\eta} < 1.48$
in a barrel region and $1.48 < \abs{\eta} < 3.0$ in two endcap
regions. In the region $\abs{\eta} < 1.74$, the HCAL cells have widths
of 0.087 in pseudorapidity and 0.087 in azimuth ($\phi$). In the
$\eta$-$\phi$ plane, and for $\abs{\eta} < 1.48$, the HCAL cells map
on to $5\times{}5$ arrays of ECAL crystals to form calorimeter towers
projecting radially outwards from close to the nominal interaction
point. For $\abs{\eta} > 1.74$, the coverage of the towers increases
progressively to a maximum of 0.174 in $\Delta\eta$ and
$\Delta\phi$. Within each tower, the energy deposits in ECAL and HCAL
cells are summed to define the calorimeter tower energies,
subsequently used to provide the energies and directions of hadronic
jets. The forward hadron (HF) calorimeter extends the pseudorapidity
coverage provided by the barrel and endcap detectors and uses steel as
an absorber and quartz fibers as the sensitive material. The two
halves of the HF are located 11.2\unit{m} from the interaction region,
one on each end, and together they provide coverage in the range $3.0
< \abs{\eta} < 5.2$. Muons are measured in gas-ionisation detectors
embedded in the steel flux-return yoke outside the solenoid.

A more detailed description of the CMS detector, together with a
definition of the coordinate system used and the relevant kinematic
variables, can be found in Ref.~\cite{Chatrchyan:2008aa}.

\section{Event reconstruction and selection}
\label{sec:event_selection}

Dijet events are collected using five single-jet high-level
triggers~\cite{Adam:2005zf,Khachatryan:2016bia}, which require at least one jet with \pt
larger than 80, 140, 200, 260, and 320\GeV, respectively. At
trigger level the jets are reconstructed with a simplified version of
the particle-flow (PF) event reconstruction described in the following
paragraph. All but the highest threshold trigger were prescaled in
the 2012 LHC run. The triggers are employed in mutually exclusive
regions of the \ptavg spectrum, cf.\ Table~\ref{tab:triggers}, in which
their efficiency exceeds 99\%.

\begin{table}[htbp]
\centering
\topcaption[]{List of single-jet trigger thresholds used in the analysis.}
\label{tab:triggers}
\begin{tabular}{cc}
\hline
Trigger threshold [$\GeV$] & \ptavg range [$\GeV$]\\\hline
\x80  & 123--192\\
140 & 192--263\\
200 & 263--353\\
260 & 353--412\\
320 & ${>} 412$\\\hline
\end{tabular}
\end{table}

The PF event algorithm reconstructs and identifies particle candidates
with an optimised combination of information from the various elements
of the CMS detector~\cite{Sirunyan:2017ulk}. The energy of
photons is directly obtained from the ECAL measurement, corrected for
zero-suppress\-ion effects. The energy of electrons is determined from a
combination of the electron momentum at the primary interaction vertex
as determined by the tracker, the energy of the corresponding ECAL
cluster, and the energy sum of all bremsstrahlung photons spatially
compatible with originating from the electron track. The energy of
muons is obtained from the curvature of the corresponding track. The
energy of charged hadrons is determined from a combination of their
momentum measured in the tracker and the matching ECAL and HCAL energy
deposits, corrected for zero-suppression effects and for the response
function of the calorimeters to hadronic showers. Finally, the energy
of neutral hadrons is obtained from the corresponding corrected ECAL
and HCAL energies. The leading primary vertex (PV) is chosen as the one with
the highest sum of squares of all associated track transverse momenta. The
remaining vertices are classified as pileup vertices, which result
from additional proton-proton collisions. To reduce the background caused
by such additional collisions, charged hadrons
within the coverage of the tracker, $|\eta| < 2.5$~\cite{CMS:2014ata},
that unambiguously originate from a pileup vertex are removed.

Hadronic jets are clustered from the reconstructed
particles with the infrared- and collinear-safe anti-\mykt
algorithm~\cite{Cacciari:2008gp} with a jet size parameter
$R$ of 0.7, which is the default for CMS jet measurements.
The jet momentum is determined as the vectorial sum of all
particle momenta in the jet, and is found in the simulation to be
within 5 to 10\% of the true momentum over the whole \pt range.
Jet energy corrections (JEC) are derived from the
simulation, and are confirmed with in situ measurements of the energy
balance of dijet, photon+jet, and Z boson+jet
events~\cite{Chatrchyan:2011ds,Khachatryan:2016kdb}. After applying
the usual jet energy corrections, a small bias in the
reconstructed pseudorapidity of the jets is observed at the edge of
the tracker. An additional correction removes this effect.

All events are required to have at least one PV that
must be reconstructed from four or more tracks.
The longitudinal and transverse distances of the PV to the nominal
interaction point of CMS must satisfy $|z_\mathrm{PV}| < 24 \cm$
and $\rho_\mathrm{PV} < 2 \cm$, respectively. Nonphysical jets are removed by
loose jet identification criteria: each jet must contain at least two PF
candidates, one of which is a charged hadron, and the jet energy
fraction carried by neutral hadrons and photons must be less than
99\%. These criteria remove less than 1\% of genuine jets.

Only events with at least two jets up to an absolute rapidity of
$|y|=5.0$ are selected and the two jets leading in \pt are required to
have transverse momenta greater than 50\GeV and $|y| < 3.0$. The
missing transverse momentum is defined as the negative vector sum of
the transverse momenta of all PF candidates in the event. Its
magnitude is referred to as \ptmiss. For consistency with previous jet
measurements by CMS, \ptmiss is required to be smaller than 30\% of
the scalar sum of the transverse momenta of all PF candidates.
For dijet events, which exhibit very little \pt imbalance, the impact is
practically negligible.

\section{Measurement of the triple-differential dijet cross section}
\label{sec:measurement}

The triple-differential cross section for dijet production is defined as
\begin{equation*}
\frac{\rd^3 \sigma}{\rd \ptavg \rd \ystar \rd \yboost} = \frac{1}{\epsilon
\mathcal{L}_{\mathrm{int}}^\mathrm{eff}} \frac{N}{\Delta \ptavg \Delta \ystar
\Delta \yboost},
\end{equation*}
where $N$ denotes the number of dijet events within a given bin,
$\mathcal{L}_{\mathrm{int}}^{\mathrm{eff}}$
the effective integrated luminosity, and $\epsilon$ the product of trigger and
event selection efficiencies, which are greater than 99\% in the phase
space of the measurement. Contributions from background processes, such as \ttbar production, are
several orders of magnitude smaller and are neglected.
The bin widths are $\Delta \ptavg$, $\Delta \ystar$, and
$\Delta \yboost$.

The cross section is unfolded to the stable-particle level (lifetime
$c\tau > 1\cm$) to correct for detector resolution effects. The
iterative D'Agostini algorithm with early
stopping~\cite{D'Agostini:1994zf,Lucy:1974yx,Richardson:72}, as
implemented in the \RooUnfold package~\cite{Adye:2011gm}, is employed
for the unfolding. The response matrix, which relates the
particle-level distribution to the measured distribution at detector
level, is derived using a forward smearing technique.
An \nlojetpp prediction, obtained with CT14
PDFs~\cite{Dulat:2015mca} and corrected for NP and EW effects, is
approximated by a continuous function to represent the distribution at
particle level. Subsequently, pseudoevents are distributed uniformly
in \ptavg and weighted according to the theoretical prediction.
These weighted events are smeared using the jet \pt resolution to
yield a response matrix and a prediction at detector level. By
using large numbers of such pseudoevents, statistical fluctuations in
the response matrix are strongly suppressed.

The jet energy (or $\pt$) resolution (JER) is determined from the CMS
detector simulation based on the \GEANTfour
toolkit~\cite{Agostinelli:2002hh} and the \pythia~6.4 Monte Carlo (MC)
event generator~\cite{Sjostrand:2006za} and is corrected for residual
differences between data and simulation following
Ref.~\cite{Khachatryan:2016kdb}. The rapidity dependence of both the
JER from simulation and of the residual differences have been taken
into account. The Gaussian \pt resolution in the interval $|y|<1$ is
about 8\% at 100\GeV and improves to 5\% at 1\TeV. Non-Gaussian tails
in the JER, exhibited for jet rapidities close to $|y|=3$, are
included in a corresponding uncertainty.

{\tolerance=700
The regularisation strength of the iterative unfolding procedure is
defined through the number of iterations, whose optimal value is
determined by performing a \chisq test between the original measured
data and the unfolded data after smearing with the response matrix.
The values obtained for \chisq per number of degrees of freedom,
\ndof, in these comparisons approach unity in four iterations and
thereafter decrease slowly for additional iterations. The optimal
number of iterations is therefore determined to be four.  The
procedure is in agreement with the criteria of
Ref.~\cite{IEEE:MI6-1987}. %
The response matrices derived in this manner for each bin in \ystar
and \yboost are nearly diagonal.  A cross check using the \pythias MC
event generator as theory and the detector simulation to construct the
response matrices revealed no discrepancies compared to the baseline
result.
\par}

Migrations into and out of the accepted phase space in \ystar
and \yboost or between bins happen only at a level below 5\%.
The net effect of these migrations has been included in the respective
response matrices and has been cross checked successfully using a
3-dimensional unfolding.

As a consequence of these migrations, small statistical correlations between
neighbouring bins of the unfolded cross sections are introduced during the
unfolding procedure. The statistical uncertainties after being propagated through
the unfolding are smaller than 1\% in the majority of the phase space,
and amount up to 20\% for highest \ptavg.

\begin{figure*}[h!tbp]
\centering
\includegraphics[width=0.4\textwidth]{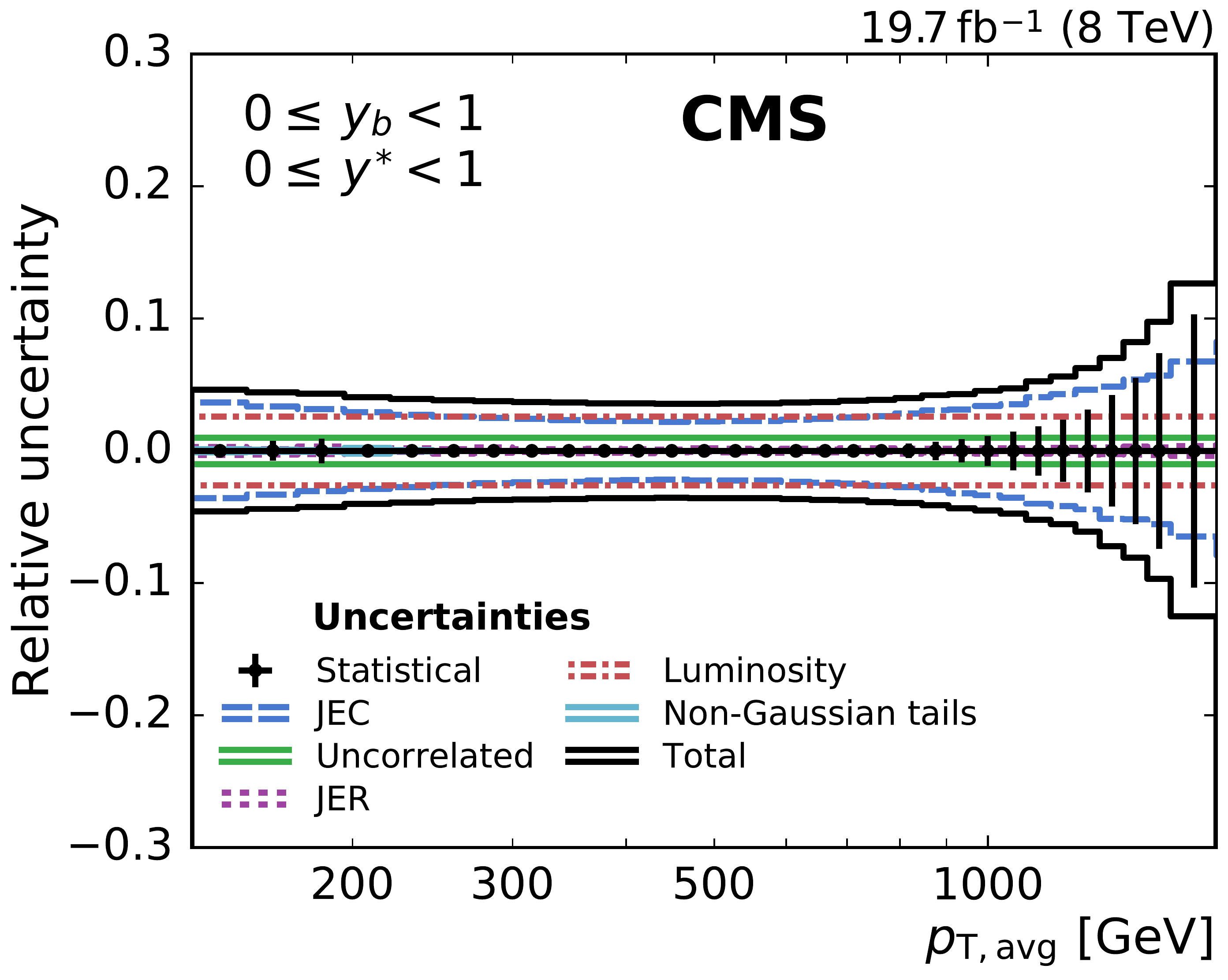}{ \hskip 0.8cm}
\includegraphics[width=0.4\textwidth]{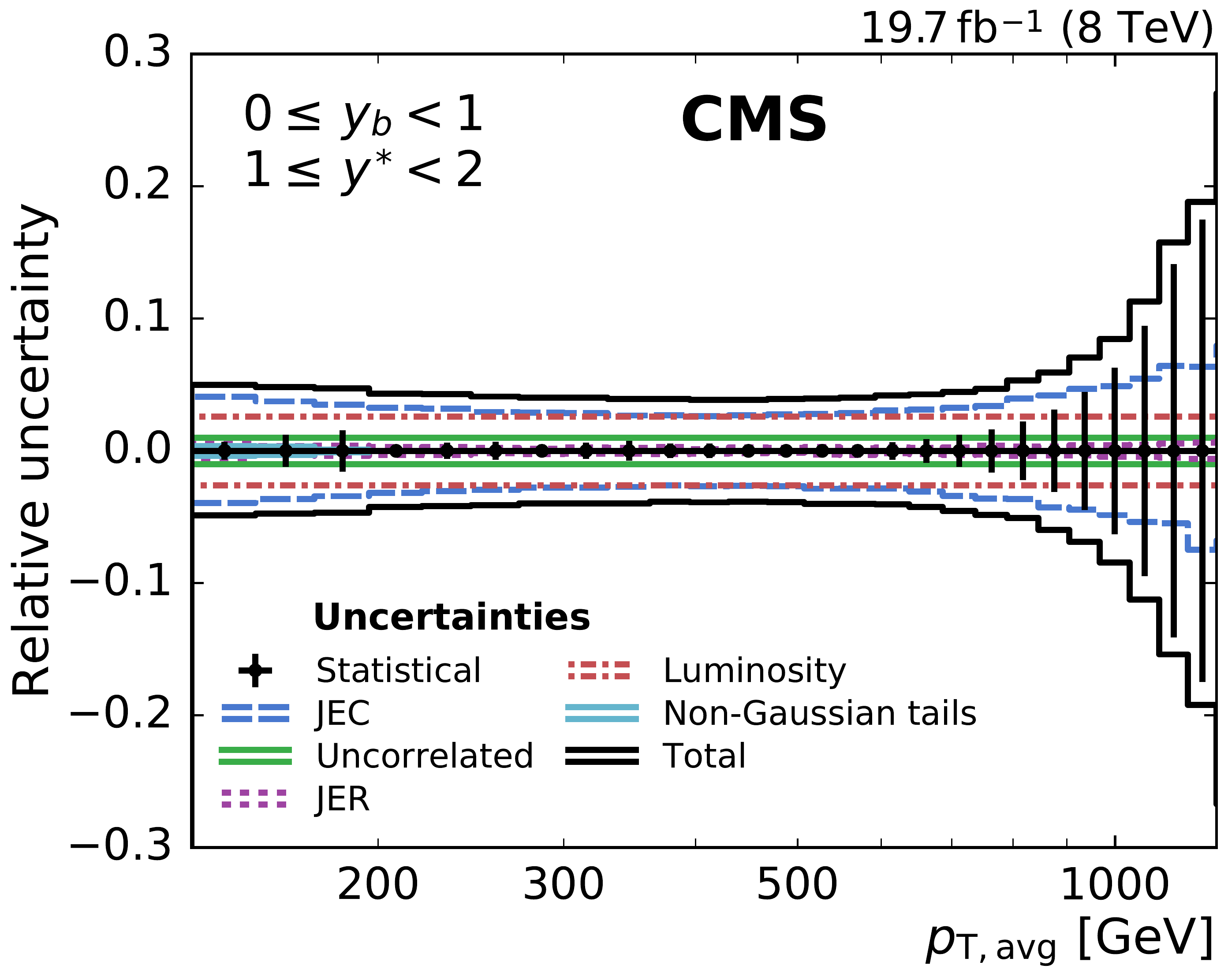}\\
\includegraphics[width=0.4\textwidth]{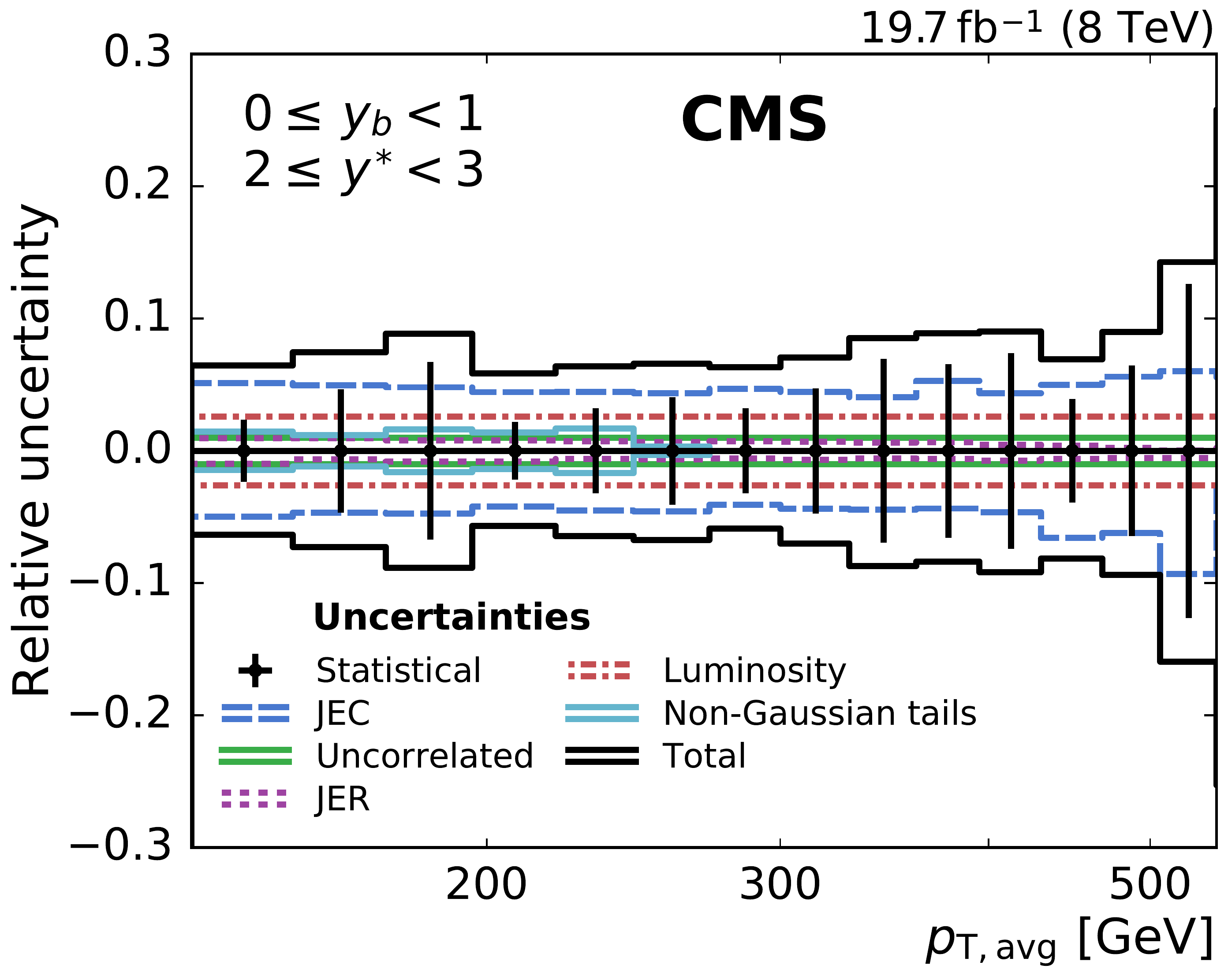}{ \hskip 0.8cm}
\includegraphics[width=0.4\textwidth]{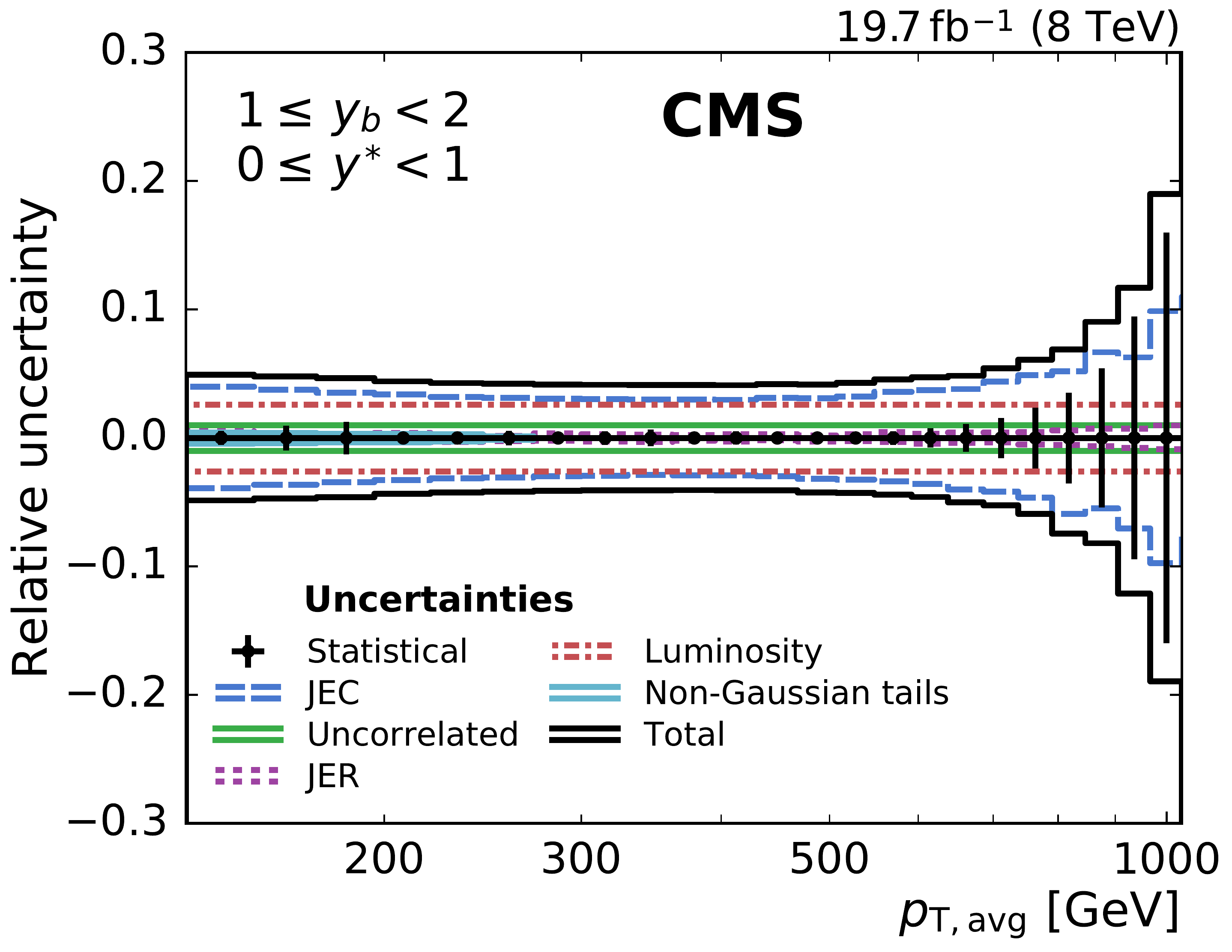}\\
\includegraphics[width=0.4\textwidth]{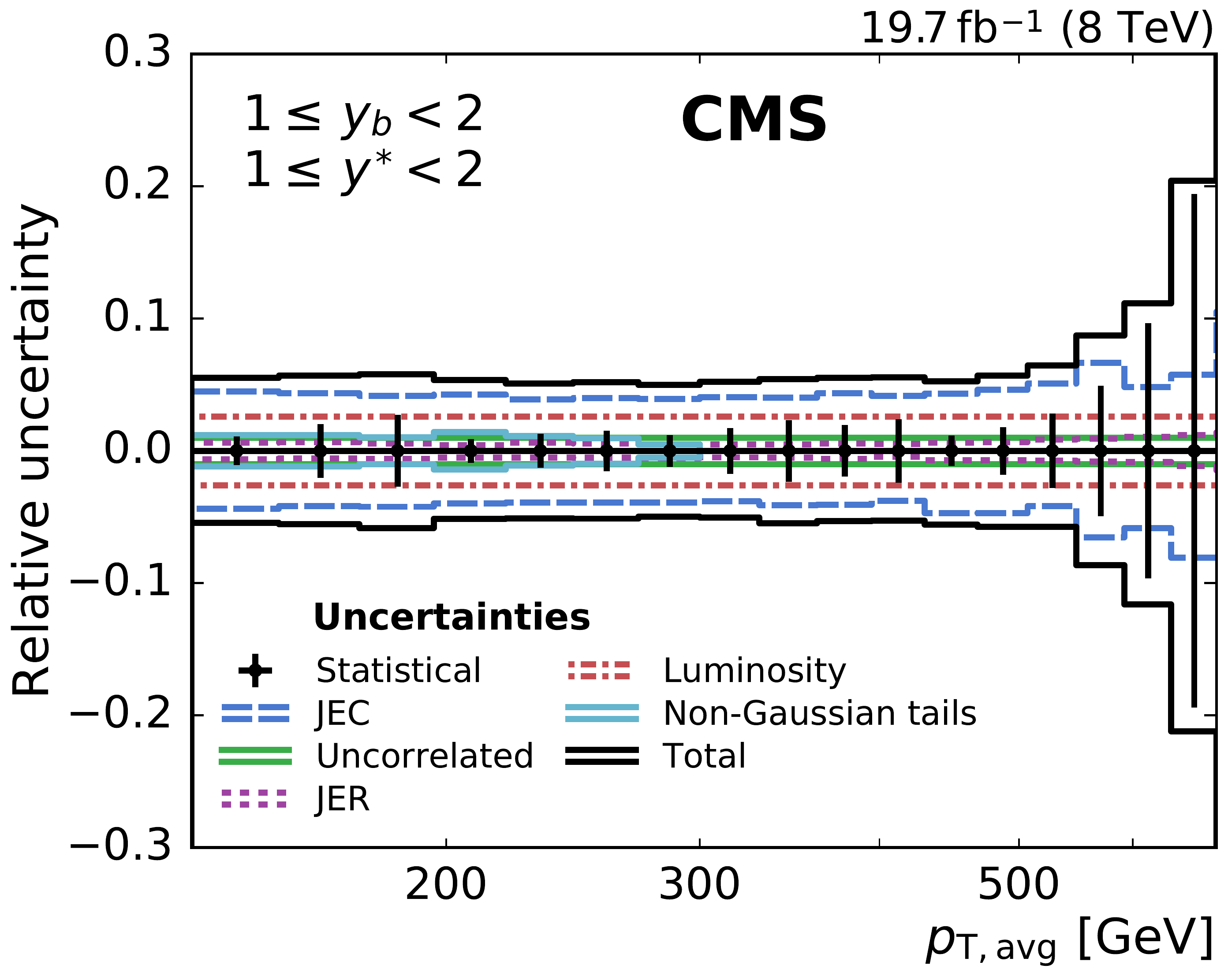}{ \hskip 0.8cm}
\includegraphics[width=0.4\textwidth]{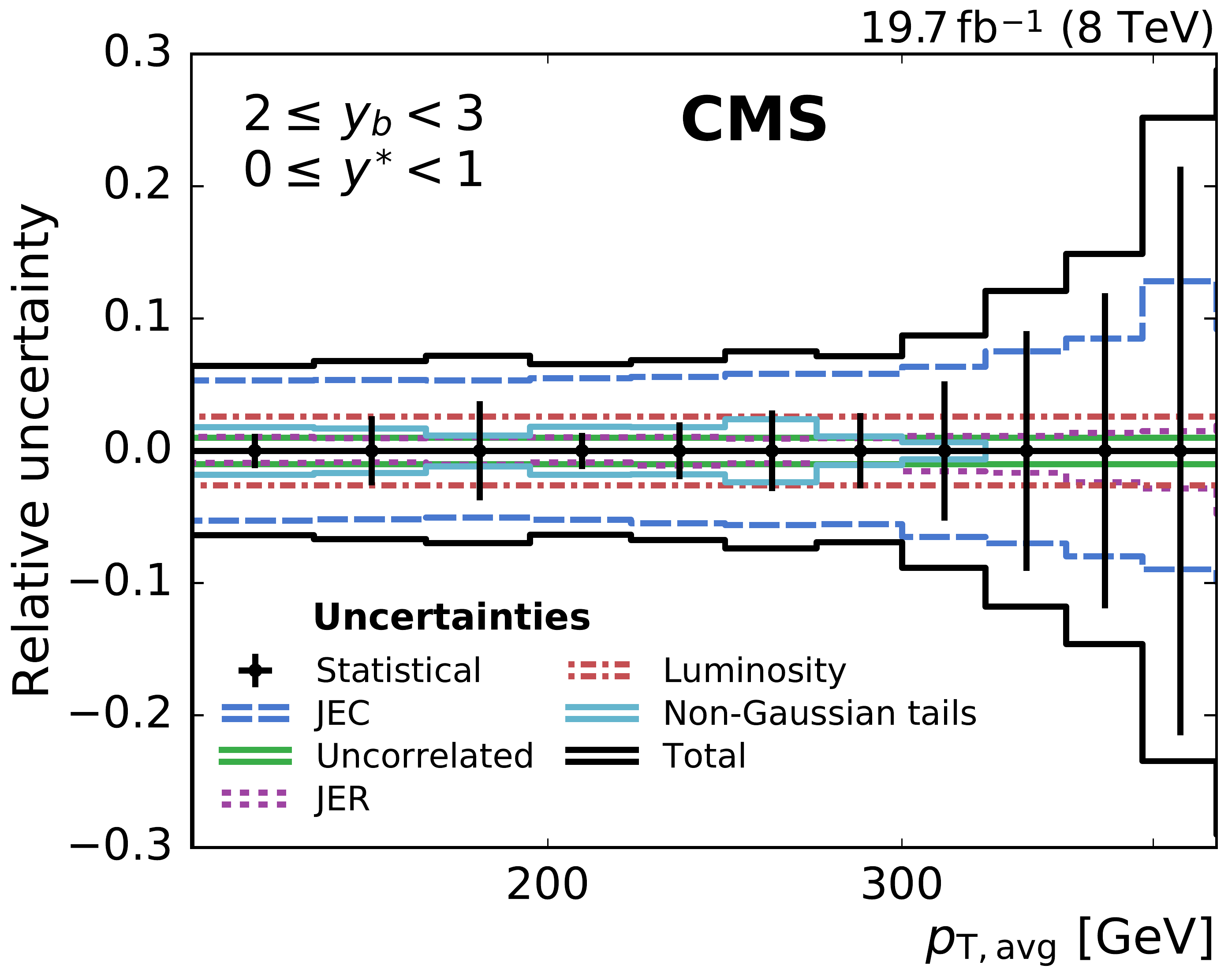}
\caption[Overview of experimental uncertainties]{Overview of all
experimental uncertainties affecting the cross section measurement
in six bins of \yboost and \ystar. The error bars
indicate the statistical uncertainty after unfolding. The
different lines show the uncertainties resulting from jet energy
corrections, jet energy resolution, integrated luminosity,
non-Gaussian tails in the resolution, and from residual effects included
in the uncorrelated uncertainty. The total uncertainty is
obtained by adding all uncertainties in quadrature.}
\label{fig:exp_uncertainties}
\end{figure*}

The dominant systematic uncertainties in the cross section measurement
arise from uncertainties in the JEC\@. Summing up quadratically all
JEC uncertainties according to the prescription given in
Ref.~\cite{Khachatryan:2016kdb}, the total JEC uncertainty amounts to about 2.5\% in the
central region and increases to 12\% in the forward regions. The 2.6\%
uncertainty in the integrated luminosity~\cite{CMS:2013gfa} is directly
propagated to the cross section. The uncertainty in the JER enters the
measurement through the unfolding procedure and results in an
additional uncertainty of 1\% to 2\% of the unfolded cross
section. Non-Gaussian tails in the detector response to jets near $|y|
= 3.0$, the maximal absolute rapidity considered in this measurement,
are responsible for an additional uncertainty of up to 2\%. Residual
effects of small inefficiencies in the jet identification and
trigger selection are covered by an uncorrelated uncertainty of
1\%~\cite{Chatrchyan:2012bja}. The total systematic experimental
uncertainty ranges from about 3 to 8\% in the central detector region
and up to 12\% for absolute rapidities near the selection limit of
3.0. Figure~\ref{fig:exp_uncertainties} depicts all experimental
uncertainties as well as the total uncertainty, which is calculated as
the quadratic sum of all the contributions from the individual
sources.

\section{Theoretical predictions}
\label{sec:theory}

\begin{figure*}[htbp]
\centering
\includegraphics[width=0.4\textwidth]{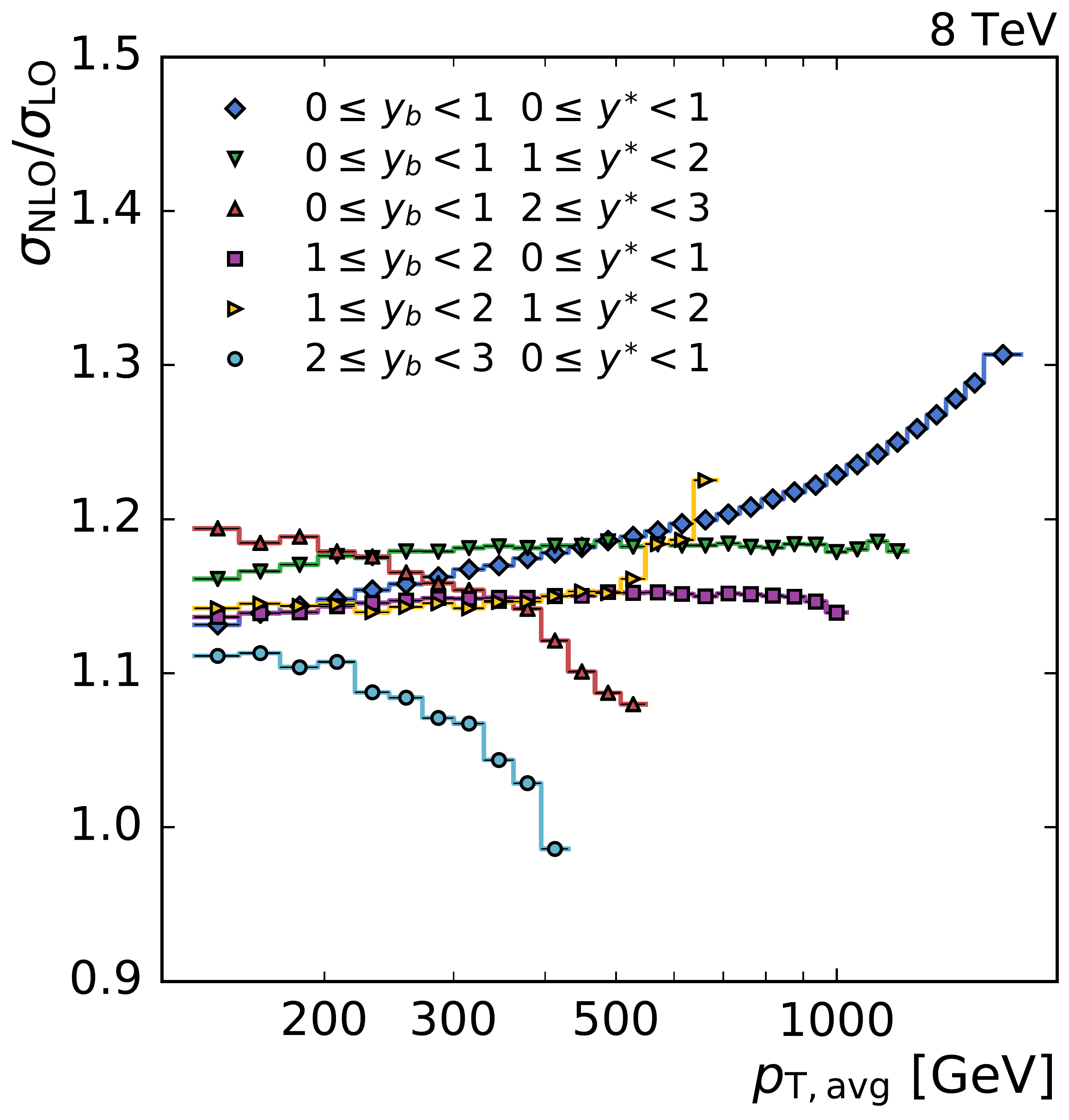}{ \hskip 0.8cm}
\includegraphics[width=0.4\textwidth]{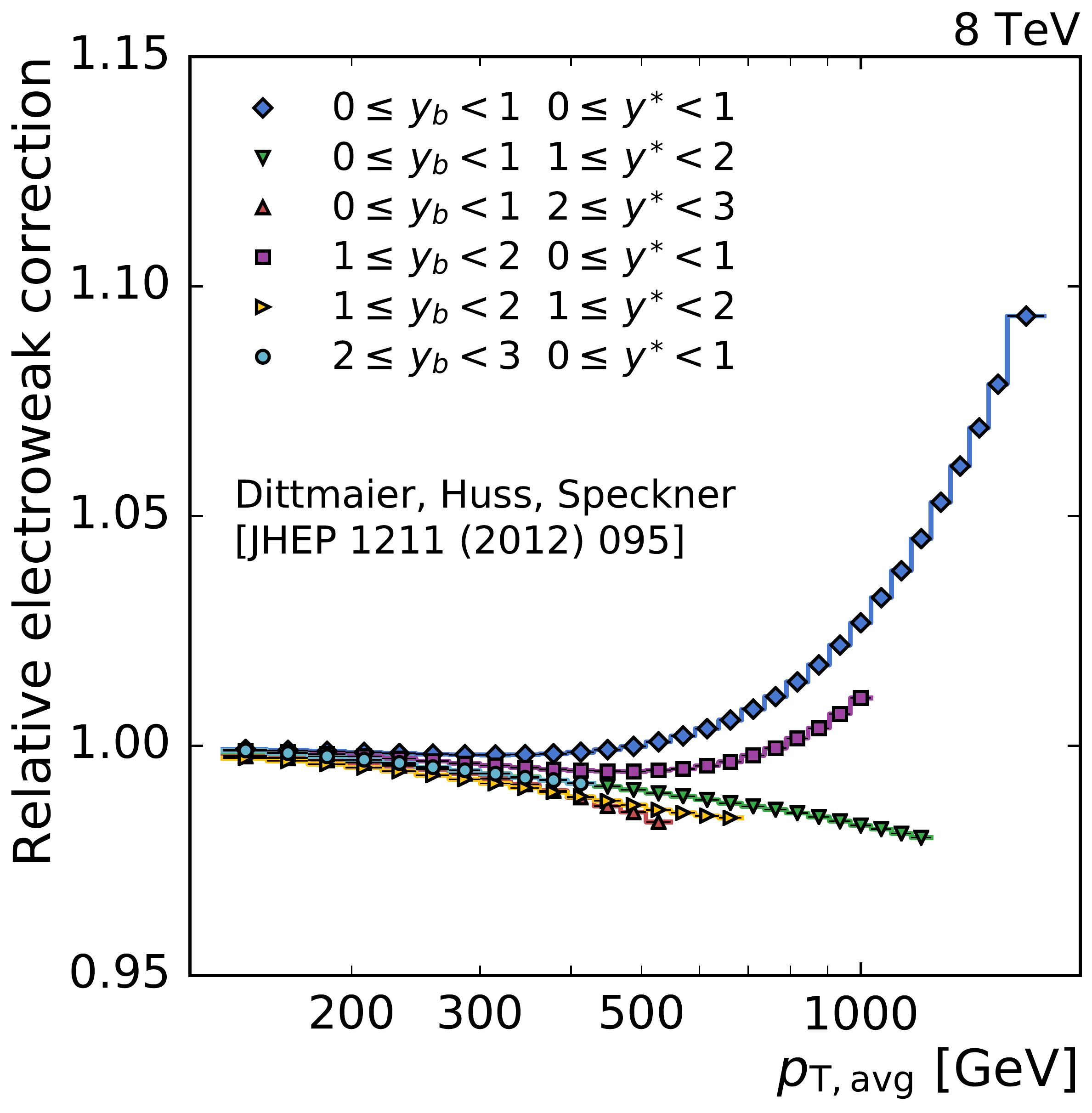}\\
\includegraphics[width=0.4\textwidth]{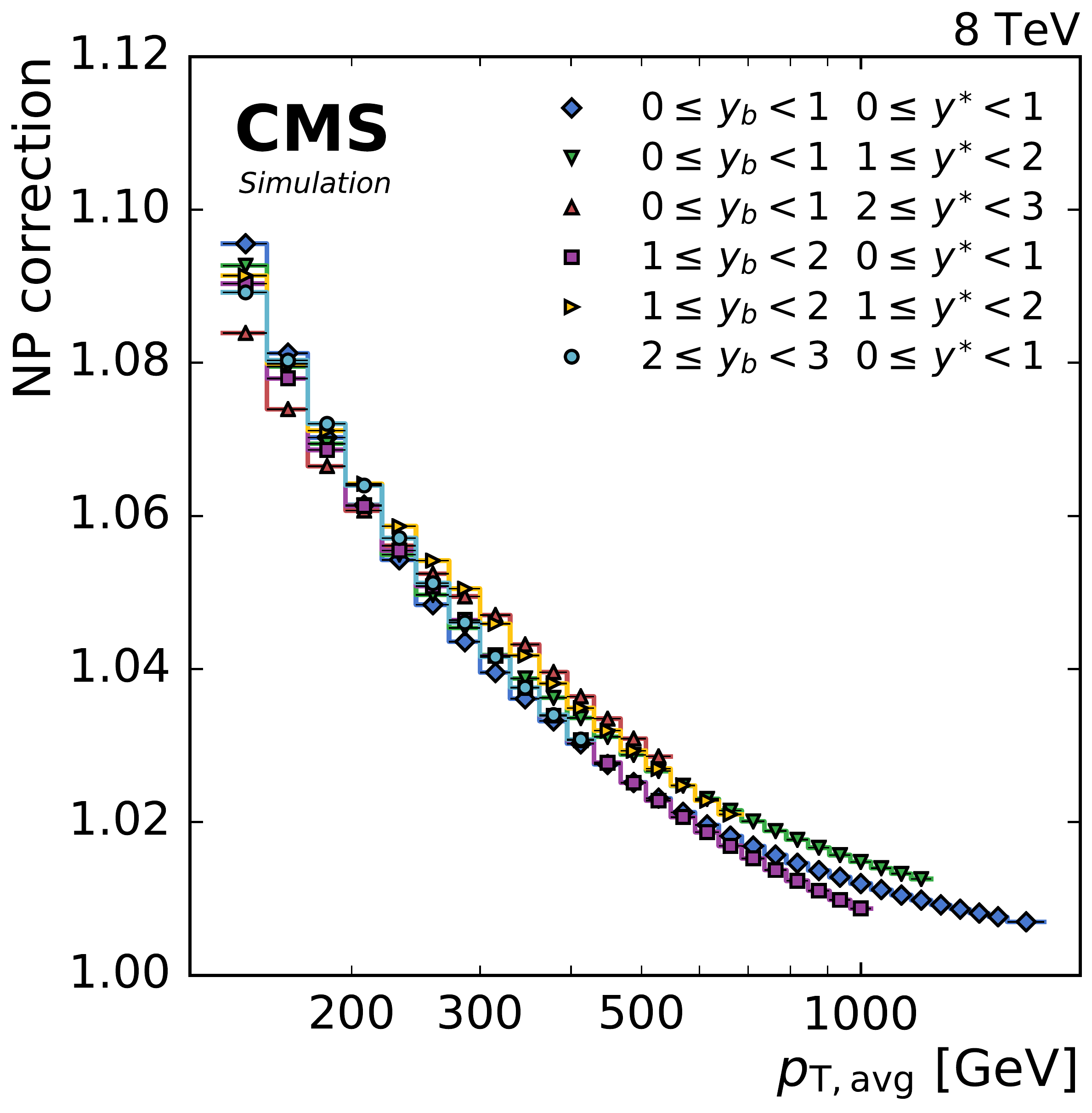}
\caption[Overview of correction factors]
{Overview of the theoretical correction factors. For each of the six
analysis bins the NLO QCD (top left), the electroweak (top right),
and the NP correction factor (bottom) are shown as a function of
\ptavg. The NLO QCD correction has been derived with the same NLO
PDF in numerator and denominator and is included in the NLO
prediction by \nlojetpp.}
\label{fig:cfactors}
\end{figure*}

The NLO predictions for the triple-differential dijet cross section are
calculated using \nlojetpp within the framework of
\fastnlo version~2.1~\cite{Kluge:2006xs,Britzger:2012bs}. The renormalisation and
factorisation scales \mur and \muf are both set to $\mu=\mun=p_{\mathrm{T,max}}
\cdot e^{0.3 \ystar}$, a scale choice first investigated in Ref.~\cite{Ellis:1992en}.
The variation of these scales by constant factors as described below
is conventionally used to estimate the effect of missing higher
orders. The scale uncertainty
is reduced in regions with large values of \yboost
with the above-mentioned choice for \mun
compared to a prediction with $\mun=p_{\mathrm{T,avg}}$.
The predictions for cross
sections obtained with different central scale choices are compatible within the
scale uncertainties. The calculation is performed using the PDF sets CT14,
ABM11~\cite{Alekhin:2012ig}, MMHT2014~\cite{Harland-Lang:2014zoa}, and
NNPDF~3.0~\cite{Ball:2014uwa} at next-to-leading evolution order, which are
accessed via the \LHAPDF~6.1.6 interface~\cite{Whalley:2005nh, Buckley:2014ana} using
the respective values of \asmz and the supplied \as evolution. The
size of the NLO correction is shown in Fig.~\ref{fig:cfactors} top
left and varies between $+10$\% and $+30$\% at high \ptavg
and low~\yboost.

The fixed-order calculations are accompanied by NP corrections,
$c_k^\mathrm{NP}$, derived from the LO MC event generators
\pythia~8.185~\cite{Sjostrand:2007gs} and
\herwigpp~2.7.0~\cite{Bahr:2008pv} with the tunes
CUETP8M1~\cite{Khachatryan:2015pea} and
UE-EE-5C~\cite{Seymour:2013qka}, respectively, and the NLO MC generator
\powheg~\cite{Nason:2004rx, Frixione:2007vw, Alioli:2010xd,
Alioli:2010xa} in combination with \pythiae and the tunes CUETP8M1
and CUETP8S1~\cite{Khachatryan:2015pea}.

The correction factor $c_{k}^{\mathrm{NP}}$ is defined as the ratio between the
nominal cross section with and without multiple parton interactions (MPI) and
hadronisation (HAD) effects

\begin{equation*}
c_{k}^{\mathrm{NP}} = \frac{\sigma_{k}^{\mathrm{PS+HAD+MPI}}}{\sigma_{k}^{\mathrm{PS}}}\,,
\label{eq:np_definition}
\end{equation*}

where the superscript indicates the steps in the simulation: the parton
shower (PS), the MPI, and the hadronisation. The corresponding
correction factor, as displayed in Fig.~\ref{fig:cfactors} bottom,
is applied in each bin~$k$ to the parton-level NLO cross section. It differs from
unity by about $+10$\% for lowest \ptavg and becomes negligible above
1\TeV.

{\tolerance=3600
To account for differences among the correction factors obtained by using
\herwigpp, \pythiae, and \powhegpluspythiae, half of the envelope of all these
predictions is taken as the uncertainty and the centre of the envelope is used as
the central correction factor.
\par}

The contribution from EW effects, which arise mainly from virtual
exchanges of massive W and Z bosons, is relevant at high jet \pt
and central rapidities~\cite{Dittmaier:2012kx,Frederix:2016ost}. These
corrections, shown in Fig.~\ref{fig:cfactors} top right,
are smaller than 3\% below 1\TeV and reach 8\% for the
highest \ptavg. Theoretical uncertainties in this correction due to
its renormalisation scheme and indirect PDF dependence are considered to
be negligible.

\begin{figure*}[htbp]
\centering
\includegraphics[width=0.4\textwidth]{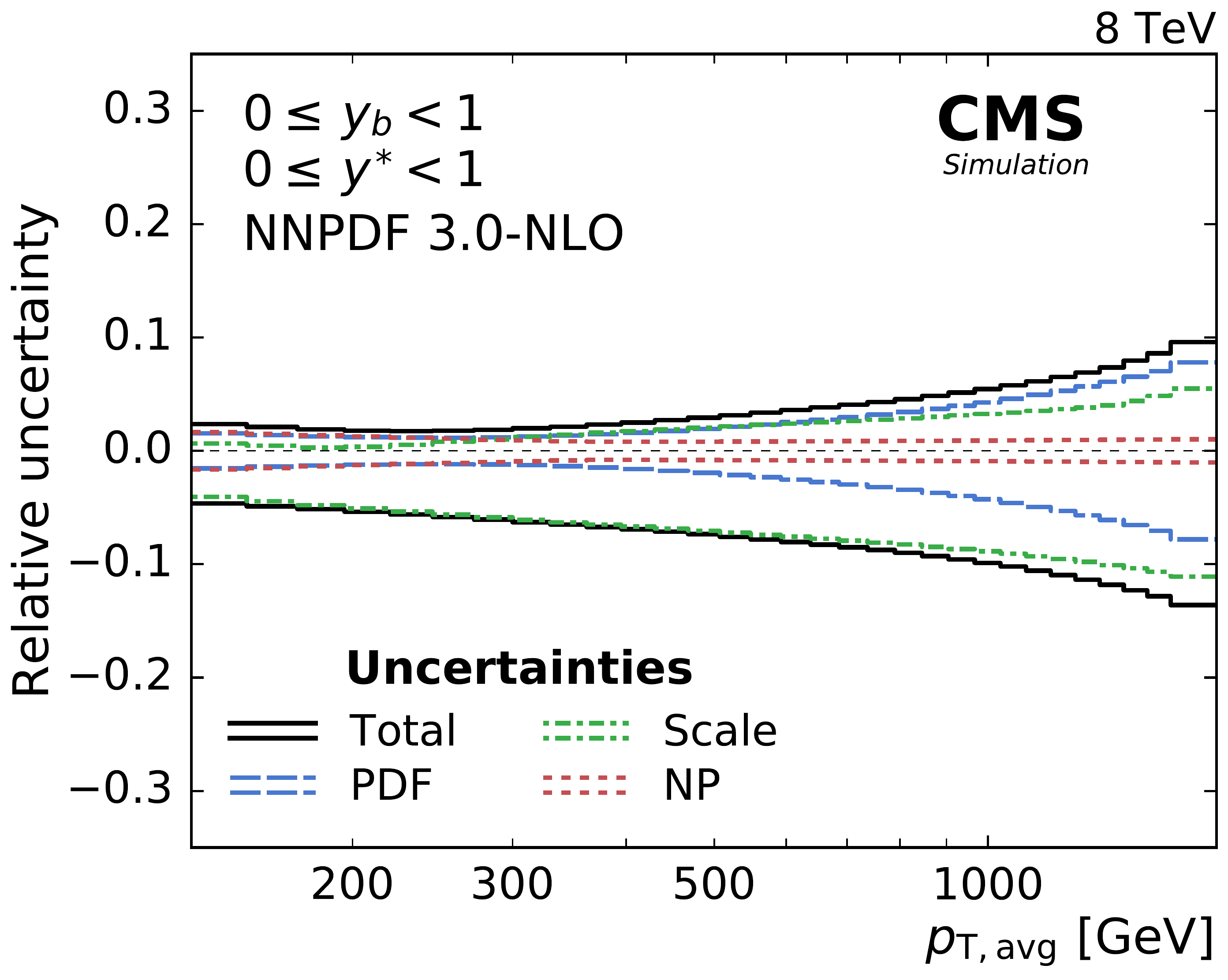}{ \hskip 0.8cm}
\includegraphics[width=0.4\textwidth]{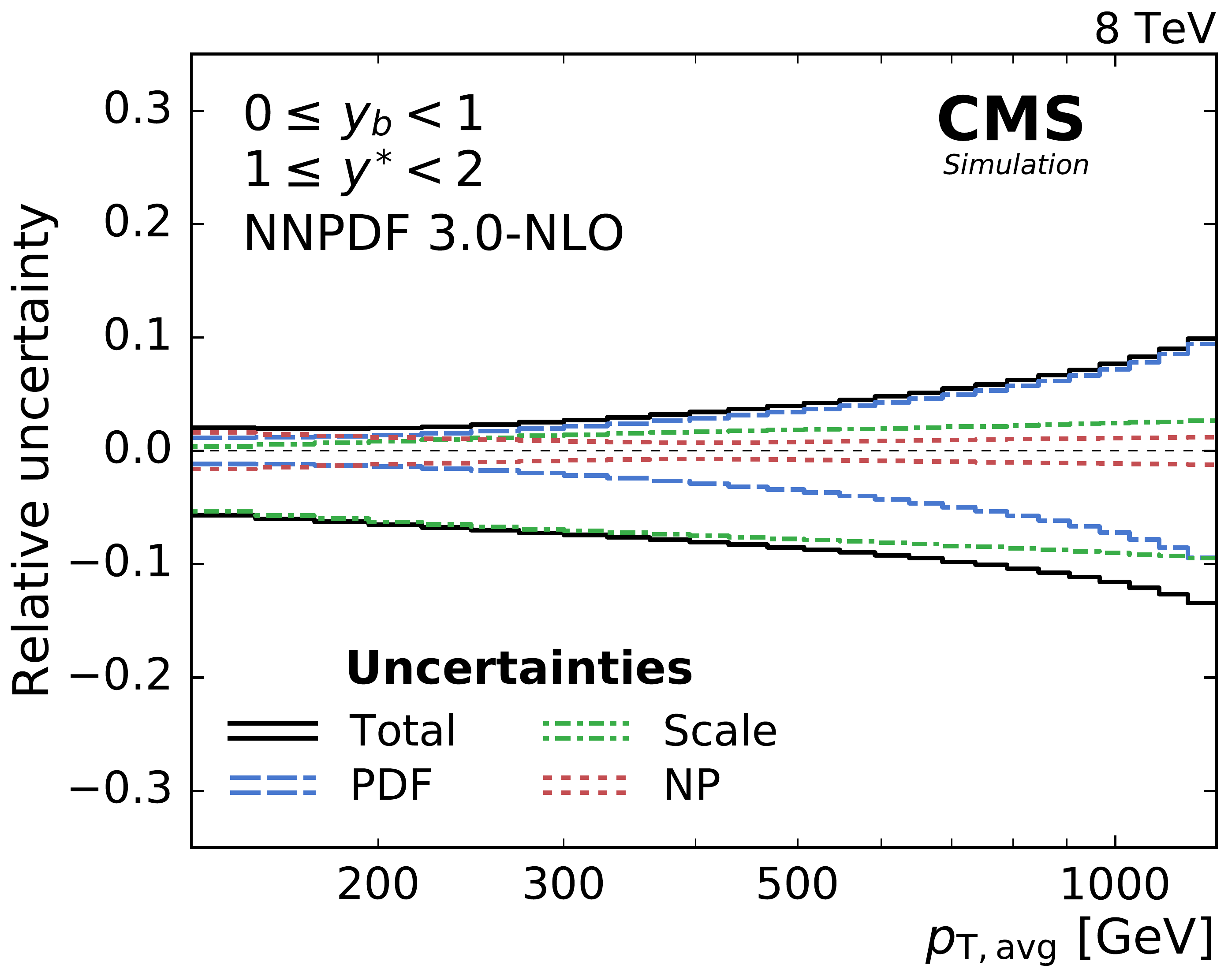}\\
\includegraphics[width=0.4\textwidth]{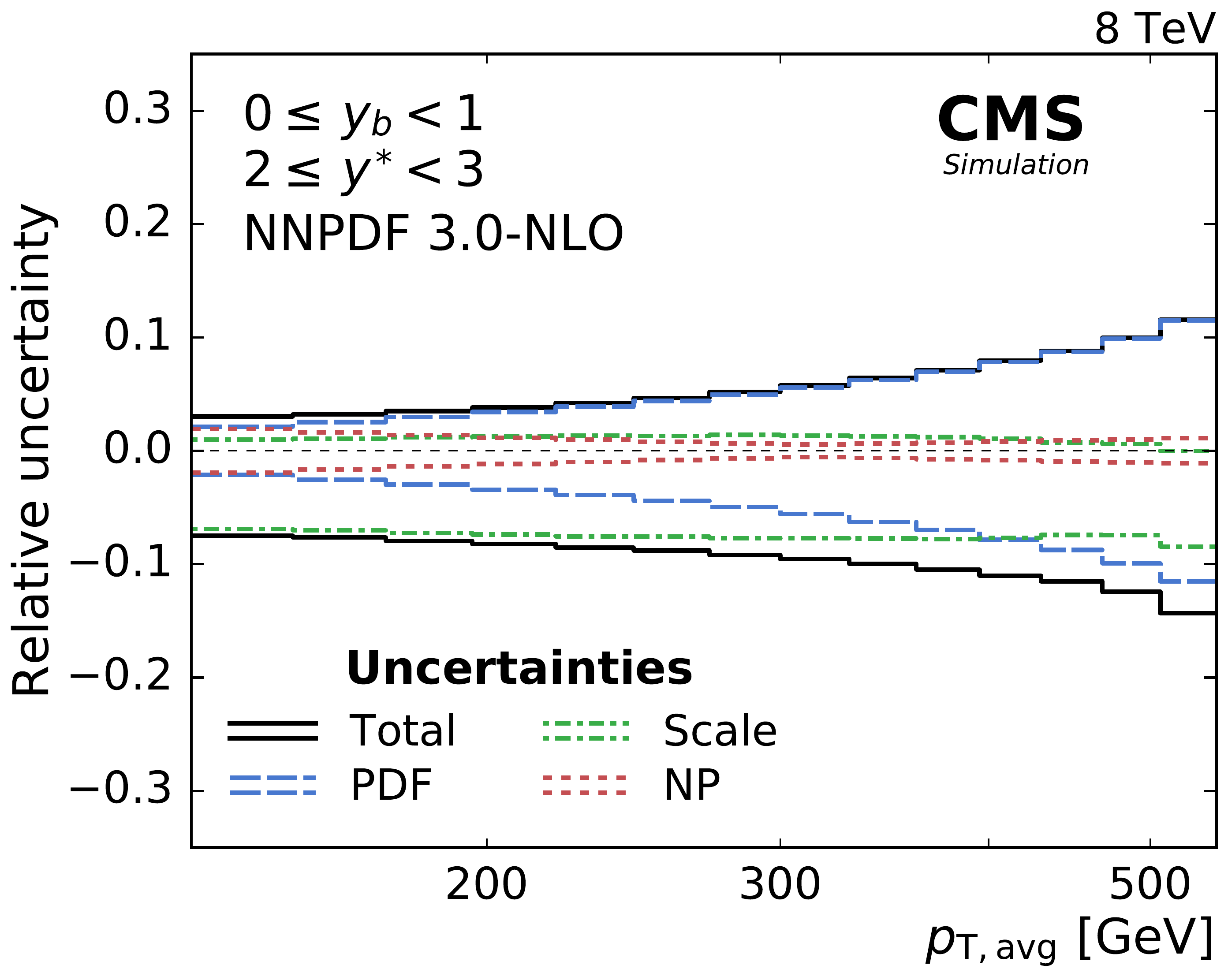}{ \hskip 0.8cm}
\includegraphics[width=0.4\textwidth]{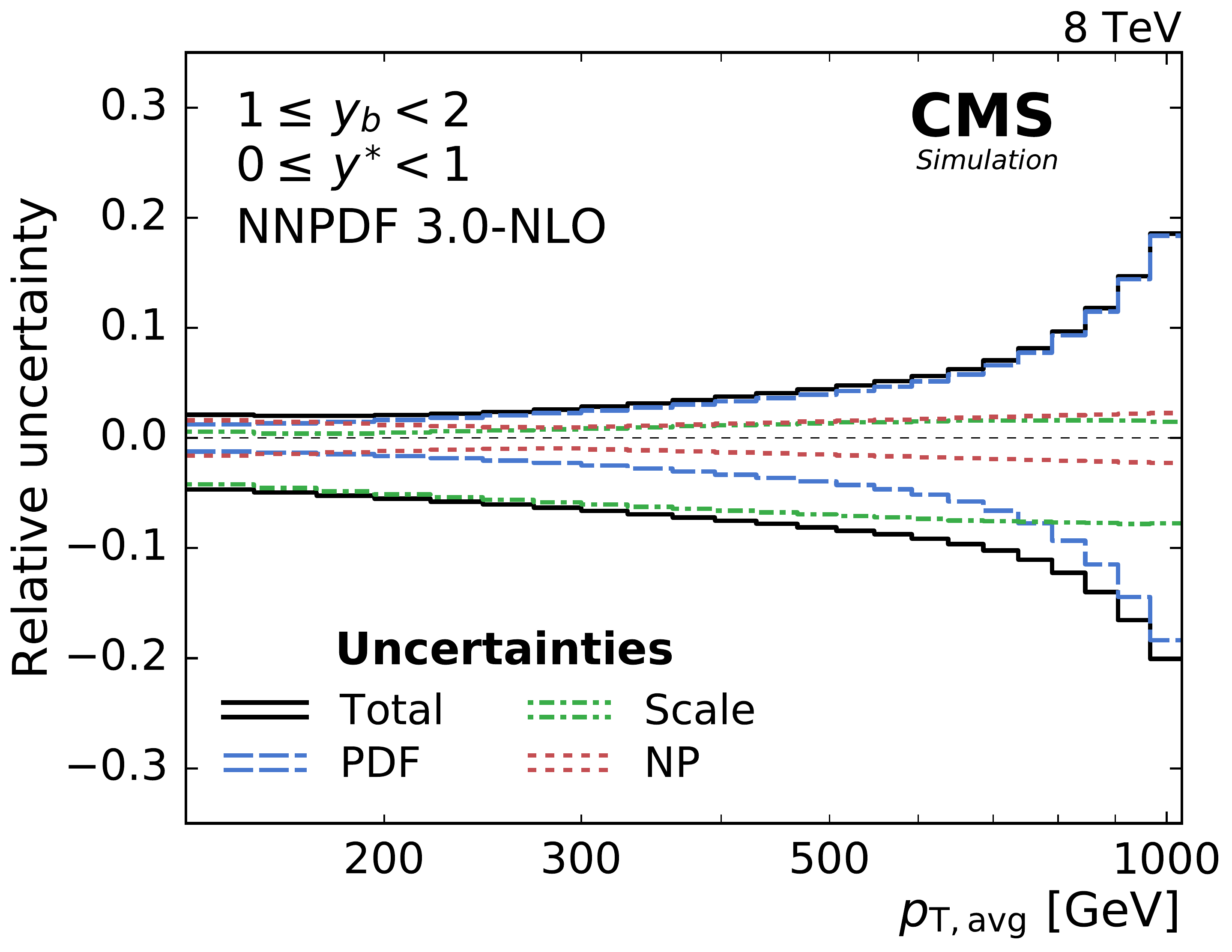}\\
\includegraphics[width=0.4\textwidth]{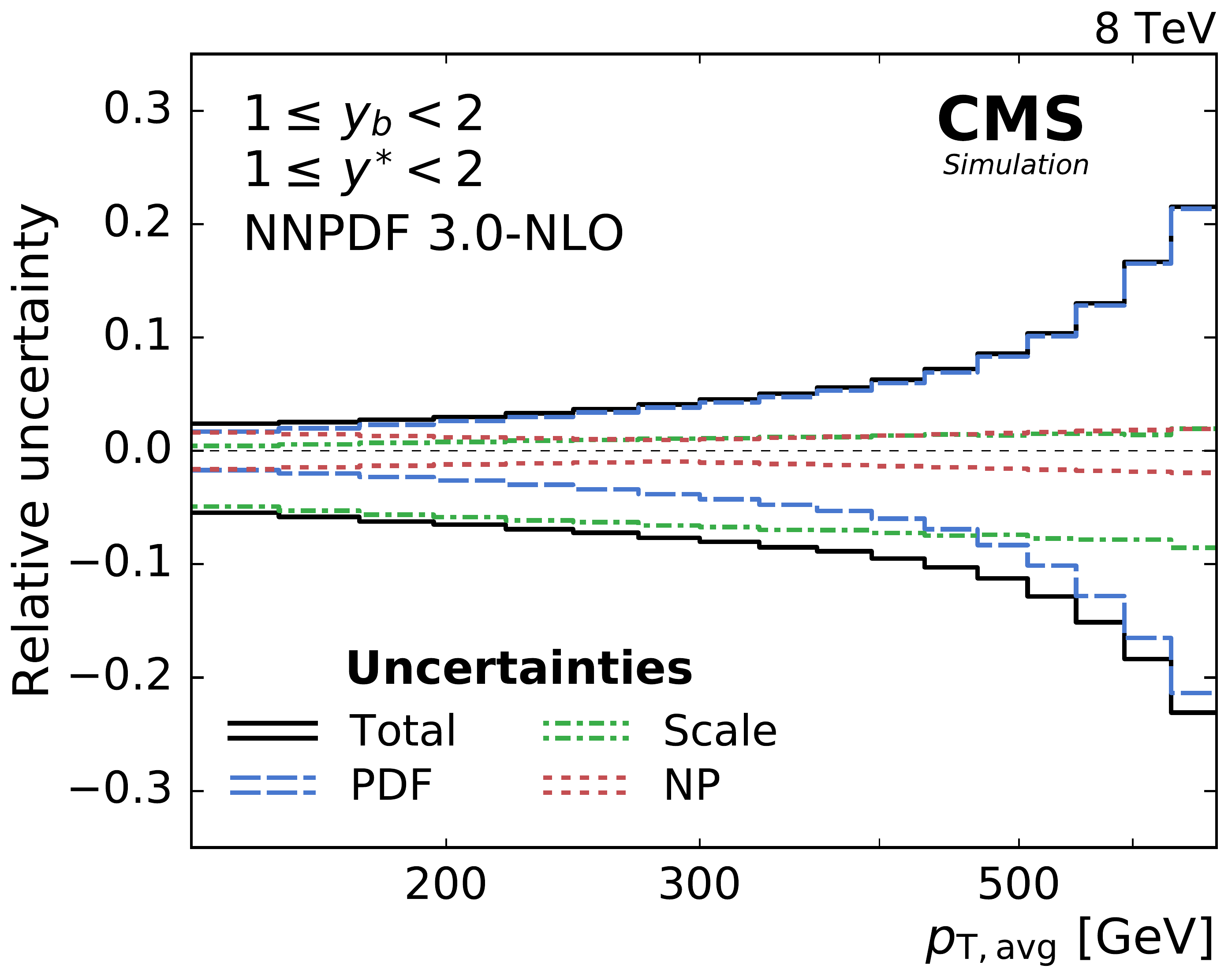}{ \hskip 0.8cm}
\includegraphics[width=0.4\textwidth]{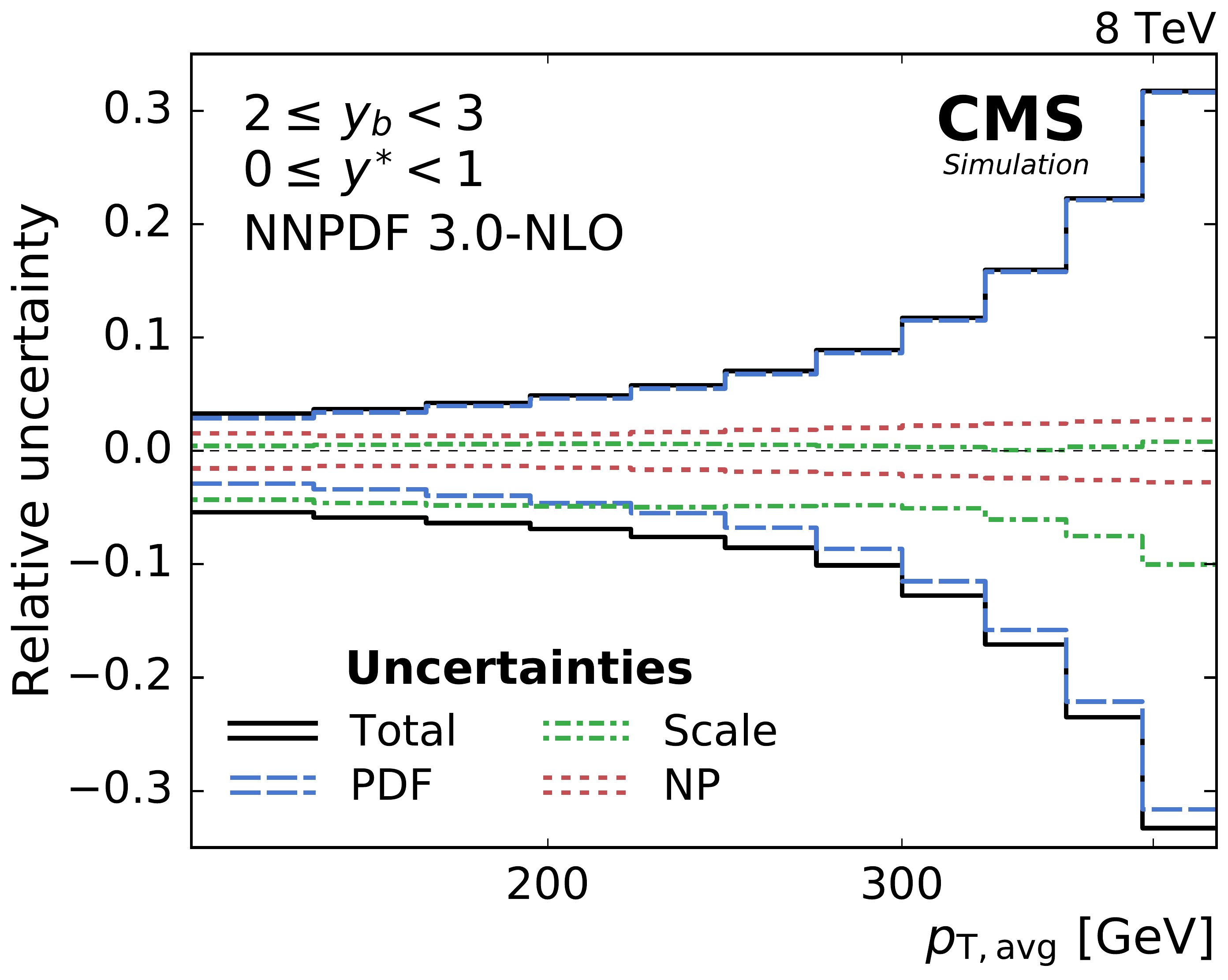}
\caption[Overview of theory uncertainties]%
{Overview of the theoretical uncertainties. The scale uncertainty dominates
in the low-\ptavg region. At high \ptavg, and especially in the
boosted region, the PDFs become the dominant source of
uncertainty.}
\label{fig:theo_uncertainties}
\end{figure*}

The total theoretical uncertainty is obtained as the quadratic sum of
NP, scale, and PDF uncertainties. The scale uncertainties are
calculated by varying \mur and \muf using multiplicative factors in
the following six combinations: $(\mur/\mun, \muf/\mun) = (1/2, 1/2)$,
$(1/2, 1)$, $(1, 1/2)$, $(1, 2)$, $(2, 1)$, and $(2, 2)$. The
uncertainty is determined as the maximal upwards and downwards
variation with respect to the cross section obtained with the nominal
scale setting~\cite{Cacciari:2003fi,Banfi:2010xy}. The PDF
uncertainties are evaluated according to the NNPDF~3.0
prescription as the standard deviation from the average
prediction. Figure~\ref{fig:theo_uncertainties} shows the relative
size of the theoretical uncertainties for the phase-space regions studied. The
scale uncertainty dominates in the low-\ptavg region. At high \ptavg,
and especially in the boosted region, the PDFs become the dominant
source of uncertainty. In total, the theoretical uncertainty increases
from about 2\% at low \ptavg to at least 10\% and up to more than 30\%
for the highest accessed transverse momenta and rapidities.

\section{Results}
\label{sec:results}

\begin{figure*}[htbp]
\centering
\includegraphics[width=0.6\textwidth]{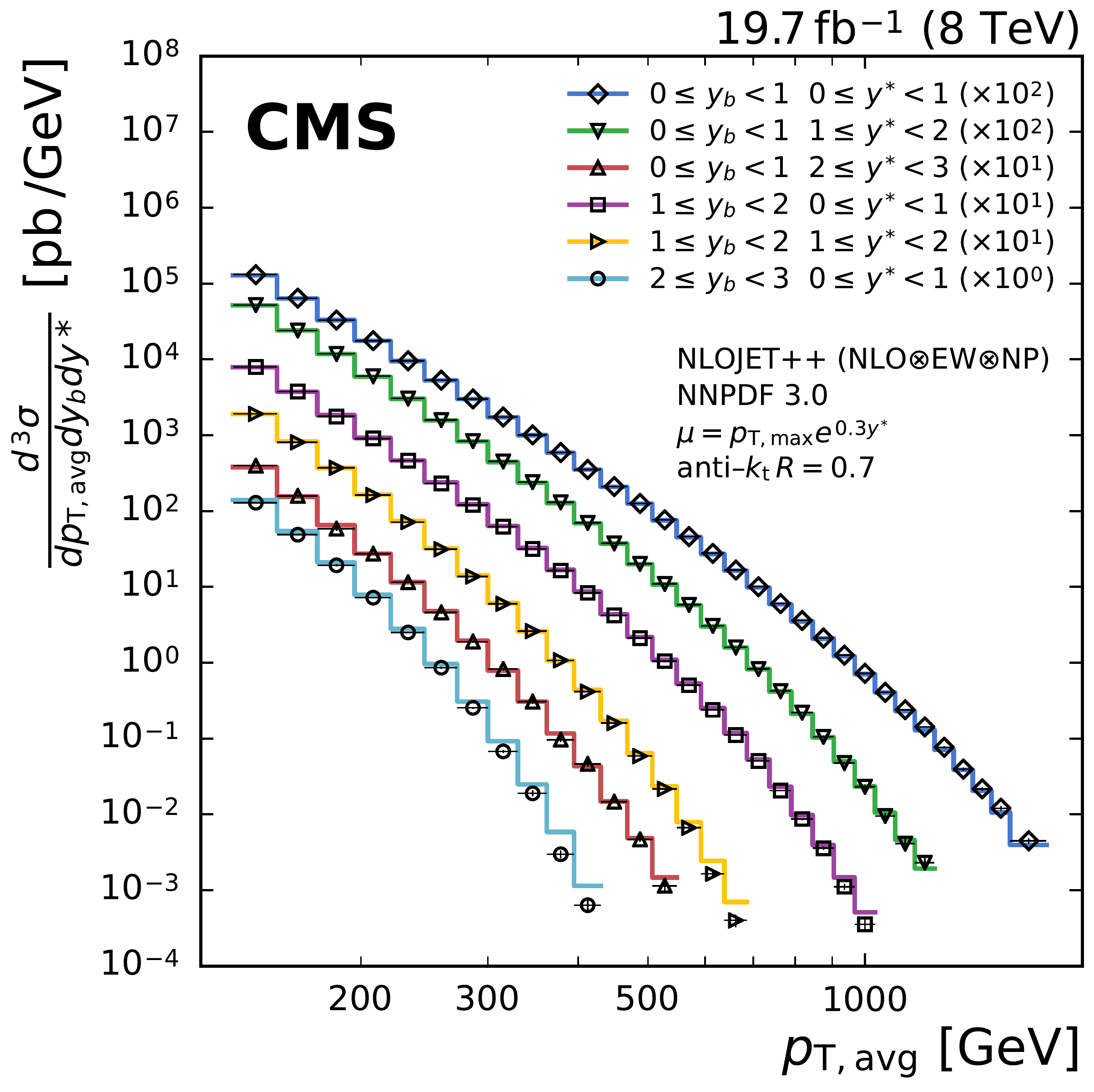}
\caption[Spectrum of the triple-differential dijet cross
section]{The triple-differential dijet cross section in six bins of
\ystar and \yboost. The data are indicated by different markers
for each bin. The theoretical predictions, obtained with \nlojetpp and NNPDF~3.0,
and complemented with EW and NP corrections, are depicted by solid
lines. Apart from the boosted region, the data are well described
by the predictions at NLO accuracy over many orders of magnitude.}
\label{fig:measurement_result}
\end{figure*}

The triple-differential dijet cross section is presented in
Fig.~\ref{fig:measurement_result} as a function of \ptavg for six
phase-space regions in \ystar and \yboost. The theoretical predictions
are found to be compatible with the
unfolded cross section over a wide range of the investigated phase
space.

\begin{figure*}[htbp]
\centering
\includegraphics[width=0.4\textwidth]{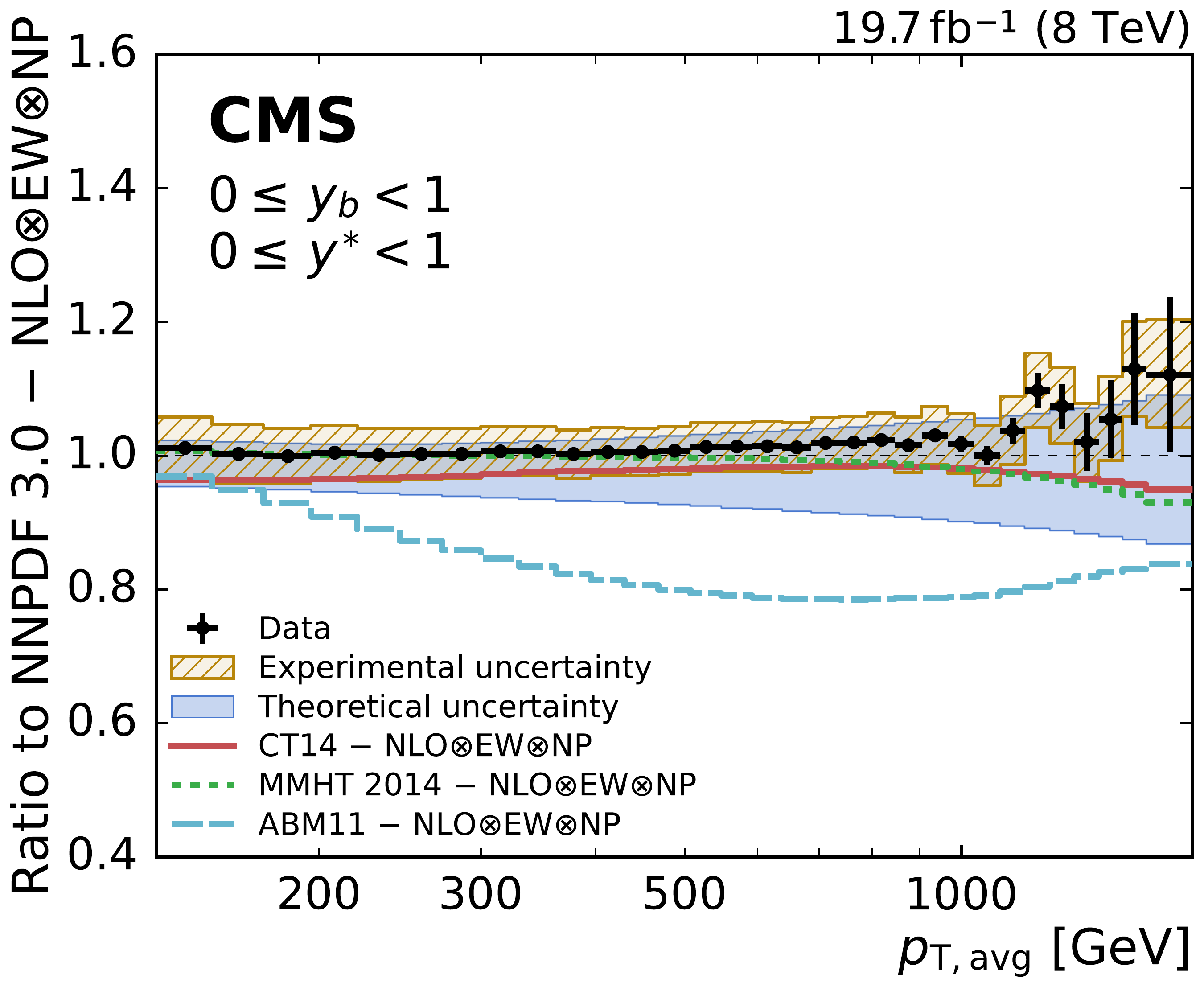}{ \hskip 0.8cm}
\includegraphics[width=0.4\textwidth]{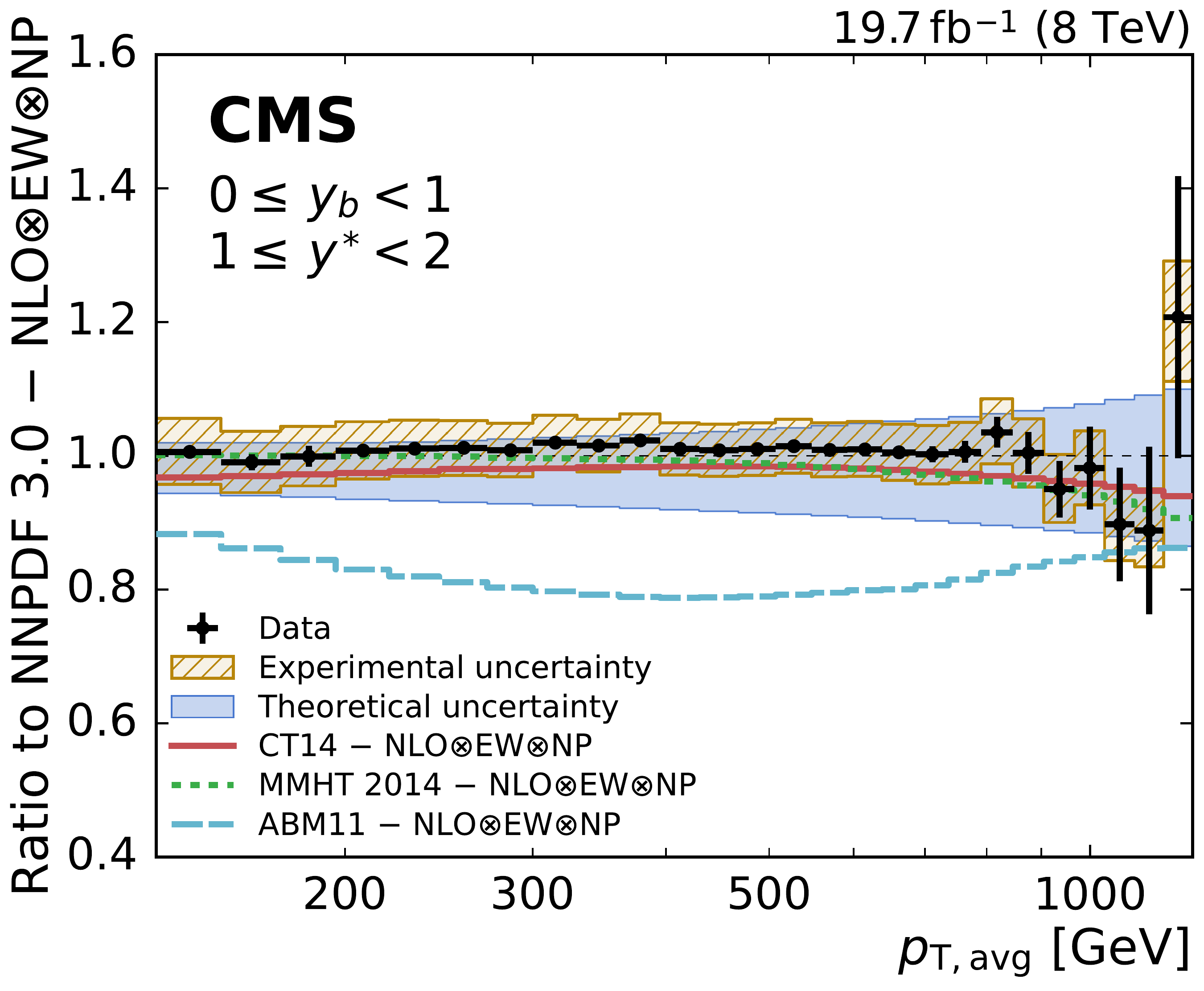}\\
\includegraphics[width=0.4\textwidth]{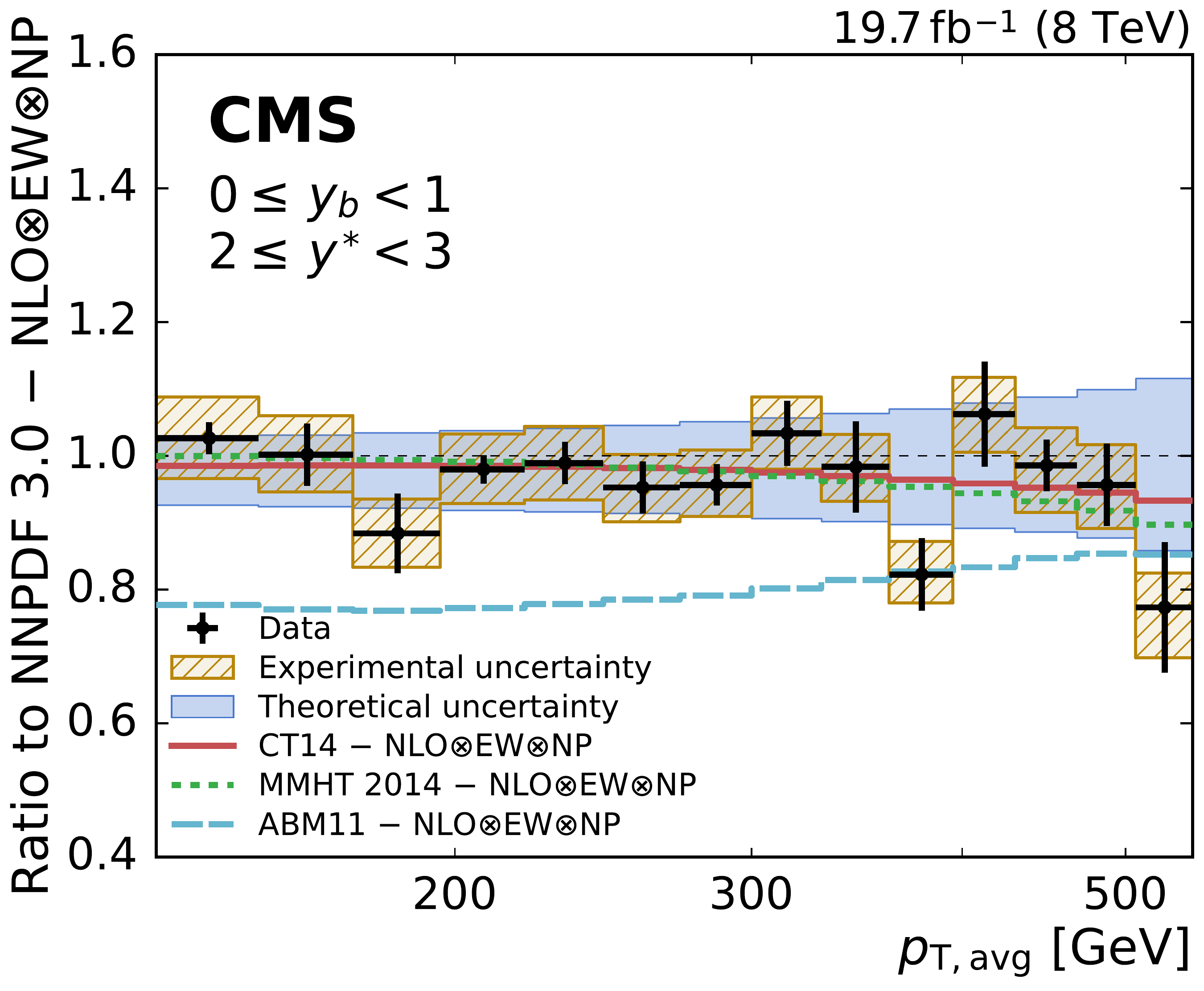}{ \hskip 0.8cm}
\includegraphics[width=0.4\textwidth]{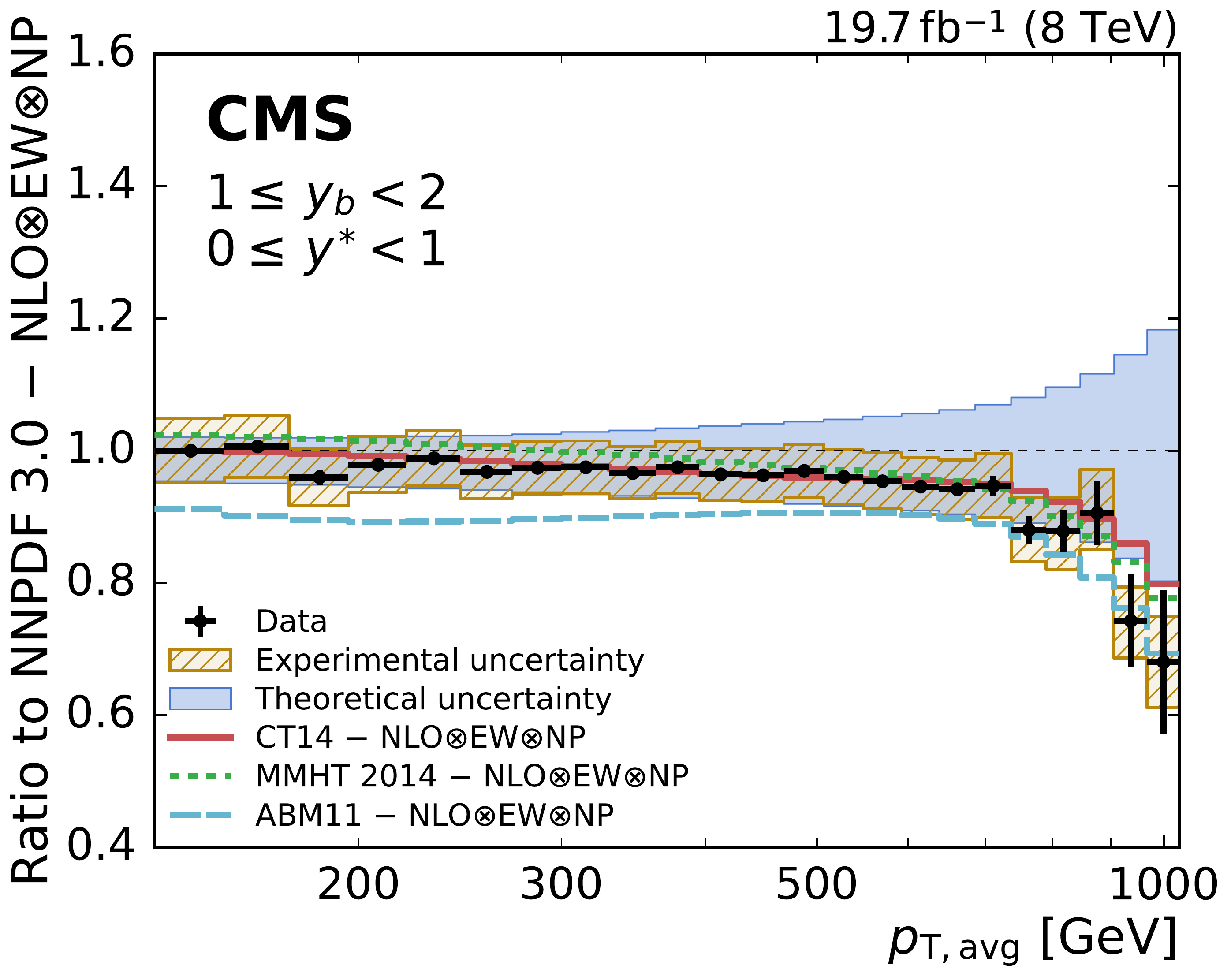}\\
\includegraphics[width=0.4\textwidth]{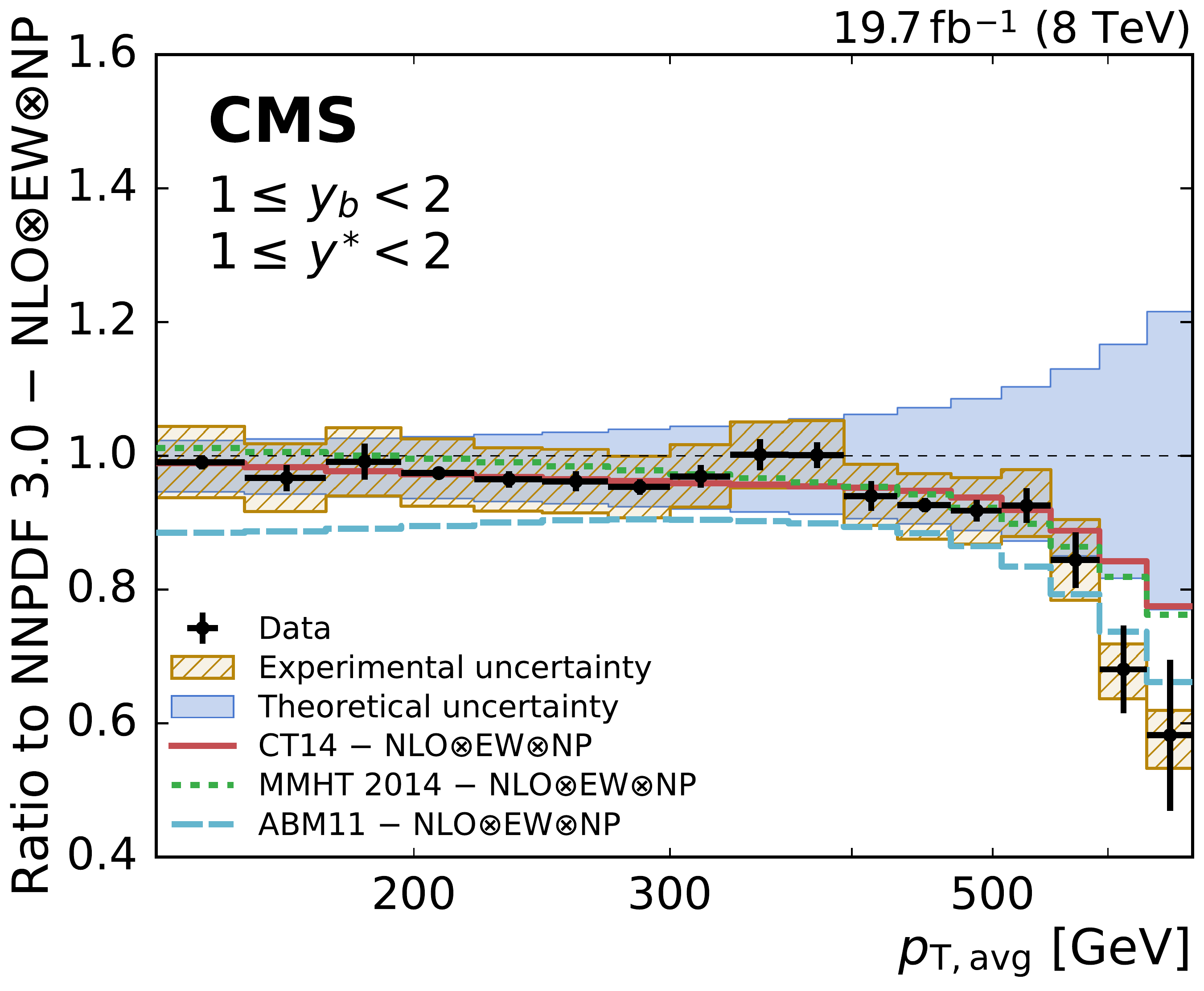}{ \hskip 0.8cm}
\includegraphics[width=0.4\textwidth]{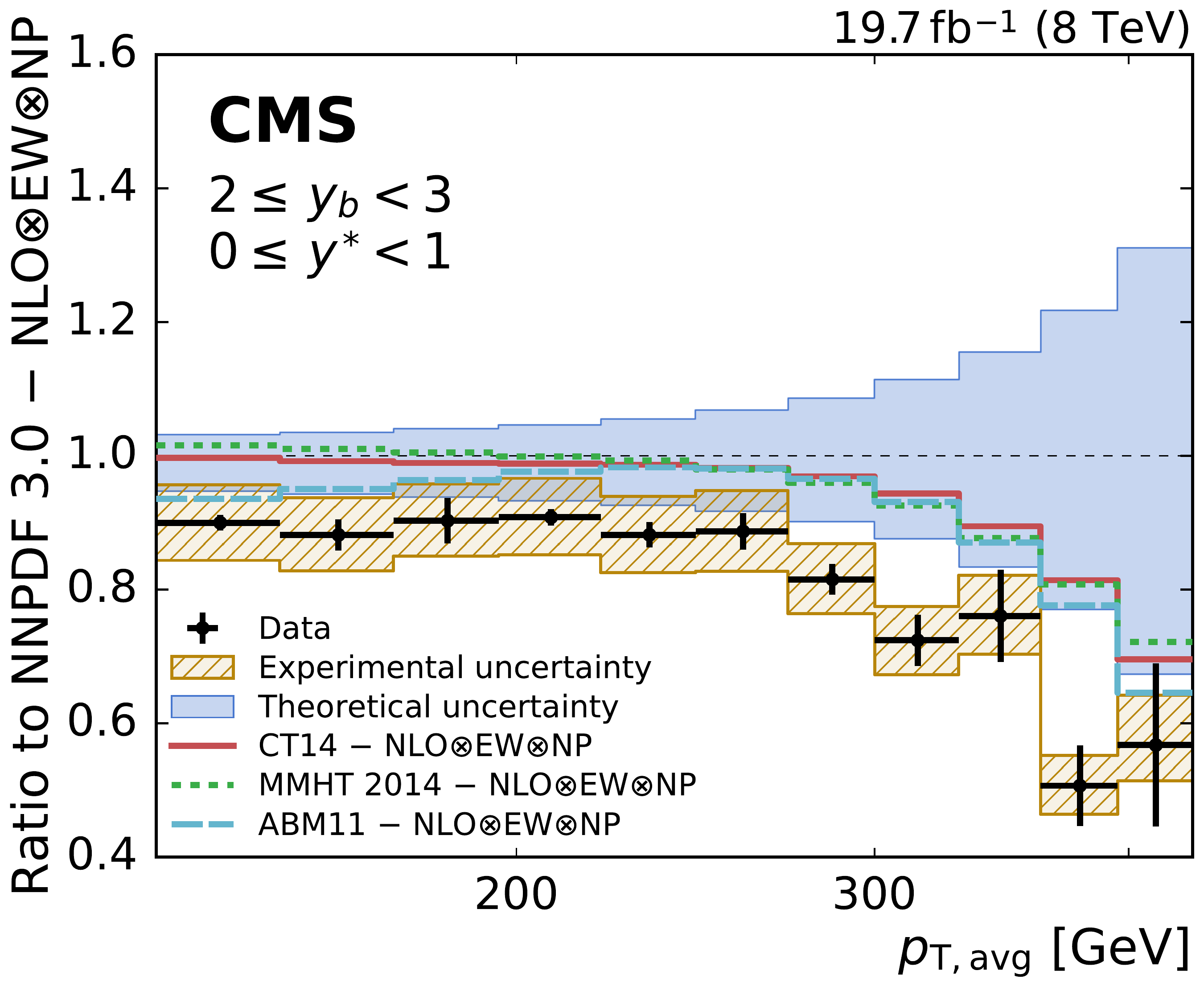}
\caption[Ratio of measured cross section to prediction using
different PDFs]{Ratio of the triple-differential dijet cross section
to the \nlojetpp prediction using the NNPDF~3.0 set. The data
points including statistical uncertainties are indicated by
markers, the systematic experimental uncertainty is represented by the
hatched band. The solid band shows the PDF, scale, and NP
uncertainties quadratically added; the solid and dashed lines give
the ratios calculated with the predictions for different PDF sets.}
\label{fig:ratio_nnpdf30_nlo}
\end{figure*}

The ratios of the measured cross section to the theoretical
predictions from various global PDF sets are shown in
Fig.~\ref{fig:ratio_nnpdf30_nlo}.
The data are well described by the predictions using the
CT14, MMHT~2014, and NNPDF~3.0 PDF sets in most of the analysed phase
space. In the boosted regions ($\yboost \geq 1$) differences between
data and predictions are observed at high \ptavg, where the less known
high-$x$ region of the PDFs is probed. In this boosted dijet topology,
the predictions exhibit large PDF uncertainties, as can be seen in
Fig.~\ref{fig:theo_uncertainties}. The significantly smaller
uncertainties of the data in that region indicate their potential to
constrain the PDFs.

Predictions using the ABM 11 PDFs systematically underestimate the
data for $\yboost<2.0$. This behavior has been observed
previously~\cite{Khachatryan:2014waa} and can be traced back to a soft
gluon PDF accompanied with a low value of \asmz.

\begin{figure*}[htbp]
\centering
\includegraphics[width=0.4\textwidth]{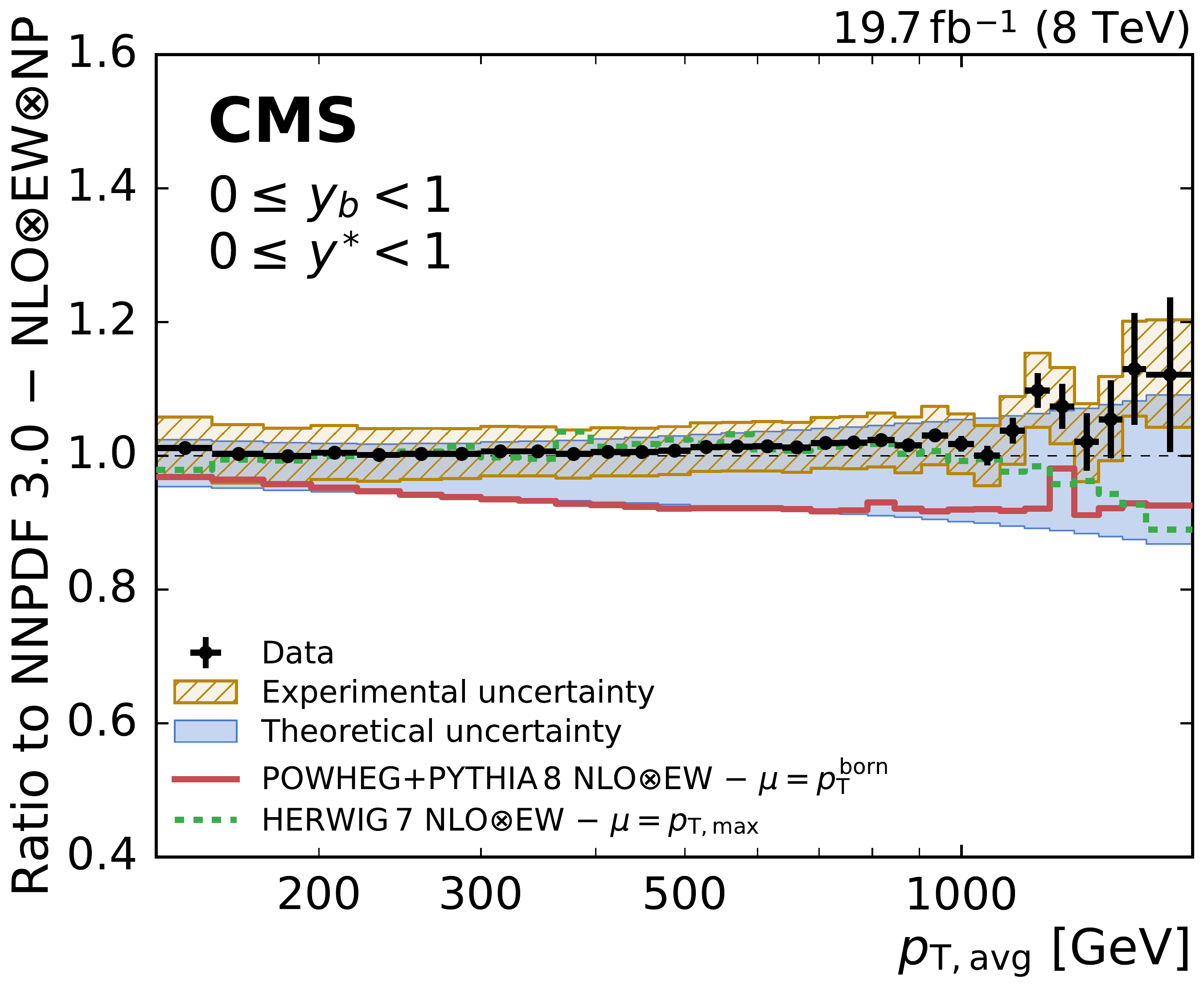}{ \hskip 0.8cm}
\includegraphics[width=0.4\textwidth]{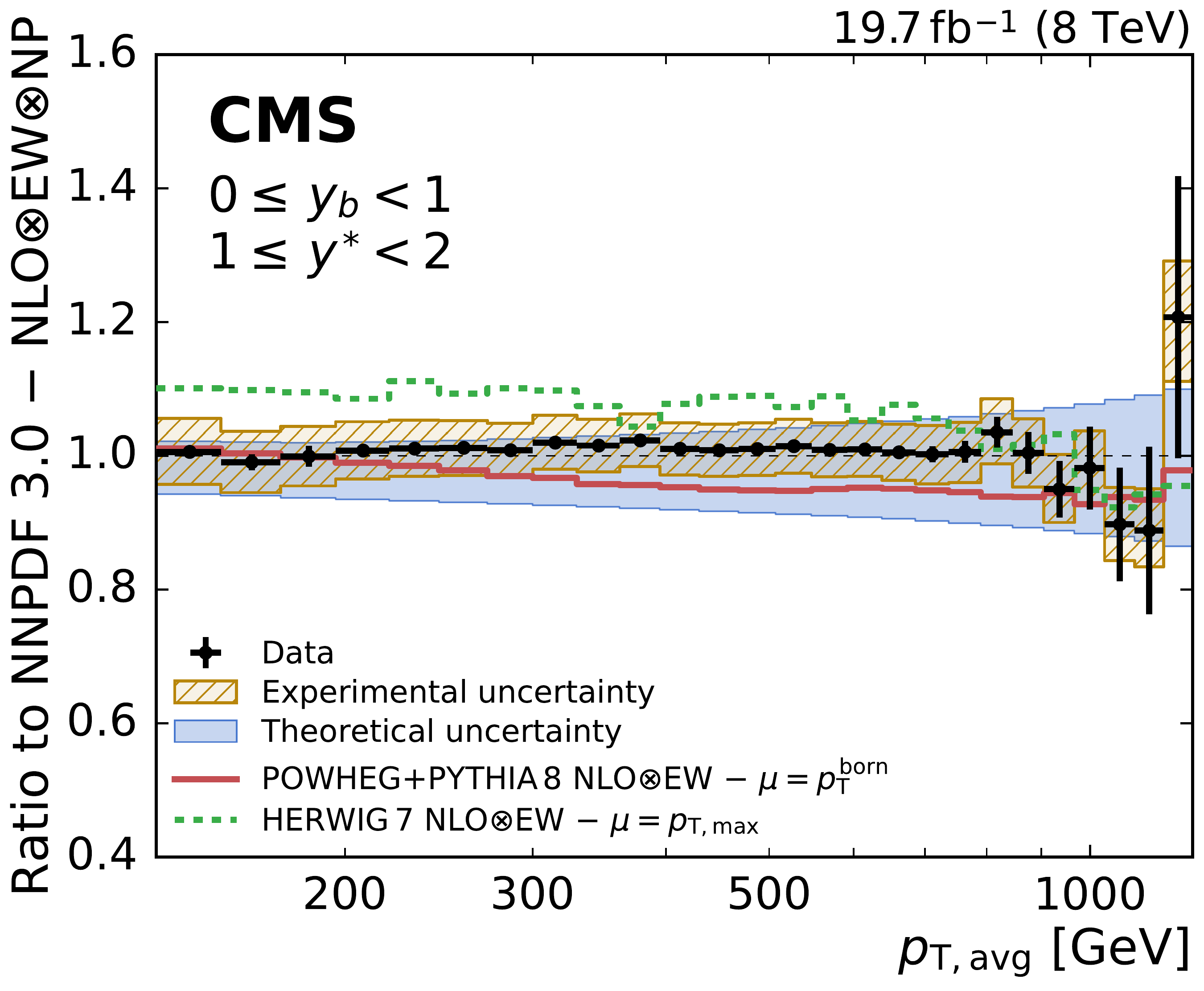}\\
\includegraphics[width=0.4\textwidth]{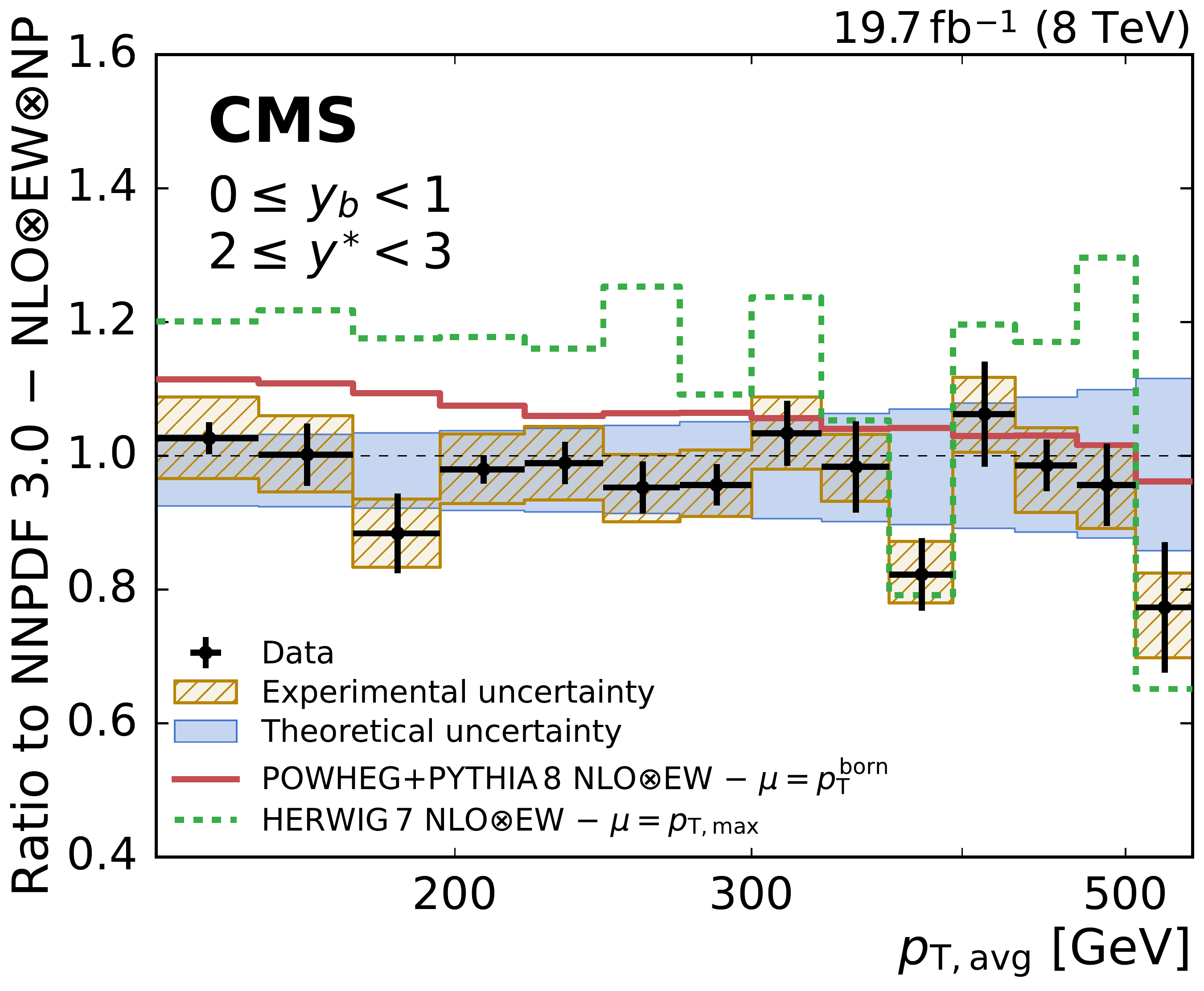}{ \hskip 0.8cm}
\includegraphics[width=0.4\textwidth]{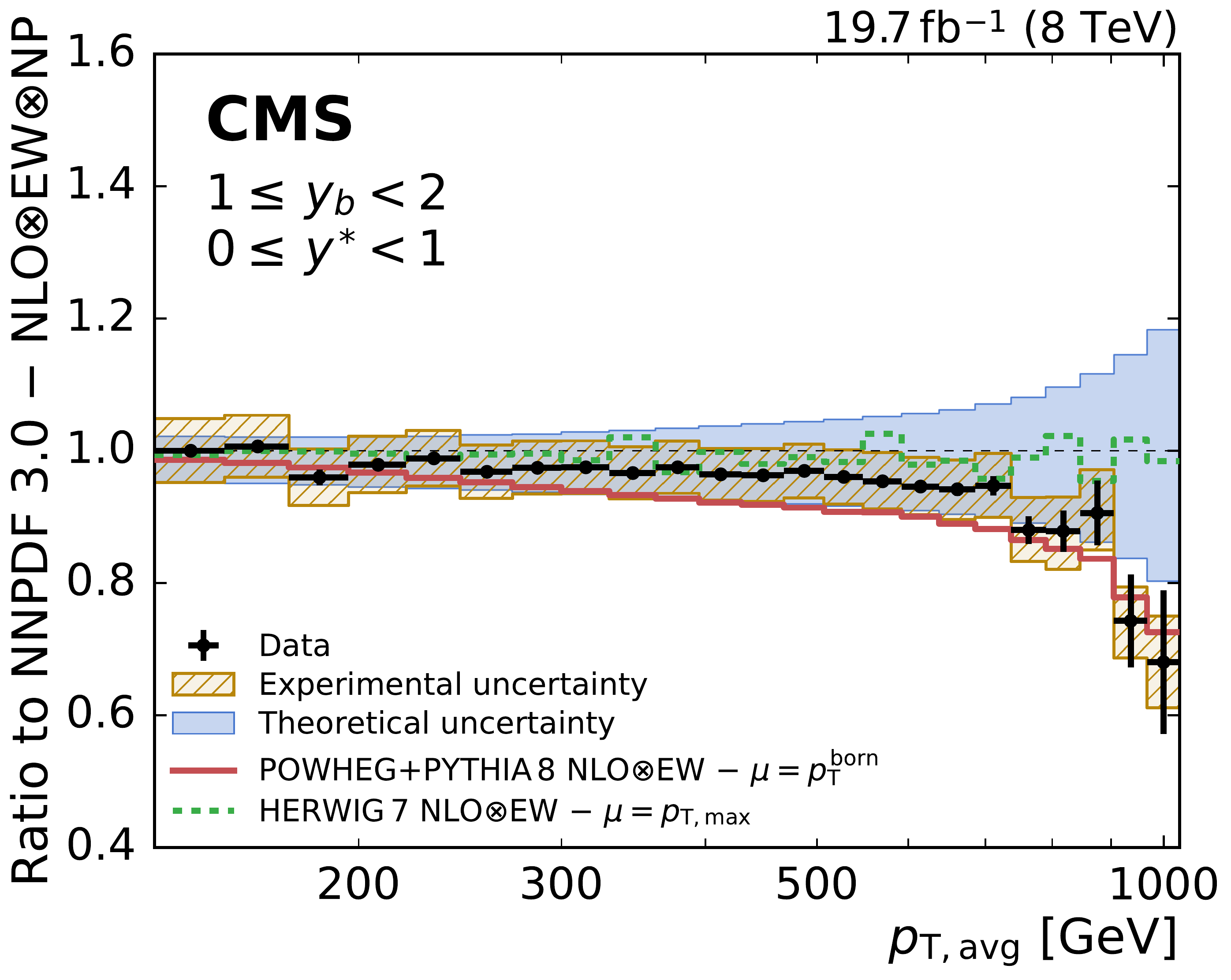}\\
\includegraphics[width=0.4\textwidth]{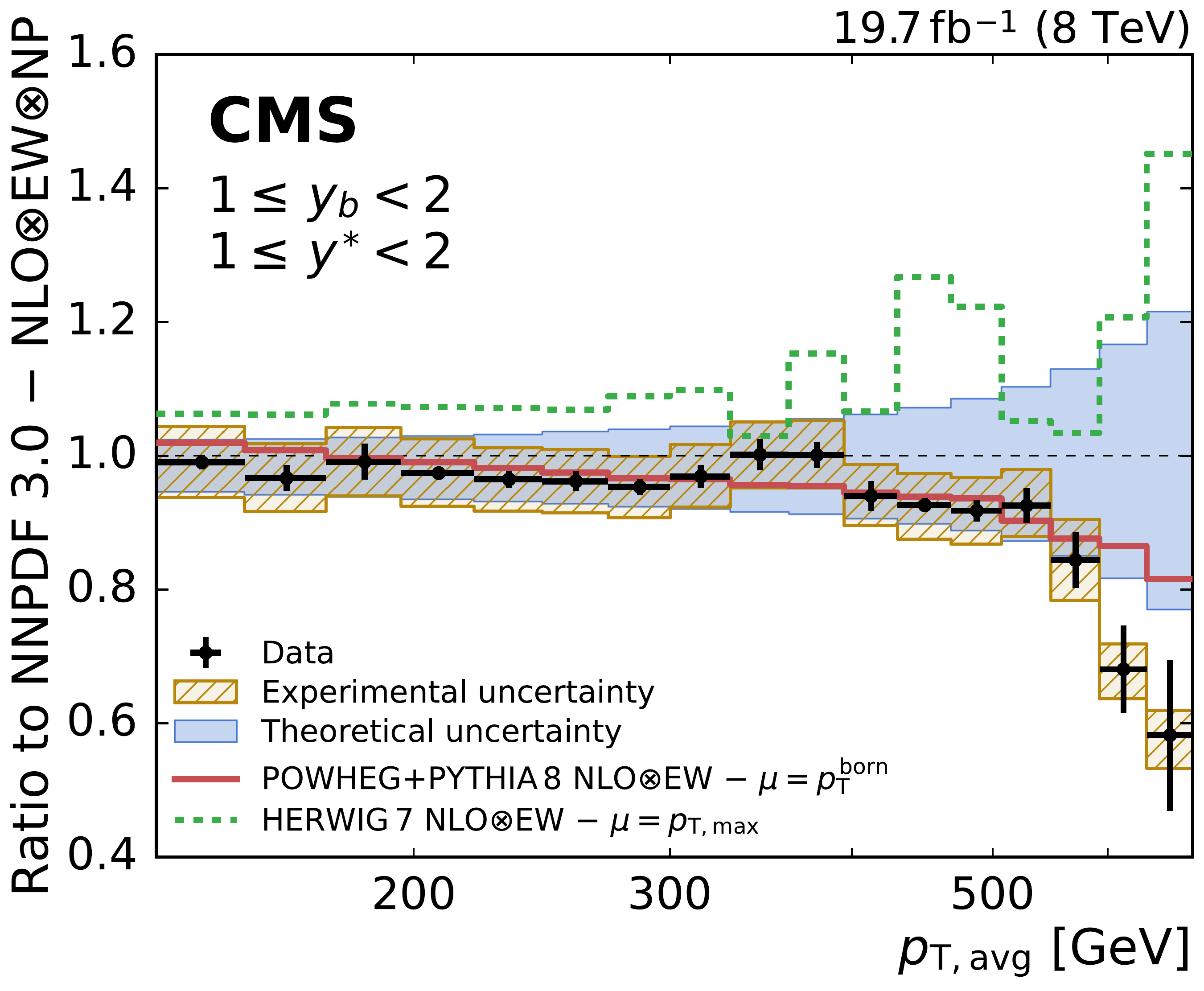}{ \hskip 0.8cm}
\includegraphics[width=0.4\textwidth]{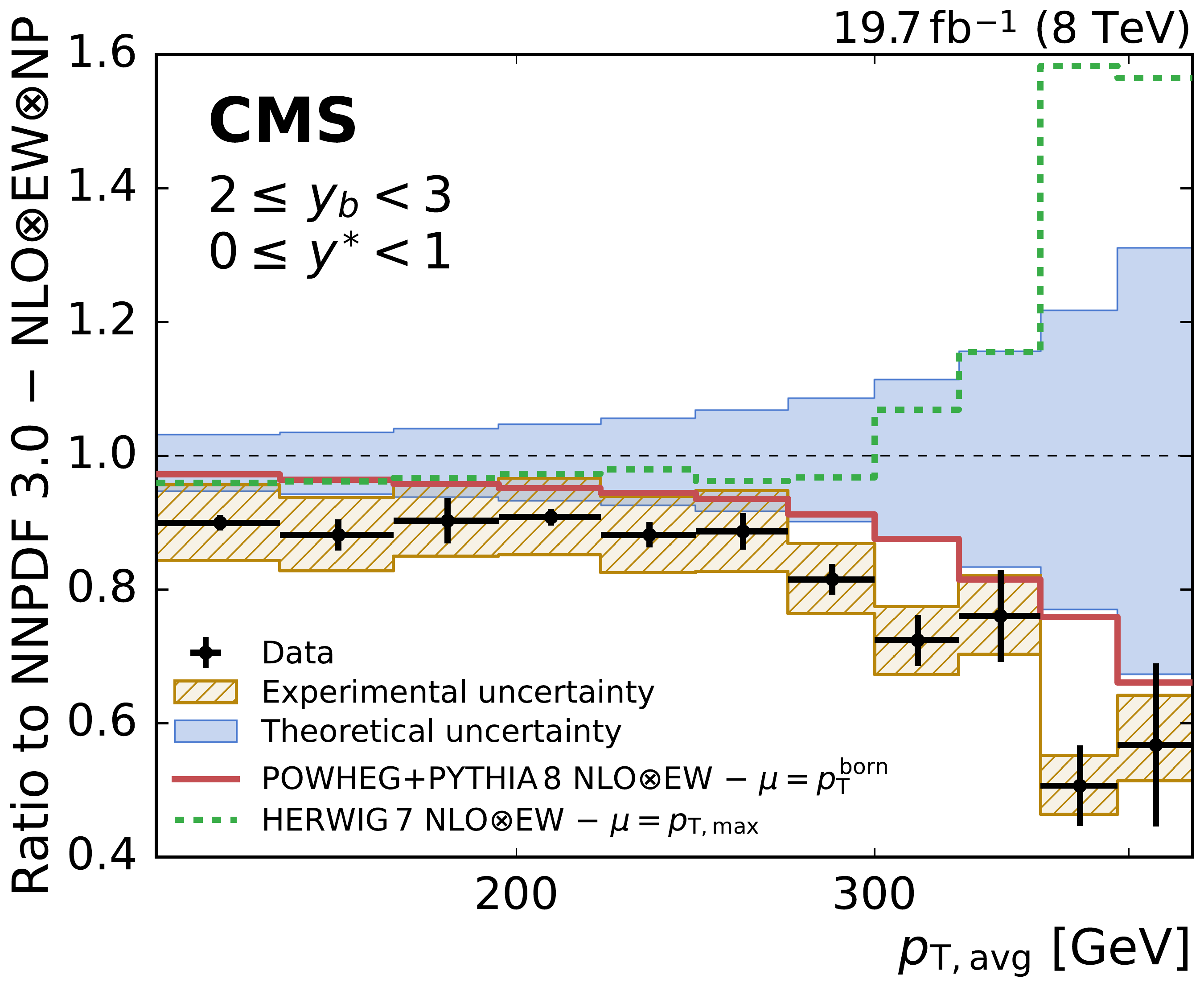}
\caption[Ratio of measured cross section to predictions using
different MC event generators]{Ratio of the triple-differential
dijet cross section to the \nlojetpp prediction using the NNPDF~3.0
set. The data points including statistical uncertainties are
indicated by markers, the systematic experimental uncertainty is
represented by the hatched band. The solid band shows the PDF,
scale, and NP uncertainties quadratically added. The predictions
of the NLO MC event generators \powhegpluspythiae and \herwigs are
depicted by solid and dashed lines, respectively.}
\label{fig:ratio_nnpdf30_mccomp_nlo}
\end{figure*}

Figure~\ref{fig:ratio_nnpdf30_mccomp_nlo} presents the
ratios of the data to the predictions of the \powhegpluspythiae and
\herwig~7.0.3~\cite{Bellm:2015jjp} NLO MC event generators. Significant
differences between the predictions from both MC event generators are
observed. However, the scale definitions and the PDF sets
are different. For \powheg and \herwigs the CT10 and MMHT 2014
PDF sets are used, respectively. In general, \herwigs describes the
data better
in the central region whereas \powheg prevails in the boosted region.

\section{PDF constraints and determination of the strong coupling constant}
\label{sec:pdf_constraints}

The constraints of the triple-differential dijet measurement on the
proton PDFs are demonstrated by including the cross section in a PDF
fit with inclusive measurements of deep-inelastic scattering
(DIS) from the H1 and ZEUS experiments at the HERA
collider~\cite{Abramowicz:2015mha}. The fit is performed with the
open-source fitting framework \xfitter
version~1.2.2~\cite{Alekhin:2014irh}. The PDF evolution is based on
the Dokshitzer--Gribov--Lipatov--Altarelli--Parisi (DGLAP) evolution
equations~\cite{Gribov:1972ri,Altarelli:1977zs,Dokshitzer:1977sg} as
implemented in the \QCDNUM~17.01.12 package~\cite{Botje:2010ay}. To
ensure consistency between the HERA DIS and the dijet cross section
calculations, the fits are performed at NLO\@.

The analysis is based on similar studies of inclusive jet data at
7\TeV~\cite{Khachatryan:2014waa} and 8\TeV~\cite{Khachatryan:2016mlc} and all
settings were chosen in accordance to the inclusive jet study at
8\TeV~\cite{Khachatryan:2016mlc}. The parameterisation of the PDFs is defined at
the starting scale $Q_0^2 = 1.9\GeVsq$. The five independent PDFs $xu_v(x)$,
$xd_v(x)$, $xg(x)$, $x\overbar{U}(x)$, and $x\overbar{D}(x)$
represent the u and d valence quarks, the gluon, and the up- and
down-type sea quarks and are parameterised as follows:
\begin{align}
xg(x) &= A_g x^{B_g} (1-x)^{C_g} - A_g' x^{B_g'}(1-x)^{C_g'}\,,\label{eq:pdfpar1}\\
xu_v(x) &= A_{u_{v}} x^{B_{u_{v}}} (1-x)^{C_{u_{v}}}(1 + D_{u_{v}}x+E_{u_{v}}x^2)\,,\\
xd_v(x) &= A_{d_v} x^{B_{d_v}} (1-x)^{C_{d_{v}}}(1 + D_{d_{v}}x)\,,\\
x\overbar U(x) &= A_{\overbar U} x^{B_{\overbar U}} (1-x)^{C_{\overbar U}}(1 + D_{\overbar U}x)\,,\\
x\overbar D(x) &= A_{\overbar D} x^{B_{\overbar D}} (1-x)^{C_{\overbar D}}\,,\label{eq:pdfpar5}
\end{align}
where $x{\overbar U}(x) = x{\overbar u}(x)$, and $x{\overbar D}(x) =
x{\overbar d}(x) + x{\overbar s}(x)$.

In these equations, the normalisation parameters $A_g$, $A_{u_{v}}$,
and $A_{d_{v}}$ are fixed using QCD sum rules. The constraints $B_{\overbar
U}=B_{\overbar D}$ and $A_{\overbar U} = A_{\overbar D}(1-f_s)$ are
imposed to ensure the same normalisation for the $\overbar U$ and
$\overbar D$ PDF for the $x \rightarrow 0$ region. The strange quark
PDF is defined to be a fixed fraction $f_s = 0.31$ of $x\overbar
D(x)$. The generalised-mass variable-flavour number scheme as
described in~\cite{Thorne:1997ga,Thorne:2006qt} is used and the strong
coupling constant is set to $\asmz= 0.1180$. The set of parameters in
Eqs.~(\ref{eq:pdfpar1})--(\ref{eq:pdfpar5}) is chosen by first
performing a fit where all $D$ and $E$ parameters are set to
zero. Further parameters are included into this set one at a
time. The improvement of \chisq of the fit is monitored and the
procedure is stopped when no further improvement is observed. This
leads to a 16-parameter fit. Due to differences in the sensitivity
of the various PDFs to dijet and inclusive jet data, the
parameterisation of the present analysis differs from that
in Ref.~\cite{Khachatryan:2016mlc}. In particular, the constraint
$B_{d_v}=B_{u_v}$ at the starting scale has been released. This
results in a d~valence quark distribution consistent with the results
obtained in Ref.~\cite{Khachatryan:2016mlc} and in a similar CMS
analysis of muon charge asymmetry in W~boson production at
8\TeV~\cite{Khachatryan:2016pev}.

The PDF uncertainties are determined using the HERAPDF
method~\cite{Abramowicz:2015mha, Alekhin:2014irh} with
uncertainties subdivided into the three categories of experimental,
model, and parameterisation uncertainty, which are evaluated
separately and added in quadrature to obtain the total PDF
uncertainty.

{\tolerance=1400
Experimental uncertainties originate from statistical and systematic
uncertainties in the data and are propagated to the PDFs using the Hessian
eigenvector method~\cite{Pumplin:2001ct} and a tolerance criterion of
$\Delta\chisq = + 1$. Alternatively, the Monte Carlo
method~\cite{Giele:1998gw} is used to determine the PDF fit uncertainties and
similar results are obtained.
\par}

The uncertainties in several input parameters in the PDF fits are combined into
one model uncertainty. For the evaluation of the model uncertainties
some variations on the input parameters are considered.
The strangeness fraction is chosen in agreement with
Refs.~\cite{Chatrchyan:2013mza, Chatrchyan:2013mza} to be $f_s=0.31$
and is varied between $0.23$ and $0.39$. %
Following Ref.~\cite{Abramowicz:2015mha}, the b quark mass, set to $4.5
\GeV$, is varied between 4.25 and $4.75\GeV$. Similarly, the
c quark mass, set by default to $1.47\GeV$, is varied between
1.41 and $1.53\GeV$. %
The minimum $Q^2$ imposed on the HERA DIS data is set in accordance
with the CMS inclusive jet analysis described
in~\cite{Khachatryan:2014waa} to $Q^2_\mathrm{min}=7.5\GeVsq$, and is
varied between $Q^2_\mathrm{min} = 5.0\GeVsq$ and $10.0\GeVsq$.

The parameterisation uncertainty is estimated by including additional
parameters in the fit, leading to a more flexible functional form of
the PDFs. Each parameter is successively added in the PDF fit, and the
envelope of all changes to the central PDF fit result is taken as
parameterisation uncertainty. The increased flexibility of the PDFs
while estimating the parameterisation uncertainty may lead to the
seemingly paradoxical effect that, although new data are included, the
total uncertainty can increase in regions, where direct constraints from data are
absent. This may happen at very low or at very high
$x$, where the PDF is determined through extrapolation
alone. Furthermore, the variation of the starting scale $Q_0^2$ to 1.6
and 2.2\GeVsq is considered in this parameterisation uncertainty.

The quality of the resulting PDF fit with and without the dijet measurement is
reported in Table~\ref{tab:fit:results}. The partial \chisq per data point for
each data set as well as the \chisqndof for
all data sets demonstrate the compatibility of the CMS dijet measurement and the
DIS data from the H1 and ZEUS experiments in a combined fit.

\begin{table*}[htbp]
\setlength\tabcolsep{5pt}
\topcaption[Fit quality in the HERA DIS and combined fit]{The partial
\chisq (\chipsq) for each data set in the HERA DIS (middle
section) or the combined fit including the CMS triple-differential dijet data
(right section) are shown. The bottom two lines show the total \chisq and
\chisqndof. The difference between the sum of all
\chipsq and the total \chisq for the combined fit is attributed to
the nuisance parameters.}
\label{tab:fit:results}
\centering
\begin{tabular}{lccccc}
\hline
\multicolumn{2}{c}{} &
\multicolumn{2}{c}{HERA data} &
\multicolumn{2}{c}{HERA \& CMS data}\rbtrr\\
{Data set} &
{\ndata} &
{\chipsq} &
{\chipsqndata} &
{\chipsq} &
{\chipsqndata}\rbthm\\
NC HERA-I+II \epp \ $E_{\mathrm{p}} = 920\GeV \ $ & 332 & 382.44 & 1.15 & 406.45 & 1.22  \rbtrr\\
NC HERA-I+II \epp \ $E_{\mathrm{p}} = 820\GeV$ \  & \x63 & \x60.62 & 0.96 & \x61.01 & 0.97  \rbtrr\\
NC HERA-I+II \epp \ $E_{\mathrm{p}} = 575\GeV \ $ & 234 & 196.40 & 0.84 & 197.56 & 0.84  \rbtrr\\
NC HERA-I+II \epp \ $E_{\mathrm{p}} = 460\GeV \ $ & 187 & 204.42 & 1.09 & 205.50 & 1.10  \rbtrr\\
NC HERA-I+II \emp & 159 & 217.27 & 1.37 & 219.17 & 1.38  \rbtrr\\
CC HERA-I+II \epp & \x39 & \x43.26 & 1.11 & \x42.29 & 1.08  \rbtrr\\
CC HERA-I+II \emp & \x42 & \x49.11 & 1.17 & \x55.35 & 1.32  \rbtrr\\
CMS triple-differential dijet & 122 & --- & --- & 111.13 & 0.91  \rbtrr\\
\\\hline
{Data set(s)} & \ndof &
{\chisq} &
{\chisqndof} &
{\chisq} &
{\chisqndof}\rbthm\\
HERA data                       & 1040\x & 1211.00\x & 1.16  &  --- &  --- \rbtrr\\
HERA \& CMS data                & 1162\x &    --- &  --- & 1372.52\x & 1.18 \rbtrr\\
\hline
\end{tabular}
\end{table*}

\begin{figure*}[tbp]
\centering
\includegraphics[width=0.4\textwidth]{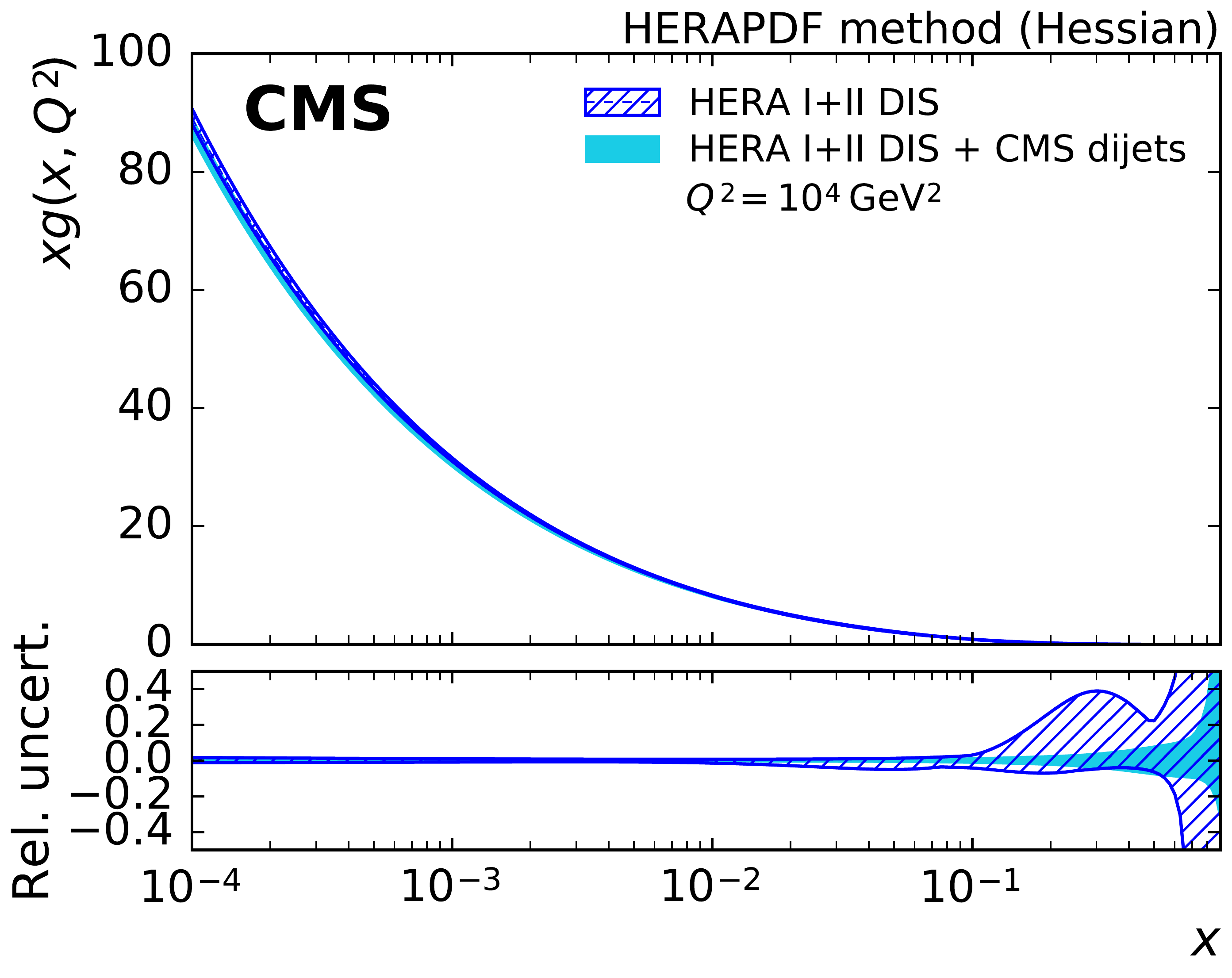}{ \hskip 0.8cm}
\includegraphics[width=0.4\textwidth]{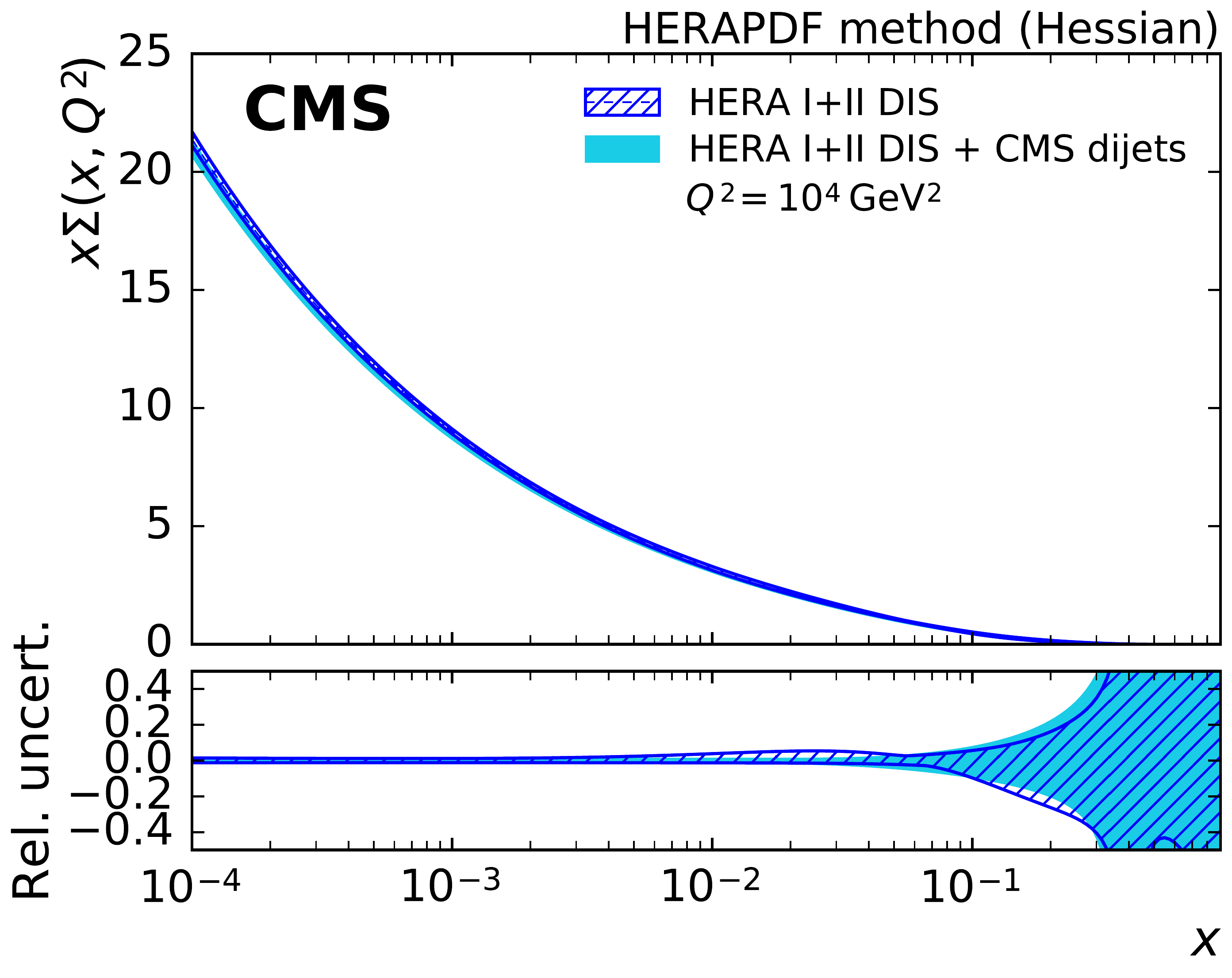}\\
\includegraphics[width=0.4\textwidth]{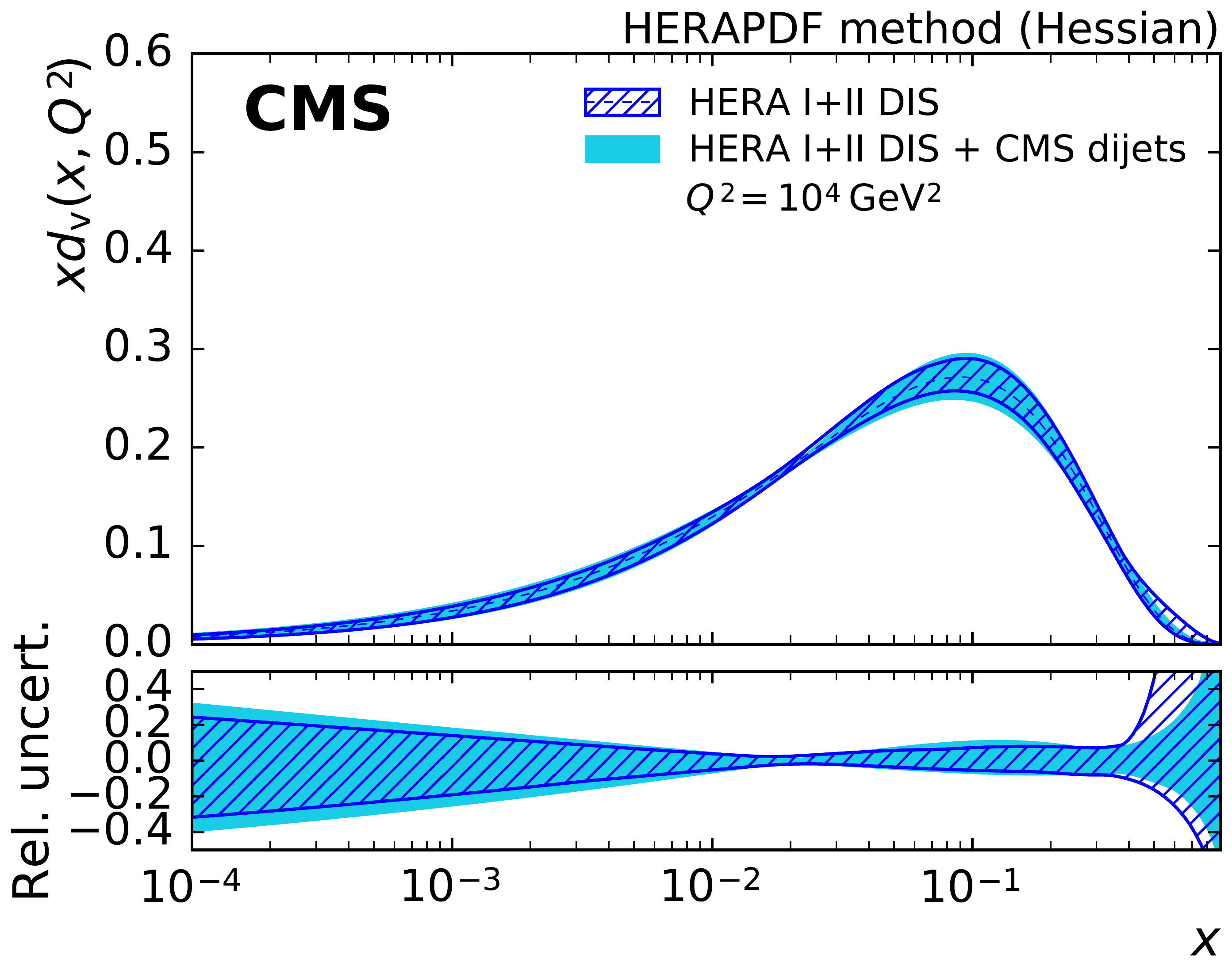}{ \hskip 0.8cm}
\includegraphics[width=0.4\textwidth]{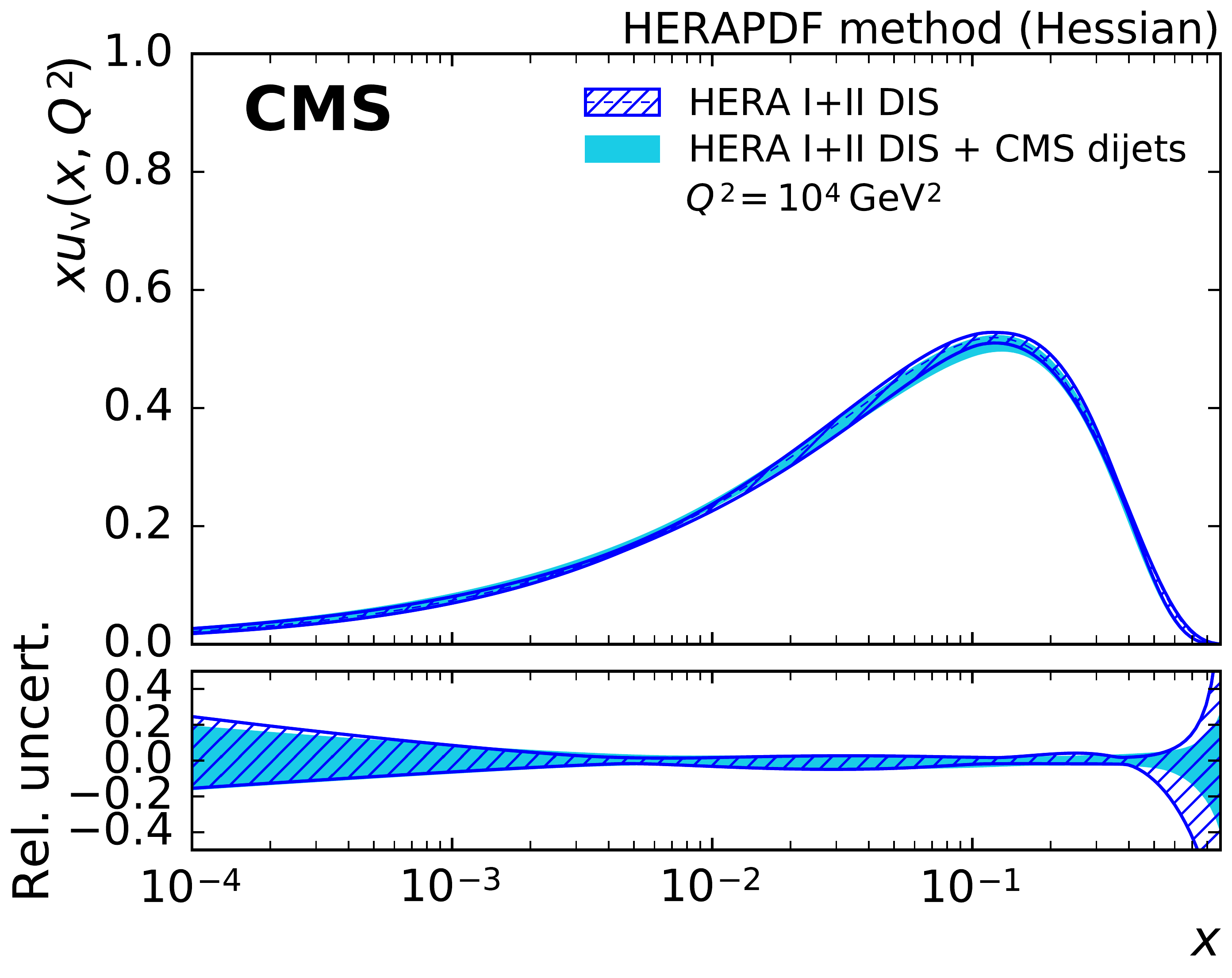}
\caption[Direct comparison of gluon and quark PDFs at the scale $Q^2 = 10^{4}\GeVsq$]%
{The gluon (top left), sea
quark (top right), d~valence quark (bottom left), and u~valence quark (bottom
right) PDFs as a function of $x$ as derived from HERA inclusive DIS data
alone (hatched band) and in combination with CMS dijet data (solid band). The PDFs
are shown at the scale $Q^2 = 10^{4}\GeVsq$ with their total uncertainties.}
\label{fig:pdfconstraints:direct:10000}
\end{figure*}

The PDFs obtained for the gluon, u~valence, d~valence, and sea quarks
are presented for a fit with and without the CMS dijet data in
Fig.~\ref{fig:pdfconstraints:direct:10000} for $Q^2=10^4\GeVsq$. The
uncertainty in the gluon PDF is reduced over a large range in $x$ with
the largest impact in the high-$x$ region, where some reduction in
uncertainty can also be observed for the valence quark and the sea
quark PDFs.
For $x$ values beyond ${\approx}0.7$ or below $10^{-3}$, the extracted
PDFs are not directly constrained by data and should be considered as
extrapolations that rely on PDF parameterisation assumptions alone.

\begin{figure*}[tbp]
\hftwo\includegraphics[width=0.4\textwidth]{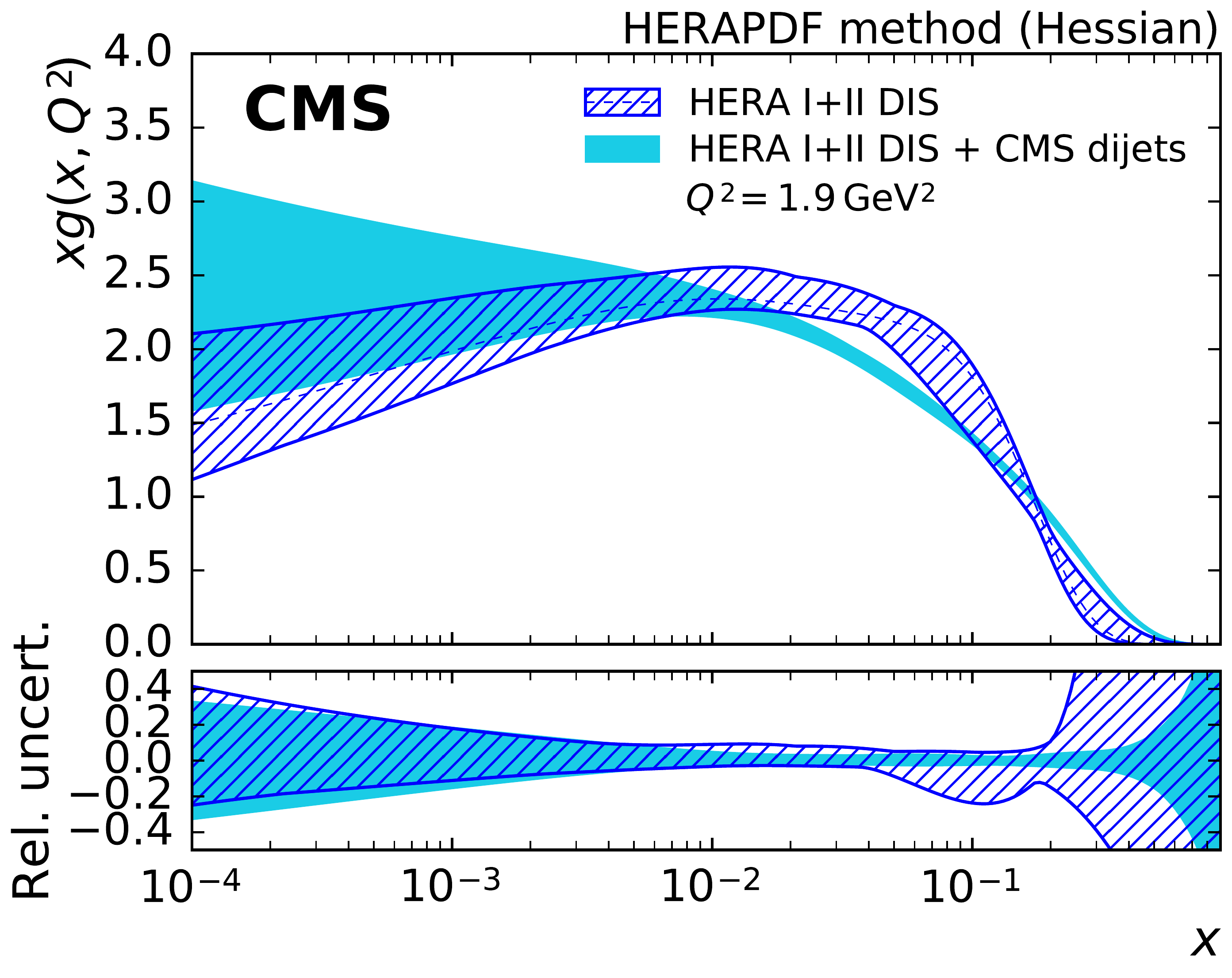}\hftwo
\caption[Direct comparison of gluon PDFs at the starting scale $Q^2
= 1.9\GeVsq$]
{The gluon PDF as a function of $x$ as derived from HERA inclusive
DIS data alone (hatched band) and in combination with CMS dijet
data (solid band). The PDF and its total uncertainty are shown at
the starting scale $Q^2 = 1.9\GeVsq$ of the PDF evolution.}
\label{fig:pdfconstraints:direct:19}
\end{figure*}

The improvement in the uncertainty of the gluon PDF is accompanied by
a noticeable change in shape, which is most visible when evolved to
low scales as shown in Fig.~\ref{fig:pdfconstraints:direct:19}.
Compared to the fit with HERA DIS data alone, the gluon PDF shrinks at
medium $x$ and increases at high $x$. A similar effect has been
observed before, \eg in Ref.~\cite{Khachatryan:2014waa}.

\begin{figure*}[tbp]
\centering
\includegraphics[width=0.4\textwidth]{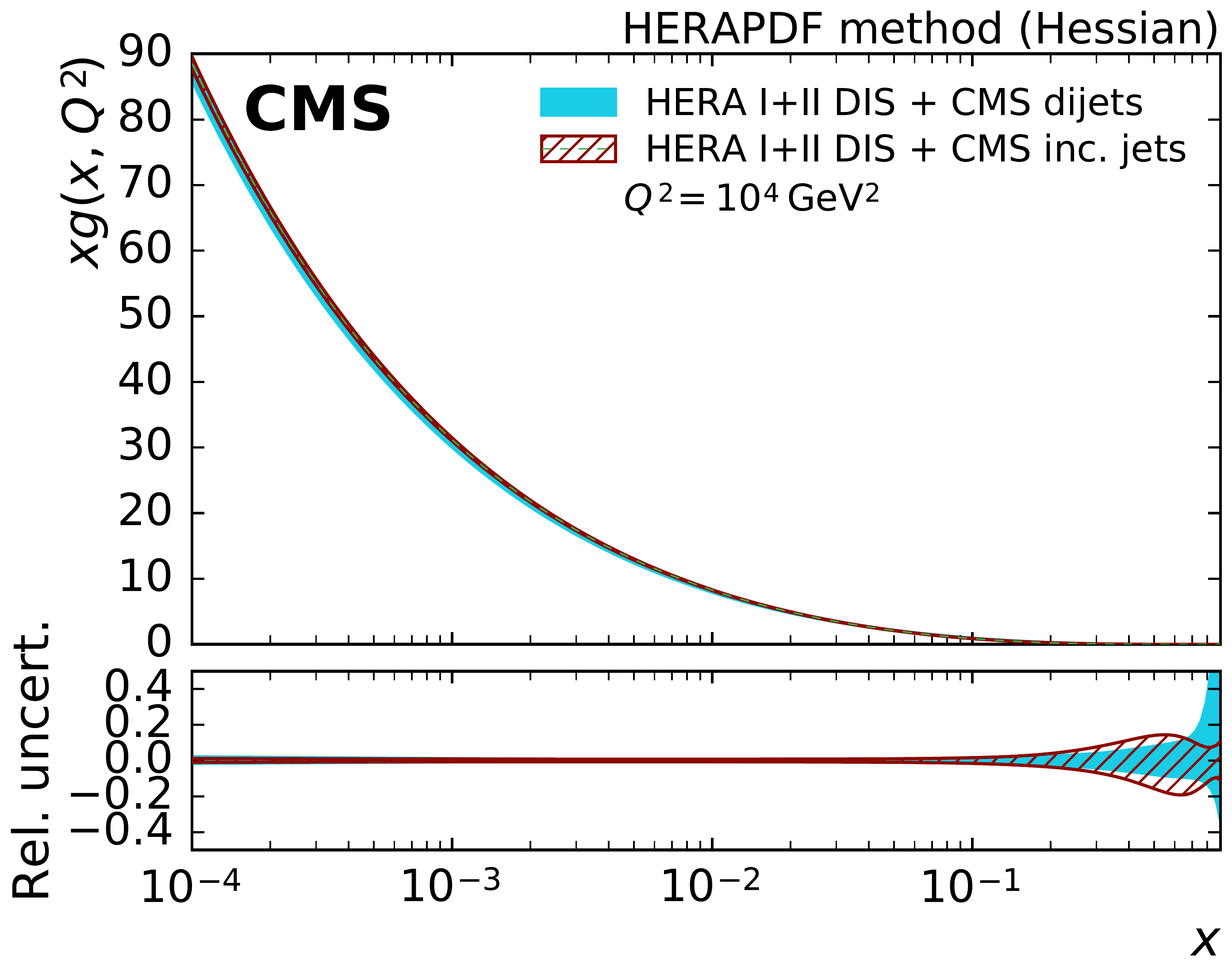}{\hskip 0.8cm}
\includegraphics[width=0.4\textwidth]{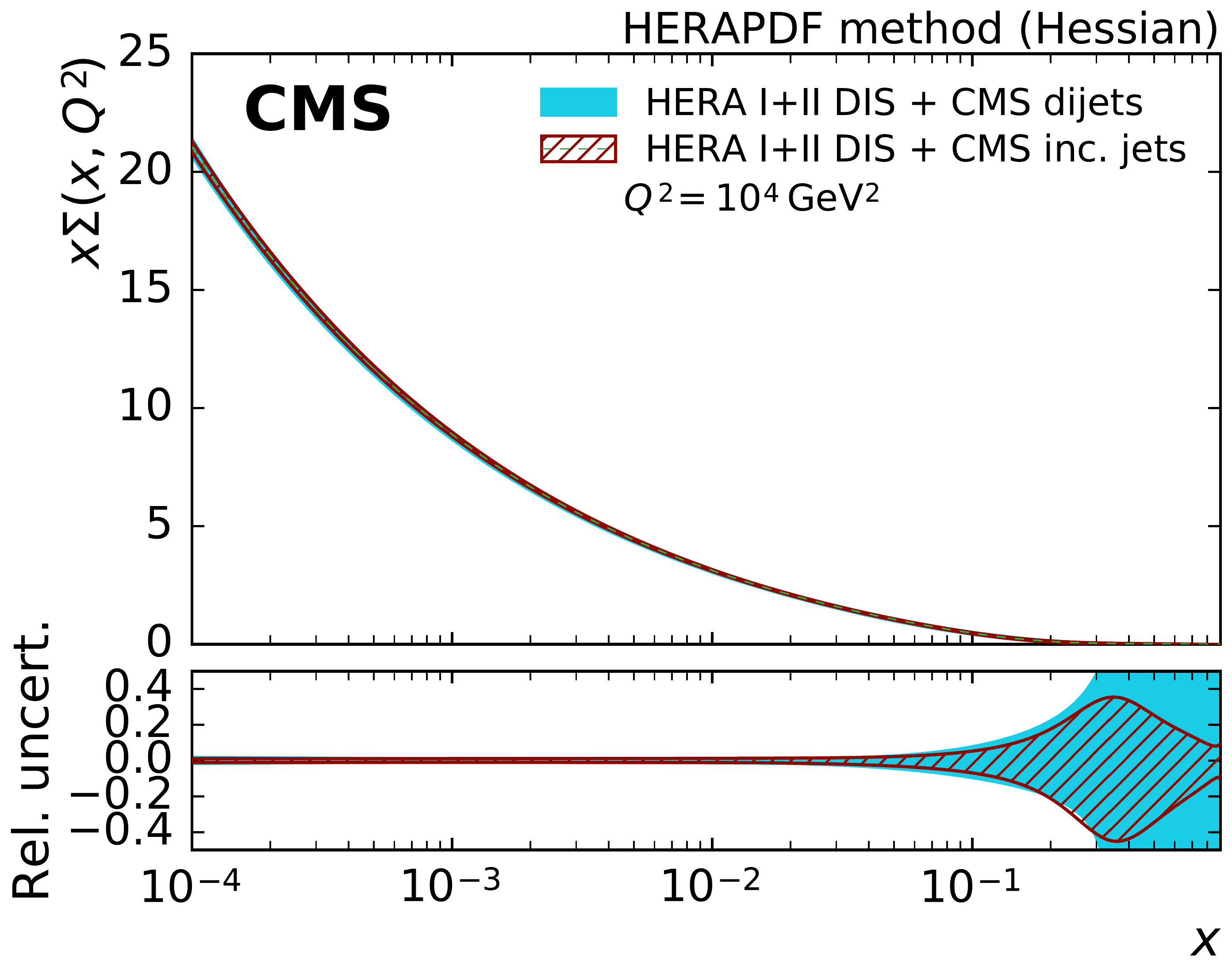}\\
\includegraphics[width=0.4\textwidth]{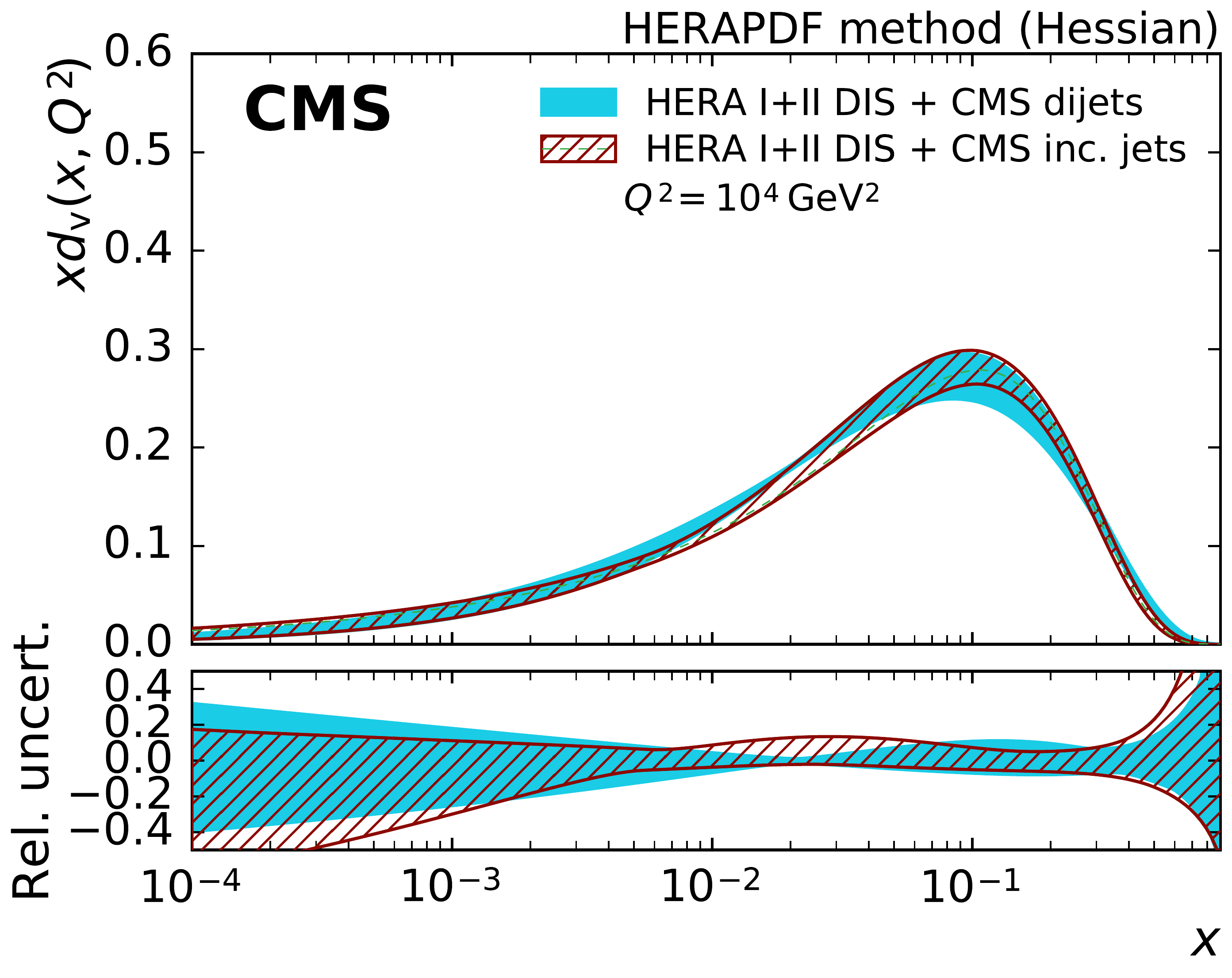}{\hskip 0.8cm}
\includegraphics[width=0.4\textwidth]{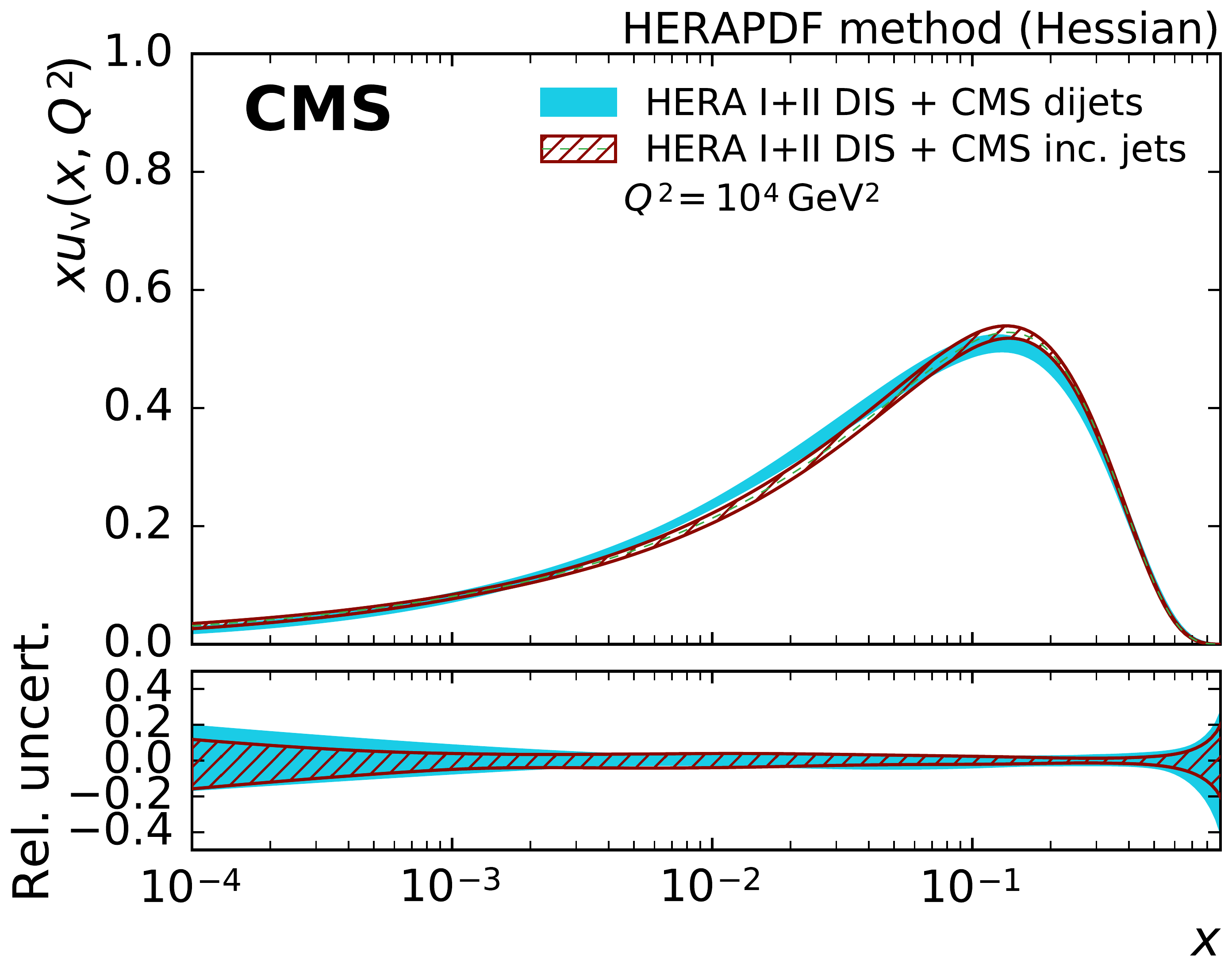}
\caption[Direct comparison of gluon and quark PDFs]
{The gluon (top left), sea
quark (top right), d~valence quark (bottom left), and u~valence quark (bottom
right) PDFs as a function of $x$ as derived from a fit of HERA inclusive DIS
data in combination with CMS inclusive jet data (solid band) and CMS dijet
data (hatched band) at 8 TeV. The PDFs
are shown at the scale $Q^2 = 10^{4}\GeVsq$ with their total uncertainties.}
\label{fig:pdfconstraints:direct:incjetcomp:10000}
\end{figure*}

The PDFs are compared in Fig.~\ref{fig:pdfconstraints:direct:incjetcomp:10000}
to those obtained with inclusive jet data at
$\sqrt{s} = 8\TeV$~\cite{Khachatryan:2016mlc}. The shapes of
the PDFs and the uncertainties are similar. Somewhat larger
uncertainties in the valence quark distributions are observed in the
fit using the dijet data with respect to those obtained from the inclusive
jet cross section. This behaviour can be explained by a stronger
sensitivity of the dijet data to the light quark distributions,
resulting in an increased flexibility of the PDF parameterisation,
however, at the cost of an increased uncertainty.

The measurement of the triple-differential dijet cross section not only
provides constraints on the PDFs, but also on the strong coupling constant.
Therefore, the PDF fit is repeated with an additional free parameter: the strong
coupling constant \asmz. The value obtained for the strong coupling constant
is
\begin{equation*} \asmz =
0.1199\,\pm\,0.0015(\mathrm{exp})_{-0.0002}^{+0.0002}(\mathrm{mod})_{-0.0004}^{+0.0002}(\mathrm{par}),
\end{equation*}
where the quoted experimental (exp) uncertainty accounts for all
sources of uncertainties in the HERA and CMS data sets, as well as the
NP uncertainties. The model (mod) and parameterisation (par)
uncertainties are evaluated in the same way as in the PDF
determination. The consideration of scale uncertainties in a global
PDF fit is an open issue in the PDF community because it is unclear how to
deal with the correlations in scale settings among the different
measurements and observables. Therefore they are not taken into
account in any global PDF fit up to now,
although an elaborate study of the effect of scale settings on dijet
cross sections has been performed in Ref.~\cite{Watt:2013oha}, which
also reports first combined PDF and \asmz fits using LHC inclusive
jet data.
Following Ref.~\cite{Khachatryan:2014waa}, where the final
uncertainties and correlations of CMS inclusive jet data at 7\TeV
are used in such combined fits, two different methods to evaluate the
scale uncertainty of the jet cross section on \asmz are studied.
First, the renormalisation and factorisation scales are varied in the
calculation of the dijet predictions. The fit is repeated for each
variation. The uncertainty is evaluated as detailed in
Section~\ref{sec:theory} and yields $\Delta\asmz =
^{+0.0026}_{-0.0016}(\textrm{scale, refit})$.

The second procedure is analogous to the method applied by CMS in
previous determinations of \asmz without simultaneous PDF fits, cf.\
Refs.~\cite{Chatrchyan:2013txa, Khachatryan:2014waa, CMS:2014mna,
Khachatryan:2016mlc}. The PDFs are derived for a series of fixed
values of \asmz and the nominal choice of \mur and \muf.
Using this series, the best fit \asmz value of the
dijet data is determined for each scale variation. Here, the evaluated
uncertainty is $\Delta\asmz = ^{+0.0031}_{-0.0019}(\textrm{scale,
\asmz series})$.

{\tolerance=700
Both results, $\asmz = 0.1199 ^{+0.0015}_{-0.0016}$ (all except scale)
with $^{+0.0026}_{-0.0016}$ (scale, refit) and $^{+0.0031}_{-0.0019}$
(scale, \asmz series), are in agreement with
Ref.~\cite{Khachatryan:2014waa}, which reports $\asmz = 0.1192
^{+0.0023}_{-0.0019}$ (all except scale) and $^{+0.0022}_{-0.0009}$
(scale, refit) respectively $^{+0.0024}_{-0.0039}$ (scale, \asmz
series). Similarly, it is observed that the second procedure leads to
somewhat larger scale uncertainties, because there is less freedom for
compensating effects between different gluon distributions and the
\asmz values. Since this latter uncertainty is the most consistent to
be compared with previous fixed-PDF determinations of \asmz, it is
quoted as the main result. The dominant source of uncertainty is of
theoretical origin and arises due to missing higher order corrections,
whose effect is estimated by scale variations.
\par}

This value of \asmz is in agreement with the results from other
measurements by CMS~\cite{Chatrchyan:2013txa, Chatrchyan:2013haa,
  Khachatryan:2014waa, CMS:2014mna, Khachatryan:2016mlc} and
ATLAS~\cite{ATLAS:2015yaa}, with the value obtained in a similar
analysis complementing the DIS data of the HERAPDF2.0 fit with HERA
jet data~\cite{Abramowicz:2015mha}, and with the world average of
$\asmz=0.1181\pm 0.0011$~\cite{Patrignani:2016xmw}. In contrast to
the other CMS results, this analysis is mainly focused on PDF
constraints. The running of the strong coupling constant was tested
only indirectly via the renormalisation group equations. No explicit
test of the running was carried out by subdividing the phase space
into regions corresponding to different values of the renormalisation
scale.

\section{Summary}
\label{sec:summary}

A measurement of the triple-differential dijet cross section is
presented for $\sqrt{s}=8\TeV$. The data are found to be well described by NLO predictions
corrected for nonperturbative and electroweak effects,
except for highly boosted event topologies
that suffer from large uncertainties in parton distribution functions (PDFs).

The precise data constrain
the PDFs, especially in the highly boosted regime that probes
the highest fractions $x$ of the proton momentum carried by a parton.
The impact of the data on the PDFs is
demonstrated by performing a simultaneous fit to cross sections
of deep-inelastic scattering
obtained by the HERA experiments and the dijet cross section measured
in this analysis.
When including the dijet data, an increased gluon PDF at
high $x$ is obtained and the overall uncertainties of the PDFs,
especially those of the gluon distribution, are significantly reduced.
In contrast to a fit that uses inclusive jet data, this
measurement carries more information on the valence-quark content of
the proton such that a more flexible parameterisation is needed
to describe the low-$x$ behaviour of the
u~and d~valence quark PDFs. This higher sensitivity is accompanied by
slightly larger uncertainties in the valence quark distributions as a consequence
of the greater flexibility in the parameterisation of the PDFs.

{\tolerance=700
In a simultaneous fit the strong coupling constant \asmz is
extracted together with the PDFs. The value obtained at the
mass of the Z boson is
\begin{align*}
\ifthenelse{\boolean{cms@external}}{
\asmz &=
0.1199\,  \pm{0.0015}\,(\mathrm{exp})\\
 & \ \ \ \ \ \pm{0.0002}\,(\mathrm{mod})
 \,{}_{-0.0004}^{+0.0002}\,(\mathrm{par})\,
{}_{-0.0019}^{+0.0031}\,(\mathrm{scale})\\
&= 0.1199\, \pm{0.0015}\,(\mathrm{exp})\,
_{-0.0020}^{+0.0031}\,(\mathrm{theo}),
}
{
\asmz &=
0.1199\,\pm{0.0015}\,(\mathrm{exp})\,\pm{0.0002}\,(\mathrm{mod})
\,{}_{-0.0004}^{+0.0002}\,(\mathrm{par})\,
{}_{-0.0019}^{+0.0031}\,(\mathrm{scale})\\
&= 0.1199\,\pm{0.0015}\,(\mathrm{exp})\,
_{-0.0020}^{+0.0031}\,(\mathrm{theo}),
}
\end{align*}
and is in agreement with previous measurements at the LHC by
CMS~\cite{Chatrchyan:2013txa, Chatrchyan:2013haa, Khachatryan:2014waa,
CMS:2014mna, Khachatryan:2016mlc} and ATLAS~\cite{ATLAS:2015yaa},
and with the world average value of
$\asmz = 0.1181 \,\pm\, 0.0011$~\cite{Patrignani:2016xmw}.
The dominant uncertainty
is theoretical in nature and is expected to be reduced significantly
in the future using pQCD predictions at next-to-next-to-leading
order~\cite{Currie:2016bfm}.
\par}

\begin{acknowledgments}

We congratulate our colleagues in the CERN accelerator departments for the excellent performance of the LHC and thank the technical and administrative staffs at CERN and at other CMS institutes for their contributions to the success of the CMS effort. In addition, we gratefully acknowledge the computing centres and personnel of the Worldwide LHC Computing Grid for delivering so effectively the computing infrastructure essential to our analyses. Finally, we acknowledge the enduring support for the construction and operation of the LHC and the CMS detector provided by the following funding agencies: BMWFW and FWF (Austria); FNRS and FWO (Belgium); CNPq, CAPES, FAPERJ, and FAPESP (Brazil); MES (Bulgaria); CERN; CAS, MoST, and NSFC (China); COLCIENCIAS (Colombia); MSES and CSF (Croatia); RPF (Cyprus); SENESCYT (Ecuador); MoER, ERC IUT, and ERDF (Estonia); Academy of Finland, MEC, and HIP (Finland); CEA and CNRS/IN2P3 (France); BMBF, DFG, and HGF (Germany); GSRT (Greece); OTKA and NIH (Hungary); DAE and DST (India); IPM (Iran); SFI (Ireland); INFN (Italy); MSIP and NRF (Republic of Korea); LAS (Lithuania); MOE and UM (Malaysia); BUAP, CINVESTAV, CONACYT, LNS, SEP, and UASLP-FAI (Mexico); MBIE (New Zealand); PAEC (Pakistan); MSHE and NSC (Poland); FCT (Portugal); JINR (Dubna); MON, RosAtom, RAS, RFBR and RAEP (Russia); MESTD (Serbia); SEIDI, CPAN, PCTI and FEDER (Spain); Swiss Funding Agencies (Switzerland); MST (Taipei); ThEPCenter, IPST, STAR, and NSTDA (Thailand); TUBITAK and TAEK (Turkey); NASU and SFFR (Ukraine); STFC (United Kingdom); DOE and NSF (USA).

\hyphenation{Rachada-pisek} Individuals have received support from the Marie-Curie programme and the European Research Council and EPLANET (European Union); the Leventis Foundation; the A. P. Sloan Foundation; the Alexander von Humboldt Foundation; the Belgian Federal Science Policy Office; the Fonds pour la Formation \`a la Recherche dans l'Industrie et dans l'Agriculture (FRIA-Belgium); the Agentschap voor Innovatie door Wetenschap en Technologie (IWT-Belgium); the Ministry of Education, Youth and Sports (MEYS) of the Czech Republic; the Council of Science and Industrial Research, India; the HOMING PLUS programme of the Foundation for Polish Science, cofinanced from European Union, Regional Development Fund, the Mobility Plus programme of the Ministry of Science and Higher Education, the National Science Center (Poland), contracts Harmonia 2014/14/M/ST2/00428, Opus 2014/13/B/ST2/02543, 2014/15/B/ST2/03998, and 2015/19/B/ST2/02861, Sonata-bis 2012/07/E/ST2/01406; the National Priorities Research Program by Qatar National Research Fund; the Programa Clar\'in-COFUND del Principado de Asturias; the Thalis and Aristeia programmes cofinanced by EU-ESF and the Greek NSRF; the Rachadapisek Sompot Fund for Postdoctoral Fellowship, Chulalongkorn University and the Chulalongkorn Academic into Its 2nd Century Project Advancement Project (Thailand); and the Welch Foundation, contract C-1845.

\end{acknowledgments}

\bibliography{auto_generated}
\clearpage
\cleardoublepage \appendix\section{The CMS Collaboration \label{app:collab}}\begin{sloppypar}\hyphenpenalty=5000\widowpenalty=500\clubpenalty=5000\textbf{Yerevan Physics Institute,  Yerevan,  Armenia}\\*[0pt]
A.M.~Sirunyan, A.~Tumasyan
\vskip\cmsinstskip
\textbf{Institut f\"{u}r Hochenergiephysik,  Wien,  Austria}\\*[0pt]
W.~Adam, E.~Asilar, T.~Bergauer, J.~Brandstetter, E.~Brondolin, M.~Dragicevic, J.~Er\"{o}, M.~Flechl, M.~Friedl, R.~Fr\"{u}hwirth\cmsAuthorMark{1}, V.M.~Ghete, C.~Hartl, N.~H\"{o}rmann, J.~Hrubec, M.~Jeitler\cmsAuthorMark{1}, A.~K\"{o}nig, I.~Kr\"{a}tschmer, D.~Liko, T.~Matsushita, I.~Mikulec, D.~Rabady, N.~Rad, B.~Rahbaran, H.~Rohringer, J.~Schieck\cmsAuthorMark{1}, J.~Strauss, W.~Waltenberger, C.-E.~Wulz\cmsAuthorMark{1}
% ...
\vskip\cmsinstskip
% ...
1:~~Also at Vienna University of Technology, Vienna, Austria\\
% ...

\end{sloppypar}
\end{document}